\documentclass{article}

\usepackage{arxiv}

\usepackage[utf8]{inputenc} 
\usepackage[T1]{fontenc}    
\usepackage{hyperref}       
\usepackage{url}            
\usepackage{booktabs}       
\usepackage{amsfonts}       
\usepackage{nicefrac}       
\usepackage{microtype}      
\usepackage{lipsum}
\usepackage{amsmath}   
\usepackage{graphicx}
\usepackage{appendix}
\usepackage{amssymb}

\hypersetup{colorlinks=true,urlcolor=blue,citecolor=green}

\title{Revisiting cosmological diffusion models in Unimodular Gravity and the $H_0$ tension}

\author{
  Francisco X. Linares Cede\~no$^{a\ b\ \dagger}$ and Ulises Nucamendi$^{a\ b\ \dagger \dagger}$\\ 
  $^{a}$Instituto de Física y Matemáticas, Universidad Michoacana de San Nicolás de Hidalgo,\\ Edificio C-3, Ciudad Universitaria, CP. 58040 Morelia, Michoacán, México. \\ $^{b}$Mesoamerican Centre for Theoretical Physics, Universidad Autónoma de Chiapas,\\ Carretera Zapata Km 4, Real del Bosque (Terán), 29040, Tuxla Gutierrez, Chiapas, México. \\ $^{\dagger}$\texttt{francisco.linares@umich.mx}\, , $^{\dagger \dagger}$\texttt{unucamendi@gmail.com} \\
}

\begin{document}
\maketitle

\begin{abstract}
Within the framework of Unimodular Gravity, we consider non--gravitational interactions between dark matter and dark energy. Particularly, we describe such interactions in the dark sector by considering diffusion models that couple the cold dark matter fluid with the dark energy component, where the latter has the form of a variable cosmological ``constant''. For the first time, we solve the cosmological evolution of these models from the radiation dominated era to the present day. We show how the diffusion processes take place by analyzing the cosmological evolution of the energy density parameters $\Omega_{cdm}$ and $\Omega_{\Lambda}$, as well as that of the Hubble parameter. Finally, we perform the statistical analysis, imposing constraints on the diffusion parameters, by using data from Planck 2018, SH0ES, Pantheon, and H0LICOW collaborations. We found that cosmological diffusion models in the framework of Unimodular Gravity can ease the current tension in the value of $H_0$. We also show that the very far future cosmological evolution for all diffusion models is eternally accelerating without future singularities.
\end{abstract}


\section{Introduction}

After more than 100 years, General Relativity (GR) remains as the theory successfully describing the gravitational interactions~\cite{Will:2014kxa,Will:2018lln}. In the cosmological context, GR offers a mathematical ground that has allowed to describe the evolution of the Universe, from the era when the first atomic nuclei form to the current phase of accelerated expansion. Such description lies within the standard cosmological model $\Lambda$CDM, which demands, however, a new particle for the so-called \textit{Cold Dark Matter} (CDM) component, responsible for the structure formation process, and which interacts mostly gravitationally with the rest of the known particles. The other ingredient of this model is the \textit{Cosmological Constant} $\Lambda$, which is required to explain the current accelerated expansion of the Universe. Thus, GR is a successful gravitational framework to describe the evolution of the Universe as long as a new \textit{dark sector} (dark matter particles as well as a cosmological constant) is added. 

With the aim of gaining some theoretical understanding of what these new components of the Universe might be, other approaches have been proposed, such as modifying or including new contributions to the Einstein-Hilbert action. In particular, one of the main motivations to consider alternative approaches to GR, is the observed current phase of accelerated expansion of the Universe~\cite{Riess:1998cb,Perlmutter:1998np,Eisenstein:2005su,Parkinson:2012vd,Aghanim:2018eyx}, which is thought it is produced by the so-called \textit{Dark Energy}  (DE). Many models to explain DE have been proposed, such as fluids with variable equation of state~\cite{Carturan:2002si,Cardone:2005ut,Nojiri:2006zh,Brevik:2007jt,Linder:2008ya,Duan:2011jj,Bini:2014pmk,Barrera-Hinojosa:2019yyh}, scalar fields~\cite{Caldwell:1999ew,Caldwell:2003vq,Nojiri:2005sx,Feng:2006ya,Linder:2007wa,Setare:2008sf,Tsujikawa:2013fta,Chiba:2012cb,Linde:2015uga,Linder:2015qxa,Durrive:2018quo,Bag:2017vjp,Leon:2018lnd,Garcia-Garcia:2018hlc,Alestas:2020mvb}, modified gravity~\cite{Lobo:2008sg,Clifton:2011jh,Dimitrijevic:2012kb,Brax:2015cla,Joyce:2016vqv,Jaime:2018ftn,Slosar:2019flp}. However, it seems that the most favored explanation, according to several observations, is the cosmological constant term $\Lambda$ in the Einstein field equations,
\begin{equation}
    R_{\mu \nu} - \frac{1}{2}Rg_{\mu \nu} + \Lambda g_{\mu \nu} = \kappa^{2}T_{\mu \nu}\, ,
    \label{efe}
\end{equation}
where $\kappa^2=8\pi G$. Historically, with the aim of having a quasi-static matter distribution in the Universe, Einstein introduced by-hand the term $\Lambda$ in his field equations~\cite{Einstein:1917ce}. After Edwin Hubble's observations of how distant galaxies are receding away from us~\cite{Hubble:1929ig}, and later with the discovery of the accelerating expansion of the Universe through supernovae observations~\cite{Riess:1998cb,Perlmutter:1998np}, the cosmological constant plays the role of the DE component responsible for such accelerated expansion that the Universe is currently experiencing. Its origin and physical interpretation have been debated and speculated since then, giving rise to the different approaches mentioned above.

One of the proposals for a possible origin of $\Lambda$ that is being currently studied with great interest, is related to the original formulation of the field equations of GR, where Einstein showed that it is always possible to consider a choice of coordinate such that the determinant of the metric tensor is fixed~\cite{Einstein:1914bx}. Specifically, when the determinant $g$ of the metric tensor $g_{\mu \nu}$ satisfies the \textit{unimodular condition} $\sqrt{-g} = 1$, the Einstein tensor gets a simplified form. Later, in trying to understand the role of gravitational forces in the constitution of matter, Einstein showed that in a formulation of the field equations freed of the scalar of curvature $R\equiv g^{\mu \nu} R_{\mu \nu}$, the cosmological constant term can arise as a constant of integration~\cite{Einstein:1919gv}.

 A link between the condition of a fixed metric determinant and the cosmological constant was made for the first time in~\cite{Anderson:1971pn}, where the authors consider unimodular coordinate mappings, this is,  $x^{\mu}\rightarrow x^{\prime \mu}$ for which ${\rm{det}}\mid \partial x^{\prime \mu}/\partial x^{\nu} \mid = 1$. Then, with the aim of building a theory considering such coordinate transformations, they add the unimodular condition to the Einstein-Hilbert action through a Lagrange multiplier $\lambda$,
\begin{equation}
    S = S_{EH} + S_{\lambda} = \frac{1}{\kappa^2}\int d^4x\sqrt{-g}R + \frac{1}{\kappa^2}\int d^4x \lambda(x)\left( \sqrt{-g} - f \right)\, ,
    \label{action}
\end{equation}
where $f=f(x)$ is a scalar density. Such scalar densities arise naturally when the theory is formulated in the language of differential forms, as it is shown in~\cite{Bonder:2018mfz}, where the authors present a formulation in which the spin density of matter fields is also included in the analysis as a source of spacetime torsion. Therefore, the full diffeomorphism invariance of GR is lost due to the term given by the fixed volume element $fd^4x$, and the action~\eqref{action} is invariant under the volume-preserving diffeomorphisms. More details on these kind of symmetries in gauge theories of gravity can be found in~\cite{Corral:2018hxi}. Variation with respect to $\lambda$ gives the constraint $\sqrt{-g} = f$, whereas variation with respect to the inverse of the metric tensor $g^{\mu \nu}$ leads to
\begin{equation}
    R_{\mu \nu} - \frac{1}{2}Rg_{\mu \nu} + \lambda g_{\mu \nu} = 0\, .
    \label{efe_lambda}
\end{equation}

Taking the trace of Eq.~\eqref{efe_lambda}, the Lagrange multiplier is determined to be
\begin{equation}
    \lambda = \frac{1}{4}R\, ,
    \label{lmr}
\end{equation}
and therefore, Eq.~\eqref{efe_lambda} is written as
\begin{equation}
    R_{\mu \nu} - \frac{1}{4}Rg_{\mu \nu} = 0\, .
    \label{efe_lambda_traceless}
\end{equation}

When considering matter content $S_M$ in the action~\eqref{action}, Eqs.~\eqref{efe_lambda} and~\eqref{lmr} are given by
\begin{equation}
    R_{\mu \nu} - \frac{1}{2}Rg_{\mu \nu} + \lambda g_{\mu \nu} = \kappa^2 T_{\mu \nu}\, ,\quad \lambda = \frac{1}{4}\left( R + \kappa^2 T  \right)\, ,
    \label{efe_lambda_matter}
\end{equation}
where $T_{\mu \nu}=-2g^{-1/2}\delta S_M/\delta g^{\mu \nu}$ is the energy--momentum tensor, and $T$ its trace. After eliminating $\lambda$, we have
\begin{equation}
    R_{\mu \nu} - \frac{1}{4}Rg_{\mu \nu} = \kappa^2 \left(T_{\mu \nu} - \frac{1}{4}Tg_{\mu \nu}\right)\, ,
    \label{efe_ug}
\end{equation}
which is the trace--free version of the Einstein field equations. This theory leading to the new set of equations~\eqref{efe_ug} for the gravitational field has been dubbed \textit{Unimodular Gravity} (UG). The first works in UG are~\cite{Anderson:1971pn,vanderBij:1981ym,Weinberg:1988cp,Buchmuller:1988wx,Buchmuller:1988yn,Ng:1990xz}, approaches of UG in the quantum regime have been explored by~\cite{Smolin:2009ti,Eichhorn:2013xr,Alvarez:2015pla,Bufalo:2015wda,Percacci:2017fsy}, and some recent cosmological applications have been studied in~\cite{Jain:2012gc,Ellis:2013uxa,Barvinsky:2017pmm,Garcia-Aspeitia:2019yni,Garcia-Aspeitia:2019yod,Barvinsky:2019qzx,Barvinsky:2019agh,Perez:2020cwa,Corral:2020lxt}. An approach within the Hamiltonian formalism can be found in \cite{Henneaux:1984ji,Henneaux:1989zc}. On the other hand, a direct equivalence between these works mentioned above and UG was made in \cite{Kuchar:1991xd}, whereas a generally-covariant construction of such works was developed in \cite{Jirousek:2018ago,Hammer:2020dqp}. It has been shown that, such formulations lead to the same degrees of freedom as those in GR, either by virtue of additional constraints or by using of auxiliary fields. Notice that, as the GR case, there are 10 equations to be solved: 9 Einstein-like equations (since Eq.~\eqref{efe_ug} is a symmetric but traceless equation), and one equation from the unimodular condition $\sqrt{-g} = f$. We can focus only in the first 9 equations, without forgetting that the metric in the equations of motion of the matter fields is subject to the constraint imposed by the unimodular condition.

One of the main features of this theory can be seen as follows: notice that we can rewrite Eq.~\eqref{efe_ug} in the following way,
\begin{equation}
    R^{\mu}_{\ \ \nu} - \frac{1}{2}R\delta^{\mu}_{\ \ \nu} + \frac{1}{4}\left(R + \kappa^2T\right)\delta^{\mu}_{\nu} = \kappa^2 T^{\mu}_{\ \ \nu}\, ,
\end{equation}
and then, applying the Bianchi identities we have
\begin{equation}
    \nabla_{\mu}\left( R^{\mu}_{\ \ \nu} - \frac{1}{2}R\delta^{\mu}_{\ \ \nu} \right) + \frac{1}{4}\nabla_{\nu}\left(R + \kappa^2T\right) = \kappa^2 \nabla_{\mu}T^{\mu}_{\ \ \nu}\, .
\end{equation}

Whereas the first term still is identically zero (as in GR), the covariant derivative of the energy--momentum tensor is no longer, in general, locally conserved,
\begin{equation}\label{eff_cc}
    \kappa^2\nabla_{\mu}T^{\mu}_{\ \ \nu} = \frac{1}{4}\partial_{\nu}\left(R + \kappa^2T\right)\quad \Rightarrow \quad \lambda(x)\equiv \Lambda +  \int_l J \, ,
\end{equation}
where $\Lambda$ is a constant of integration, and $J_{\nu}\equiv \kappa^2\nabla_{\mu}T^{\mu}_{\ \ \nu}$ is the energy--momentum current violation to be integrated on some arbitrary path $l$. Replacing this result into Eq.~\eqref{efe_lambda_matter}, we have
\begin{equation}
    R_{\mu \nu} - \frac{1}{2}Rg_{\mu \nu} + \left( \Lambda + \int_l J(x) \right) g_{\mu \nu} = \kappa^2 T_{\mu \nu}\, .
    \label{EFE_UG_gen}
\end{equation}

The physical interpretation of the Lagrange multiplier $\lambda$ is now apparent: it plays the role of an effective cosmological ``constant''. In fact, in the particular case when the energy--momentum tensor is conserved ($J = 0$), the integration constant $\Lambda$ is identified as the cosmological constant term in the Einstein field equations~\eqref{efe}. Thus, within the framework of UG, the cosmological constant $\Lambda$ is not a term introduced by--hand, but it arises naturally as an integration constant when considering the Einstein--Hilbert action with volume-preserving diffeomorphisms\footnote{Theoretical analysis on the viability of UG for astrophysical and cosmological applications, as well as discussions about the integrability of $J_{\nu}$ are discussed in~\cite{Ellis:2010uc} and~\cite{Josset:2016vrq} respectively.}. However, there will be in general a non--null energy--momentum current violation. Thus, assuming that ordinary matter (photons, neutrinos, baryons) interacts only gravitationally with the dark sector, the non-conservation of the energy--momentum tensor leads to a non-gravitational interaction between cold dark matter and the cosmological constant.

\begin{center}
\begin{figure}[htp!]
    \centering
   \includegraphics[width=0.7\linewidth]{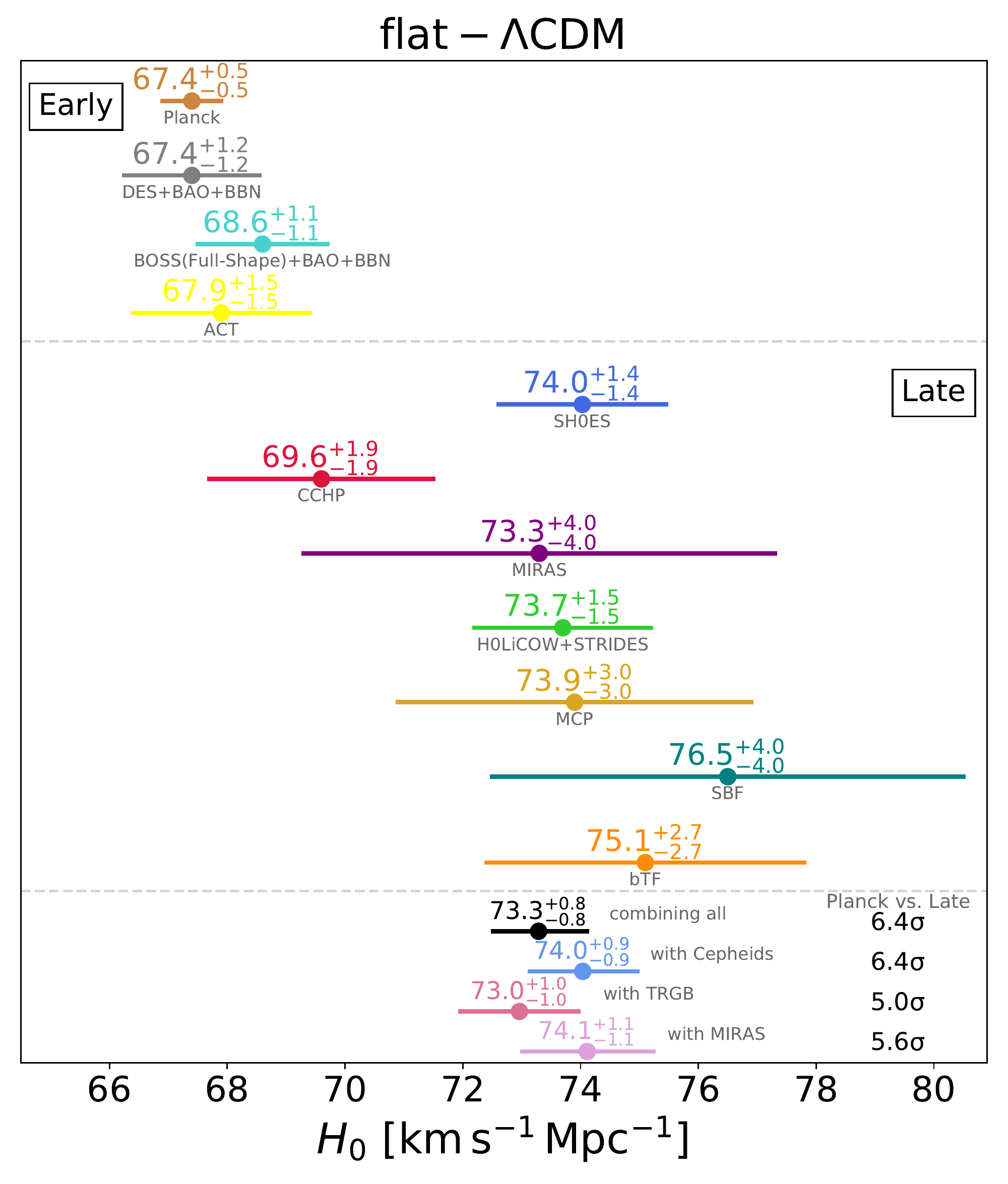}
    \caption{Value of $H_0$ from different observations. Top: $H_0$ from CMB observations. Middle: $H_0$ from late time observations. Bottom: combined late time observations and the corresponding $H_0$ tension with early time measurements. Data from CMB~\cite{Aghanim:2018eyx,Aiola:2020azj}, BAO and BBN~\cite{Abbott:2017smn,Philcox:2020vvt}, SNe Ia and Cepheids~\cite{Riess:2019cxk}, SNe Ia and TRGB~\cite{Freedman:2019jwv,Freedman:2020dne}, SNe Ia and Mira variables~\cite{Huang:2019yhh}, lensed quasars~\cite{Wong:2019kwg}, water megamasers~\cite{Pesce:2020xfe}, SBF and Cepheids~\cite{potter2018calibrating}. This is an updated version of Figure 1 from~\cite{Verde:2019ivm}. Credits to Vivien Bonvin and Martin Millon~\cite{vivien_bonvin_2020_3635517}.}
    \label{h0_tension_all}
\end{figure}
\end{center}

Interactions between the components of the dark sector have been studied broadly in the literature~\cite{Amendola:1999dr,Amendola:1999er,Billyard:2000bh,Olivares:2005tb,Olivares:2007rt,CalderaCabral:2008bx,LopezHonorez:2010ij,Pan:2012ki,Salvatelli:2013wra,Bolotin:2013jpa,Pan:2016ngu,CarrilloGonzalez:2017cll,Barros:2018efl,Yang:2019uzo,Benetti:2019lxu,Yang:2020tax,Asghari:2020ffe,Johnson:2020gzn,Benetti:2021div}, and they have shown to be useful as alternative models to $\Lambda$CDM, to address the discrepancies found in the measurements of the current value of the Hubble parameter $H_0$, when it is inferred from early and late Universe observations~\cite{DiValentino:2017iww,Yang:2018uae,Yang:2018euj,DiValentino:2019ffd,Pan:2019gop,Gomez-Valent:2020mqn,Lucca:2020zjb,Pan:2020bur}. We will refer to such discrepancy as the $H_0$ \textit{tension} (for recent discussions see~\cite{Freedman:2017yms,Verde:2019ivm}). In Figure~\ref{h0_tension_all} it can be seen the current value of the Hubble parameter according different experiments. Whereas from observations of the early Universe, based on the Cosmic Microwave Background (CMB), a value of $H_0^{early}\simeq 68$ km s$^{-1}$Mpc$^{-1}$ is inferred, the combined late Universe observations indicate that $H_0^{late} \simeq 73$ km s$^{-1}$Mpc$^{-1}$. This discrepancy may be due to systematic errors in the observations, but it can also be suggesting to look for extensions of the $\Lambda$CDM model with the aim of exploring new physics.

Non-gravitational interactions between dark matter and dark energy have been studied through \textit{diffusion models} in the cosmological context within the framework of General Relativity~\cite{Calogero:2013zba,Haba:2016swv}, as well as in the context of Two Measure Theories and dynamical spacetime theories~\cite{Benisty:2017rbw,Benisty:2017eqh,Benisty:2018oyy}. We are particularly interested in diffusion models as those presented in~\cite{Perez:2020cwa,Corral:2020lxt}, where the authors explore cosmological diffusion processes in the framework of Unimodular Gravity. Specifically, we will go further in the analysis by studying these models not only at late times, but from very early times deep within the radiation dominated era. This requires to consider the radiation component into account, which will be important in order to constraint the diffusion models we are interested in with CMB data.

Thus, in this work we will study diffusion processes between the dark sector components within the framework of Unimodular Gravity. For the first time, these diffusion models are studied in a more realistic cosmological scenario, in which the component of radiation due to the presence of photons and ultra--relativistic neutrinos at early times, is taken into account. Since we will be focused on how these models can alleviate the $H_0$ tension, it is important to include the radiation component in order to use data from the CMB, and thus being able to test each diffusion model at the last scattering surface ($z_*\simeq 1100$). By performing a statistical analysis, we will infer the most likely values of the diffusion parameters in the light of current astrophysical and cosmological data. This will allow us to analyze the viability of such diffusion models as a possible solution to the $H_0$ tension. As we will show, the inferred value of $H_0$ at early times can be in agreement with that obtained from late time observations.

The outline of the present work is the following: in Section~\ref{back} we show the cosmological equations for the background evolution within the framework of UG. We analyze the cosmological evolution for each of the diffusion models of interest, studying the diffusion process in terms of the energy density parameters $\Omega_{cdm}$ and $\Omega_{\Lambda}$. Considering that the current values for each energy density parameter $\Omega_{0,i}$, and for $H_0$ are those determined by CMB observations (and thus $H_0 = H_0^{early}$ for $\Lambda$CDM), we show how the evolution of the Hubble parameter $H(z)$ for each diffusion model gives a current value $H_0$ consistent with the reported value from local observations, i.e., $H_0 = H_0^{late}$. In particular, when the parameters of the diffusion models are set to zero, we recover the $\Lambda$CDM results. We perform the statistical analysis in Section~\ref{stat}, where we: 1) consider only CMB observations from the Planck Compressed 2018 data, and then 2) we include in the analysis observations from the local Universe as those from Cepheids, Supernovae and lensed quasars. In the first analysis, the tension on the $H_0$ value is eased because the anticorrelation between $\Omega_{cdm}$ and $H_0$ gets a broader range of values due to the presence of the new diffusion parameters. In the second analysis, we obtain that the local observations allow the diffusion models to ease the $H_0$ tension by shifting the mean value from $H_0^{early}$ to $H_0^{late}$, and thus, the diffusion models make CMB and late time observations to be in agreement on the value of the Hubble parameter at $z=0$. Finally, in Section~\ref{final} we discuss our results and give some conclusions of our analysis.

\section{Background cosmological equations with diffusion}\label{back}

We start by considering a spatially-flat Friedmann-Robertson-Walker (FRW) line element,
\begin{equation}
ds^2 = -dt^2 + a^2(t)\left[ dr^2 + r^2\left(d\theta^2 + \sin^2\theta d\phi^2\right) \right]\, ,
\label{frw}
\end{equation}
where the scale factor $a(t)$ is function of the cosmic time $t$. When considering the FRW line element~\eqref{frw}, we are assuming that at large scales the Cosmological Principle is valid, and then both homogeneity and isotropy imply that the effective cosmological constant~\eqref{eff_cc} is now a function of the cosmic time only,
\begin{equation}\label{eff_cc_time}
    \Lambda(t)\equiv \Lambda +  \int_l J(t) \, ,
\end{equation}
where we have replaced the notation $\lambda(t) \rightarrow \Lambda(t)$.
The Einstein field equations~\eqref{EFE_UG_gen} for the background evolution are given by,
\begin{subequations}\label{bg}
\begin{eqnarray}
    H^2 &=& \frac{\kappa^2}{3} \left( \rho_{\gamma} + \rho_{\nu} + \rho_{b} + \rho_{cdm} + \rho_{\Lambda(t)} \right) \, ,\label{friedmann}\\
    \dot{H} &=& - \frac{\kappa^2}{2}\left[ (\rho_{\gamma} +
      p_{\gamma} ) + (\rho_{\nu} +
      p_{\nu} ) + (\rho_{b} +
      p_{b} ) + (\rho_{cdm} +
      p_{cdm} ) \right]\, ,\label{fried_acc} \\
      \dot{\rho}_{\gamma} &=& - 3 H (\rho_{\gamma} + p_{\gamma} )\, ,\quad \dot{\rho}_{\nu} = - 3 H (\rho_{\nu} + p_{\nu} )\, ,\quad \dot{\rho}_{b} = - 3 H (\rho_{b} + p_{b} )\, , \\
      \dot{\rho}_{cdm} &=& - 3 H (\rho_{cdm} + p_{cdm} ) - \frac{\dot{\Lambda}(t)}{\kappa^2}\, .\label{diff_eq}
\end{eqnarray}
\end{subequations}

The dot denotes derivative with respect to cosmic time $t$, and $H=\dot{a}/a$ is the Hubble parameter. We will consider that baryons ($b$) and cold dark matter ($cdm$) behave as dust, and then they have vanishing pressure $p_b = p_{cdm} = 0$, whereas for photons ($\gamma$) and ultra--relativistic neutrinos ($\nu$) we have $p_{\gamma} = \rho_{\gamma}/3$ and $p_{\nu} = \rho_{\nu}/3$.

\subsection{Diffusion models}
Given a particular form of the cold dark matter energy density $\rho_{cdm}(t)$, Eq.~\eqref{diff_eq} can be integrated to find the function $\Lambda(t)$. With this approach, here we analyze two diffusion models previously studied in~\cite{Perez:2020cwa}: \textit{Sudden Transfer Model} (STM) and \textit{Anomalous Decay of the Matter Density} (ADMD). Both models are described in terms of two parameters, one regulating the amplitudes of $\rho_{cdm}$ and $\rho_{\Lambda}$ ($\alpha$ for STM and $\gamma$ for ADMD), and another one for the characteristic redshift $z^{\star}$ at which the diffusion process takes place.

Another way to solve~\eqref{diff_eq} is by explicitly proposing a particular form for the \textit{diffusion function} $Q$, which is defined as follows, 
\begin{equation}
    Q(x) \equiv \frac{1}{\kappa^2}\int_l J(x)\, ,
\end{equation}
and thus, the integrated energy--momentum current violation is expressed in terms of the diffusion function $Q(x)$. Given homogeneity and isotropy, this function will depend only on the cosmic time $t$. Therefore, what we have denoted by $\Lambda(t)$ in Eq.~\eqref{eff_cc_time} can now be written as
\begin{equation}
    \Lambda(t) \equiv \Lambda + \kappa^2Q(t)\, ,
\end{equation}
in whose case, Eq.~\eqref{diff_eq} reads
\begin{equation}
    \dot{\rho}_{cdm} = - 3 H (\rho_{cdm} + p_{cdm} ) - \dot{Q}(t)\, .\label{diff_eq_Q}
\end{equation}

An explicit functional form for the diffusion function $Q$ can be given, in order to find the CDM energy density by solving Eq.~\eqref{diff_eq_Q}. This was considered by the authors in~\cite{Corral:2020lxt}, where two phenomenological models are studied: \textit{Barotropic Model} (BM) and \textit{Continuous Spontaneous Localization} (CSL). Different from the previous two models, only one parameter characterizes the BM and CSL models ($x_{cdm}$ for BM and $\xi_{CSL}$ for CSL).

In this Section we will analyze the cosmological evolution of these four models mentioned above, from the radiation dominated era until the present day. To do so, we will solve the background equations~\eqref{bg} with the Boltzmann code \textsc{class}~\cite{Lesgourgues:2011re}. 

\subsubsection{Model 1: Sudden Transfer Model}
The energy density proposed for the CDM component is given by
\begin{subequations}
\begin{equation}
    \rho_{cdm}(z) = \rho_{0, cdm}(1+z)^3\times \left\lbrace \begin{matrix} 
1 & {\rm{if}}\quad z\geq z^{\star}\, , \\
& \\
1 - \alpha & {\rm{if}}\quad z< z^{\star}\, ,
\end{matrix}\right.
\label{rho_diff_1}
\end{equation}
where $\alpha$ is the dimensionless diffusion constant controlling the amplitude of the energy transfer, and $z^{\star}$ is the characteristic redshift at which the sudden diffusion process takes place. After integration of Eq.~\eqref{diff_eq} (with $p_{cdm} = 0$), the effective cosmological constant is given by
\begin{equation}
    \Lambda(z) = \left\lbrace \begin{matrix} 
\Lambda & {\rm{if}}\quad z\geq z^{\star}\, , \\
 & \\
\Lambda + 3H_0^2\alpha(1+z^{\star})^3\Omega_{0, cdm} & {\rm{if}}\quad z< z^{\star}\, ,
\end{matrix}\right.
\label{Lambda_diff_1}
\end{equation}
and then, once $z<z^{\star}$, the Friedmann equation~\eqref{friedmann} can be written as
\begin{equation}\label{ez_1}
    E(z)\equiv \frac{H(z)}{H_0} =\sqrt{ \Omega_{0,r}(1+z)^4 + \Omega_b(1+z)^3 + \Omega_{cdm}(1+z)^3\left[ 1 - \alpha + \alpha\left(\frac{1+z^{\star}}{1+z}\right)^3 \right] + \Omega_{\Lambda}}\, .
\end{equation}
\end{subequations}

Thus, the diffusion process allows to have different present values of $\Omega_{\Lambda}$ and $\Omega_{cdm}$ for given values of $\alpha$ and $z^{\star}$. This implies that the current Hubble parameter $H_0$ can also change its value since it is defined implicitly in the energy density parameters. It can be seen that the standard Friedmann equation is recovered when $\alpha = 0$.

\begin{center}
\begin{figure}[htp!]
    \centering
   \includegraphics[width=0.6\linewidth]{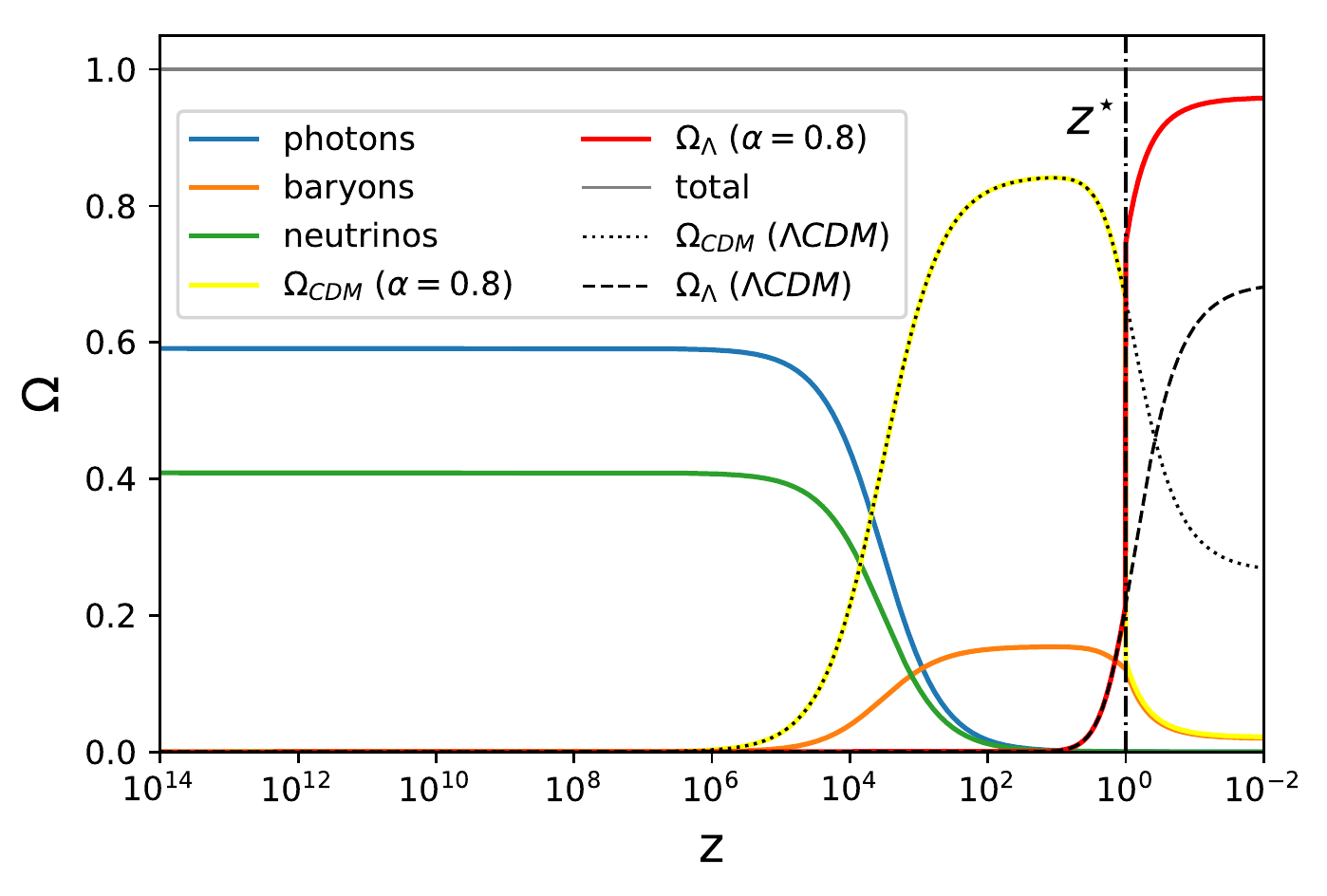}
    \caption{Sudden diffusion between $\rho_{cdm}$ and $\Lambda$ (Model 1). Whereas photons (blue line), baryons (orange line) and neutrinos (green line) evolve as usual, $\Omega_{cdm}$ (yellow line) and $\Omega_{\Lambda}$ (red line) evolve by a diffusion process for which $\alpha = 0.8$. The characteristic redshift was settled to $z^{\star} = 1$ (vertical dash--dotted line).}
    \label{omegas_diff_1}
\end{figure}
\end{center}

In order to illustrate the diffusion process between CDM and $\Lambda$, we have considered the diffusion parameters given by $\left\lbrace \alpha, z^{\star} \right\rbrace = \left\lbrace 0.8, 1 \right\rbrace$. As expected, the energy density of cold dark matter (yellow line) drops suddenly at $z^{\star}$ (vertical dash--dotted black line), and the energy density for $\Lambda$ (red line) increases, as it is shown in Figure~\ref{omegas_diff_1}. The rest of the matter components evolve as usual. Only for comparison, we also have included the standard evolution of CDM and $\Lambda$ (black dotted and dashed lines respectively). The Friedman constraint is satisfied during all the cosmological evolution (horizontal gray line), this is, $1 = \sum_i\Omega_i(z)\, .$ 

\begin{center}
\begin{figure}[htp!]
    \centering
   \includegraphics[width=0.6\linewidth]{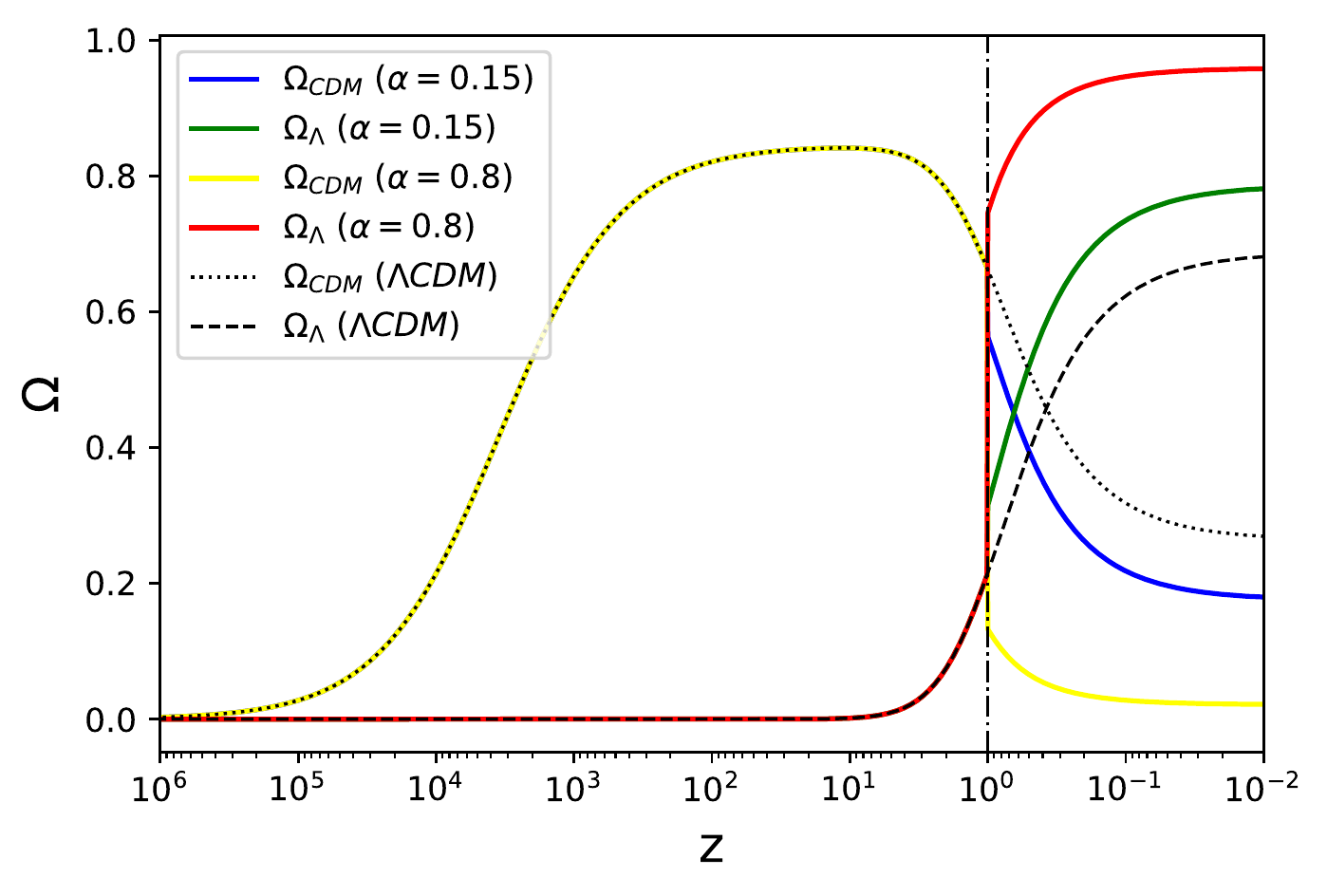}
    \caption{Sudden diffusion between $\rho_{cdm}$ and $\Lambda$ (Model 1). Black lines show the standard evolution without diffusion ($\alpha = 0$) for CDM (dotted line) and cosmological constant (dashed line). Other lines correspond to a diffusion process with $\alpha = 0.15$ (blue and green lines for CDM and $\Lambda$ respectively), and $\alpha = 0.8$ (yellow and red lines for CDM and $\Lambda$ respectively). The characteristic redshift was fixed to $z^{\star} = 1$ (vertical dash--dotted black line).}
    \label{omegas_diff_alphas}
\end{figure}
\end{center}

\begin{center}
\begin{figure}[htp!]
    \centering
   \includegraphics[width=0.6\linewidth]{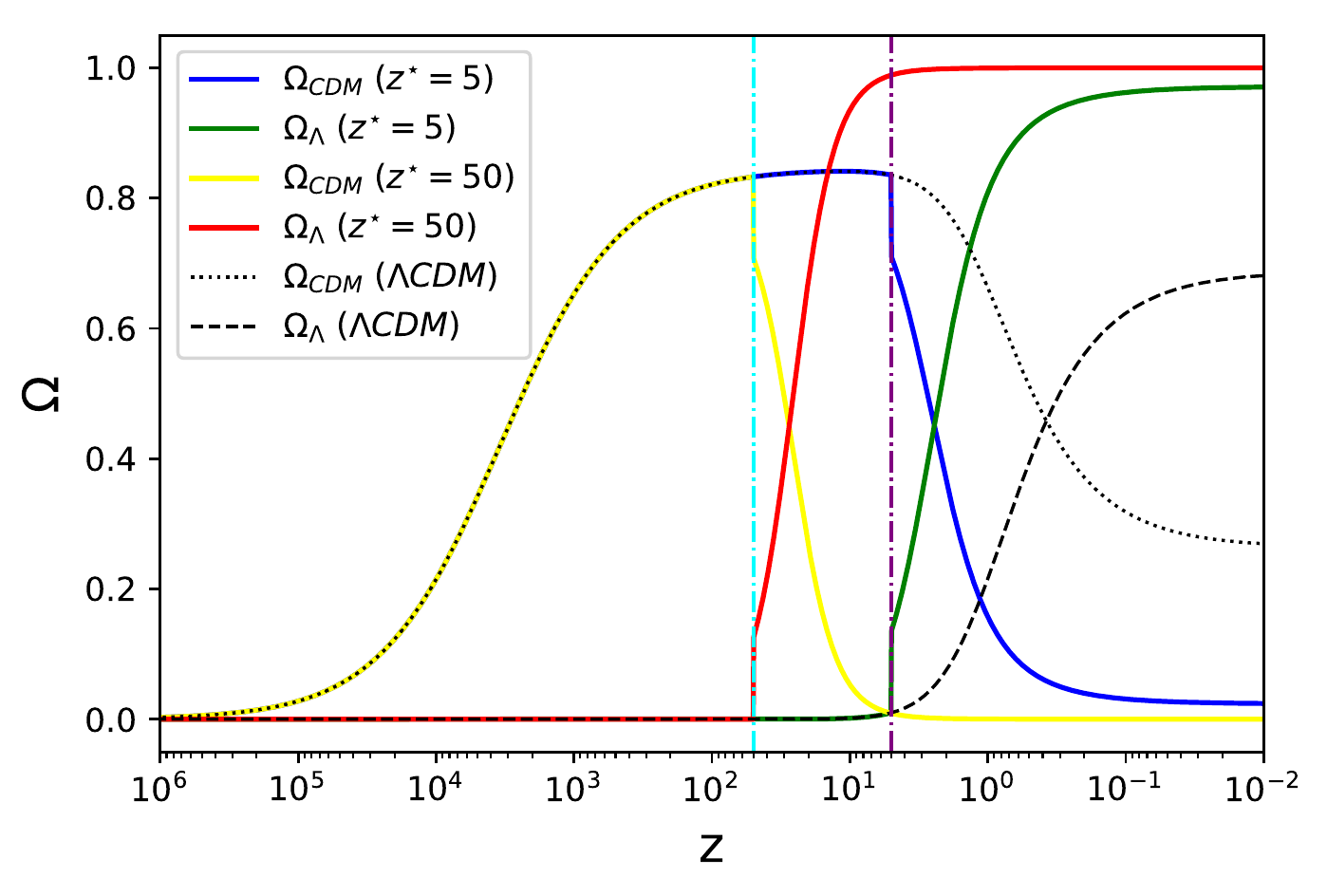}
    \caption{Sudden diffusion between $\rho_{cdm}$ and $\Lambda$ (Model 1) at different characteristic redshifts: $z^{\star} = 5$ (vertical dash--dotted purple line), and $z^{\star} = 50$ (vertical dash--dotted cyan line). Black lines show the standard evolution without diffusion ($\alpha = 0$) for CDM (dotted line) and cosmological constant (dashed line)). For the diffusion processes between CDM and $\Lambda$ at $z
   ^{\star} = 5$ (blue and green lines), as well as $z
   ^{\star} = 50$ (yellow and red lines), the diffusion constant was settled to $\alpha = 0.15\, .$}
    \label{omegas_diff_z}
\end{figure}
\end{center}

With the aim of exploring the effects of the diffusion process between the dark sector components, different cosmological evolution with several values of the diffusion constant $\alpha$ are shown in Figure~\ref{omegas_diff_alphas}, where the sudden energy transfer between $\Omega_{cdm}$ and $\Omega_{\Lambda}$ can be observed. Particularly, it can be seen that given a characteristic redshift $z^{\star}=1$ (vertical black line), the diffusion constant $\alpha$ will change the difference between the current values of the energy density parameters for the dark sector. This is expected since $\alpha$ regulates the amount of energy density that the CDM component transfers to the cosmological constant (see
Eq.~\eqref{rho_diff_1} and~\eqref{Lambda_diff_1}). On the other hand, the effect of different characteristic redshifts $z^{\star}$ on the diffusion process can be seen in Figure~\ref{omegas_diff_z}. We observe that higher values of $z^{\star}$ will lead to lower (larger) values of the cold dark matter (cosmological constant) energy density parameter today. 

\subsubsection{Model 2: Anomalous Decay of the Matter Density}
\begin{center}
\begin{figure}[htp!]
    \centering
   \includegraphics[width=0.6\linewidth]{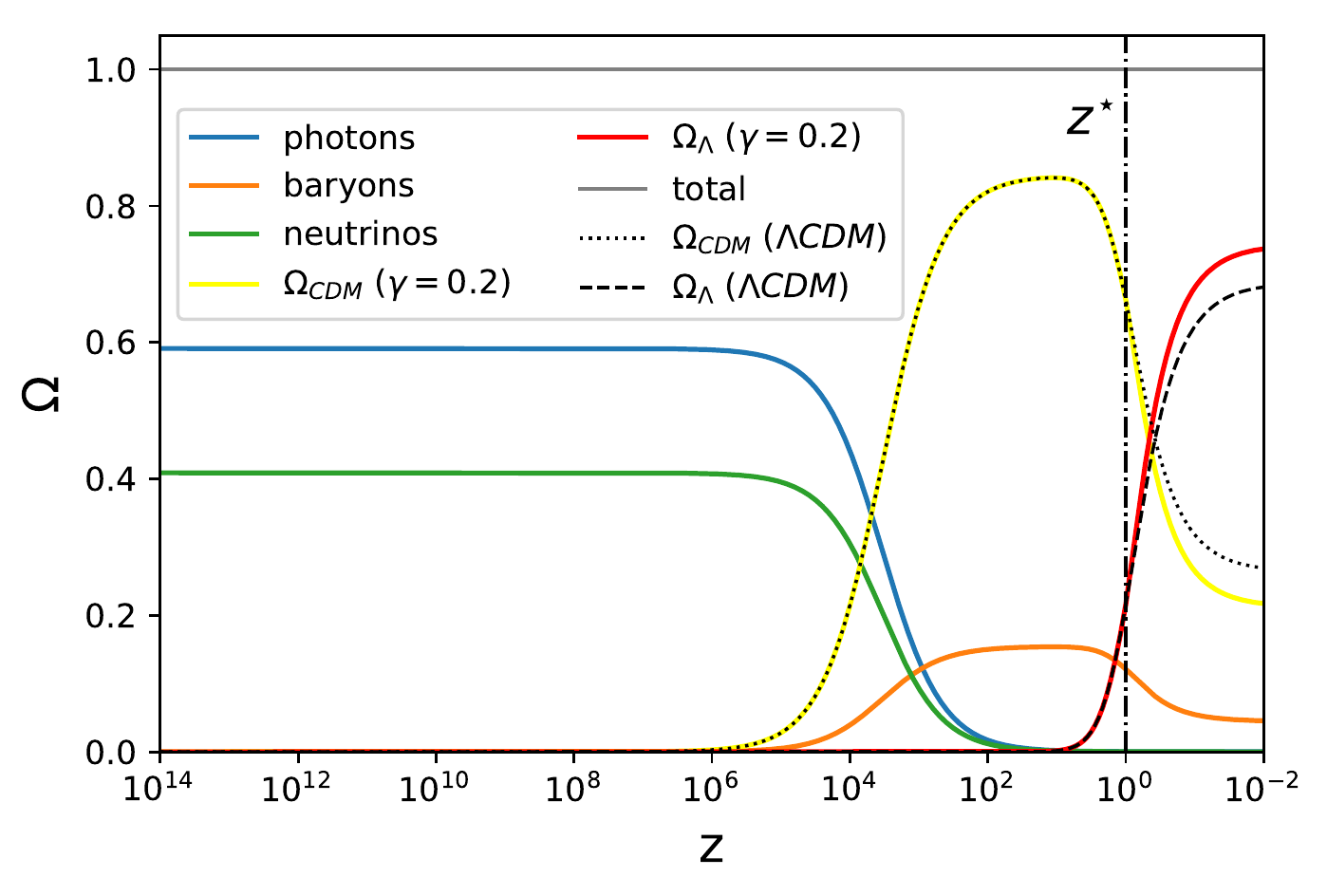}
    \caption{ADMD model for the diffusion between $\rho_{cdm}$ and $\Lambda$ (Model 2). Photons (blue line), baryons (orange line) and neutrinos (green line) evolve as usual, whereas $\Omega_{cdm}$ (yellow line) and $\Omega_{\Lambda}$ (red line) evolve by a diffusion process for which $\gamma = 0.2$. The characteristic redshift was set to $z^{\star} = 1$ (vertical dashdotted line). The horizontal gray line indicates the Friedman constraint $\sum_i\Omega_i(z) =1$.}
    \label{omegas_diff_2}
\end{figure}
\end{center}

In this case, the mathematical model for the CDM energy density is proposed to be
\begin{subequations}
\begin{equation}
    \rho_{cdm}(z) = \rho_{0, cdm}(1+z)^3\times \left\lbrace \begin{matrix} 
1 & {\rm{if}}\quad z\geq z^{\star}\, , \\
& \\
\left( \frac{1 + z}{1 + z^{\star}} \right)^\gamma & {\rm{if}}\quad z< z^{\star}\, ,
\end{matrix}\right.
\label{rho_diff_2}
\end{equation}
where $\gamma$ is the dimensionless diffusion constant controlling the power of the energy transfer term, and $z^{\star}$ is again the characteristic redshift, this time indicating when the anomalous decay of dark matter occurs. After integration of Eq.~\eqref{diff_eq}, the energy density for $\Lambda$ is given by
\begin{equation}
    \Lambda(z) = \left\lbrace \begin{matrix} 
\Lambda & {\rm{if}}\quad z\geq z^{\star}\, , \\
 & \\
\Lambda - \frac{3\gamma}{\gamma + 3}H_0^2\left[\left( \frac{1 + z}{1 + z^{\star}} \right)^{\gamma}(1+z)^3 - (1+z^{\star})^3\right]\Omega_{0, cdm} & {\rm{if}}\quad z< z^{\star}\, .
\end{matrix}\right.
\label{Lambda_diff_2}
\end{equation}

Once $z<z^{\star}$, the Friedmann equation~\eqref{friedmann} can be written as
\begin{equation}\label{ez_2}
    E(z)\equiv \frac{H(z)}{H_0} =\sqrt{ \Omega_{0,r}(1+z)^4 + \Omega_b (1+z)^3 + \Omega_{cdm}(1+z)^3\left[ \frac{3}{3+\gamma}\left(\frac{1+z}{1+z^{\star}}\right)^{\gamma} + \frac{\gamma}{3+\gamma}\left(\frac{1+z^{\star}}{1+z}\right)^3 \right] + \Omega_{\Lambda}}\, ,
\end{equation}
\end{subequations}
where it can be seen that the $\Lambda$CDM scenario is recovered when $\gamma = 0$. Figure~\ref{omegas_diff_2} shows the evolution of all the matter components present in the Universe, considering a diffusion process described by the ADMD model with $\gamma = 0.2$ at a characteristic redshift $z^{\star} = 1\, .$

\begin{center}
\begin{figure}[htp!]
    \centering
   \includegraphics[width=0.6\linewidth]{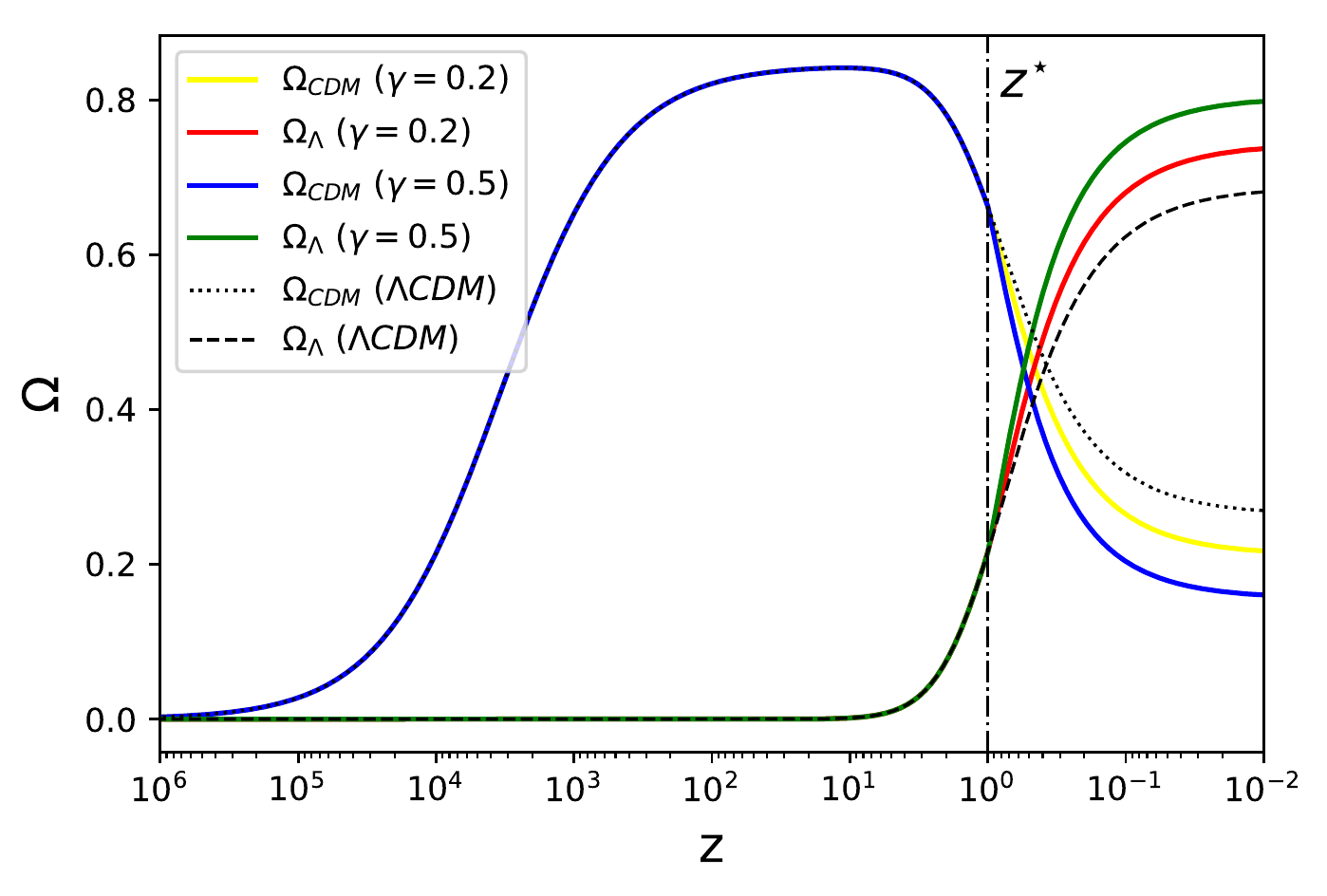}
    \caption{Anomalous decay of cold dark matter density $\rho_{cdm}$ into dark energy $\Lambda$ (Model 2). Black lines show the standard evolution without diffusion ($\gamma = 0$). Solid lines correspond to a diffusion process with $\gamma = 0.2$ (yellow and red for $\Omega_{cdm}$ and $\Omega_{\Lambda}$ respectively), and $\gamma = 0.5$ (blue and green for $\Omega_{cdm}$ and $\Omega_{\Lambda}$ respectively). The characteristic redshift was fixed to $z^{\star} = 1$.}
    \label{omegas_alphas}
\end{figure}
\end{center}

In general, depending on the values of $\left\lbrace \gamma\, , z^{\star} \right\rbrace$, the energy transfer between cold dark matter and $\Lambda$ will be smoother, or steeper than model 1. This can be seen in Figure~\ref{omegas_alphas}, where we have settled $z^{\star} = 1$ to explore the effect of $\gamma$. We observe that, at this redshift, the transition is smooth for different values of $\gamma$ (the reader can compare this to the case of sudden energy transfer of the Model 1 in Figure~\ref{omegas_diff_alphas}). The cold dark matter fluid diffuses and the cosmological constant captures this energy, which allows to $\Omega_{\Lambda}$ to reach larger values in the present day in comparison with the $\Lambda$CDM.

\begin{center}
\begin{figure}[htp!]
    \centering
   \includegraphics[width=0.6\linewidth]{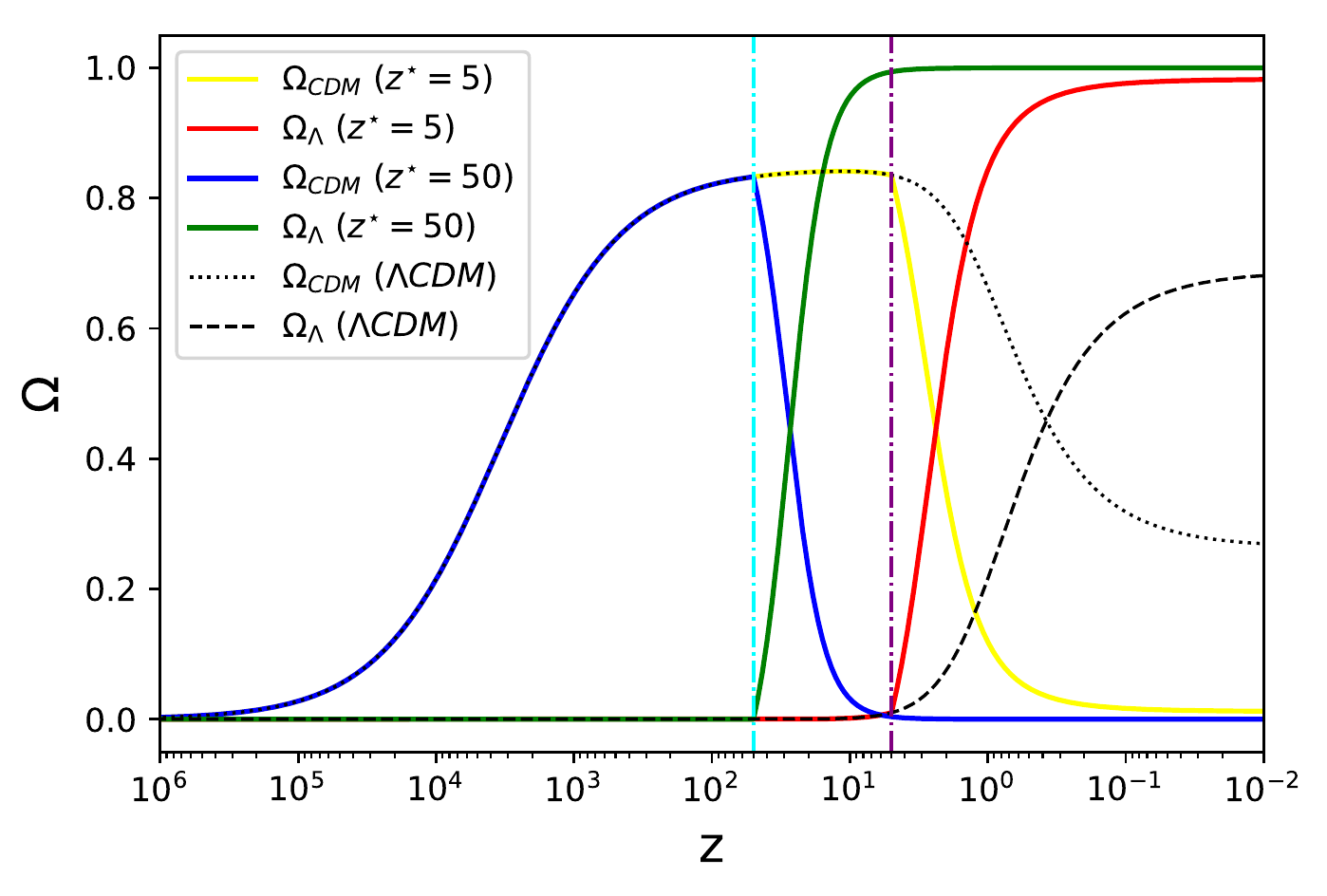}
    \caption{Anomalous decay of cold dark matter density $\rho_{cdm}$ into dark energy $\Lambda$ (Model 2) at different characteristic redshifts: $z^{\star} = 5$ (vertical dash--dotted purple line), and $z^{\star} = 50$ (vertical dash--dotted cyan line). Black lines show the standard evolution for $\Omega_{cdm}$ and $\Omega_{\Lambda}$ without diffusion ($\gamma = 0$). For the ADMD model, the diffusion constant was settled to $\gamma = 0.5$.}
    \label{omegas_z}
\end{figure}
\end{center}

The effects of this model on the dark sector energy density parameters due to the characteristic redshift $z_{\star}$ are shown in Figure~\ref{omegas_z}. We can see that the diffusion process induces a steeper fall of the cold dark matter energy density when $z^{\star} = 5\, , 50$, in comparison to lower characteristic redshifts (for example, at $z^{\star} = 1$ in Figure~\ref{omegas_alphas}). Besides, the present value of $\Omega_{cdm}\ (\Omega_{\Lambda})$ decreases (increases) much more than the cases for lower redshifts.    

\subsubsection{Model 3: Barotropic Model}
\begin{center}
\begin{figure}[htp!]
    \centering
   \includegraphics[width=0.6\linewidth]{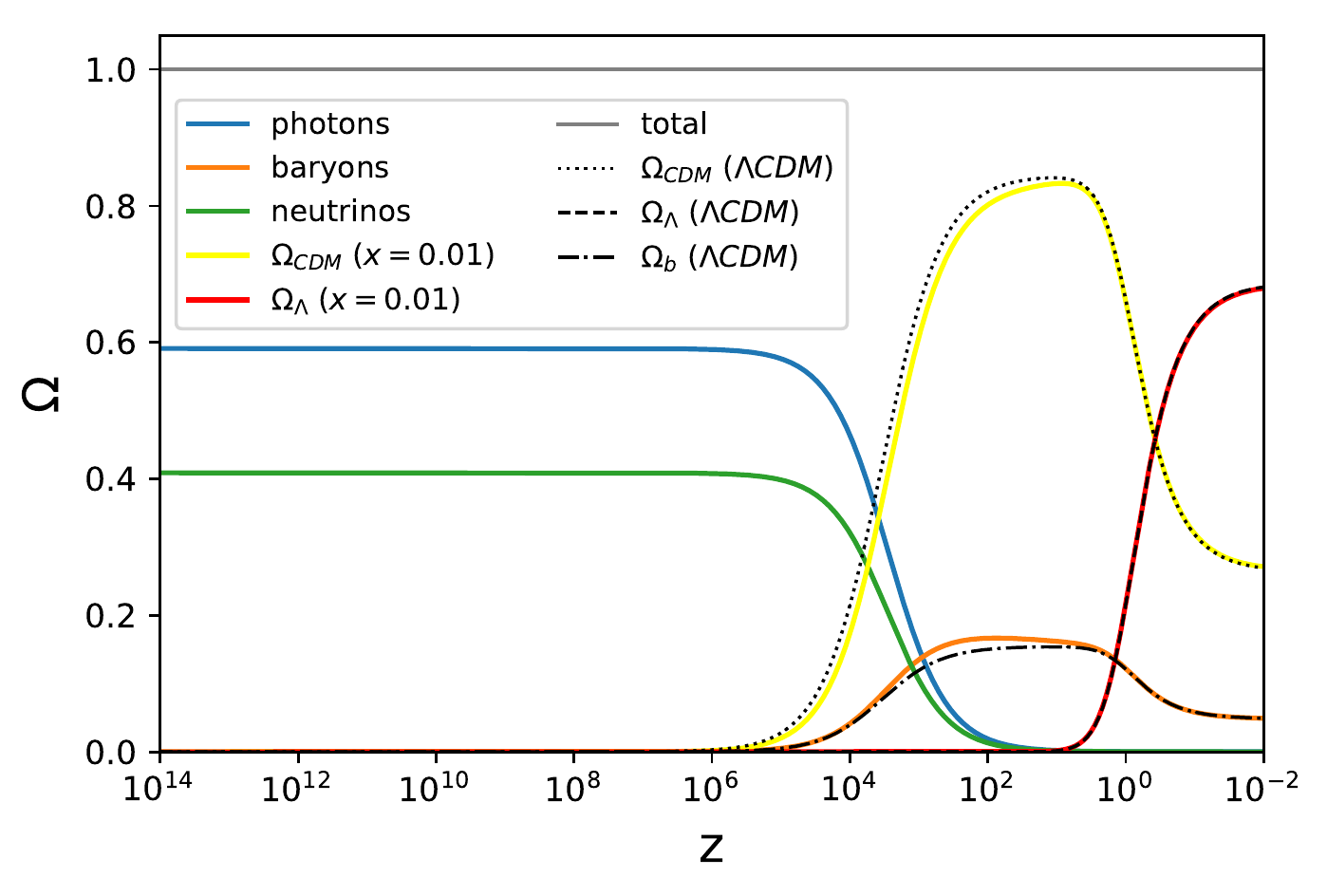}
    \caption{Cosmological evolution of the energy density parameter of each matter component of the Universe considering the diffusion process of Model 3: Barotropic model. The diffusion between CDM and $\Lambda$ is mediated by the barotropic equation of state $Q=x_{cdm}\rho_{cdm}$. }
    \label{omegas_diff3_all}
\end{figure}
\end{center}

\begin{subequations}
For the following two models we will solve Eq.~\eqref{diff_eq_Q}, for which a diffusion function $Q$ has to be given. One of the forms of this function that has been explored in~\cite{Corral:2020lxt} is
\begin{equation}
    Q\equiv x_i\rho_i\, ,
\end{equation}
this is, a constant barotropic equation of state $x_i$ relating the energy density of the different matter components $\rho_i$ ($i=b\, ,\gamma\, ,\nu\, , cdm\, , \Lambda$) and the diffusion function $Q$. In our case, the diffusion process will be only due to the CDM component, i.e., $i=cdm$, and thus we have from Eq.~\eqref{diff_eq_Q} that
\begin{equation}
    \rho_{cdm}(z) = \rho_{cdm}(1+z)^{\frac{3(\omega_{cdm} + 1)}{x_{cdm} + 1}}\, .
\end{equation}

The normalized Friedmann equation for this model is given by,
\begin{equation}\label{ez_3}
    E(z)\equiv \frac{H(z)}{H_0} =\sqrt{ \Omega_{0,r}(1+z)^4 + \Omega_{0,b}(1+z)^3 + (1 + x_{cdm})\Omega_{cdm}(1+z)^{\frac{3}{x_{cdm} + 1}} + \Omega_{\Lambda}}\, ,
\end{equation}
\end{subequations}
where we have considered the standard dust--like behavior for CDM ($\omega_{cdm} \simeq 0$). The cosmological evolution of all matter components is shown in Figure~\ref{omegas_diff3_all}, where it can be seen that the diffusion process driven by this model affects to $\Omega_{cdm}$ from the moment when its amplitude starts to grow, and during most of the matter domination era. As in the previous models, for comparison, we also show the standard $\Lambda$CDM case for $\Omega_{cdm}$ (black dotted line) and $\Omega_{\Lambda}$ (black dashed line), but this time we also show the standard evolution of the baryon energy density $\Omega_b$ (black dash--dotted line) to show that in the case of the diffusion model (orange line) it increases in the right proportion to balance the CDM contribution (yellow line) such that the total matter budget still satisfies the Friedmann constraint $\Omega_{tot} = 1$ (gray horizontal line) during all the cosmological evolution.

\begin{center}
\begin{figure}[htp!]
    \centering
   \includegraphics[width=0.6\linewidth]{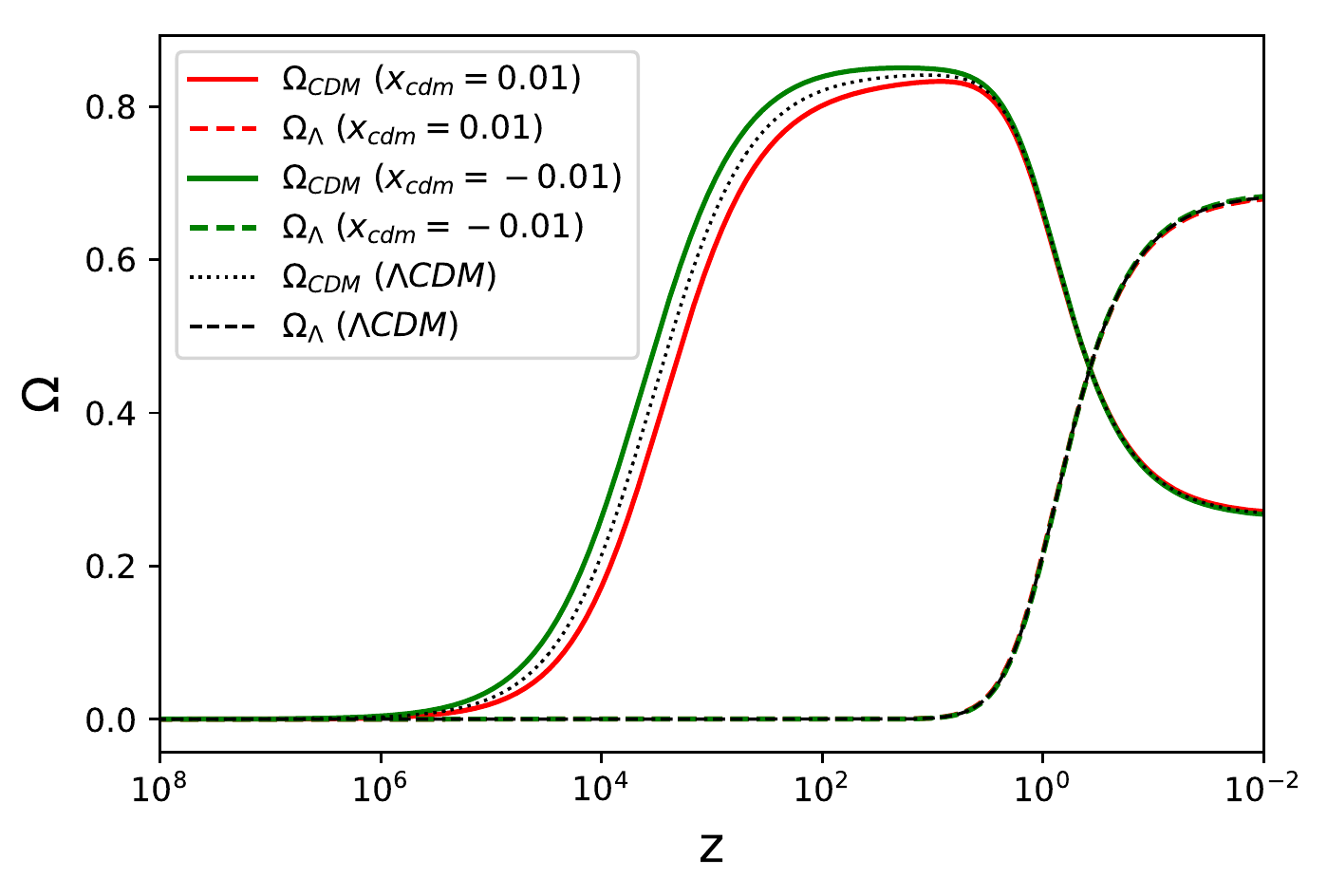}
    \caption{Diffusion between CDM and $\Lambda$ mediated by the barotropic equation of state $Q=x_{cdm}\rho_{cdm}$. The effect of different values of $x_{cdm}$ can be summarized as an increment (reduction) on $\Omega_{cdm}$ for $x_{cdm}>0$ ($x_{cdm}<0$). Dotted and dashed black lines represent the standard evolution of $\Omega_{cdm}$ and $\Omega_{\Lambda}$ respectively.}
    \label{omegas_diff3}
\end{figure}
\end{center}

In Figure~\ref{omegas_diff3} we show the evolution of $\Omega_{cdm}$ and $\Omega_{\Lambda}$ for several values of $x_{cdm}$. Positive values of $x_{cdm}$ (red lines) lead to larger values of $\Omega_{cdm}$ at $z=0$, whereas the current value of the cosmological constant energy density parameter $\Omega_{0,\Lambda}$ decreases (red dashed line). The opposite occurs when negative values of $x_{cdm}$ are considered (green lines).

\begin{center}
\begin{figure}[htp!]
    \centering
   \includegraphics[width=0.6\linewidth]{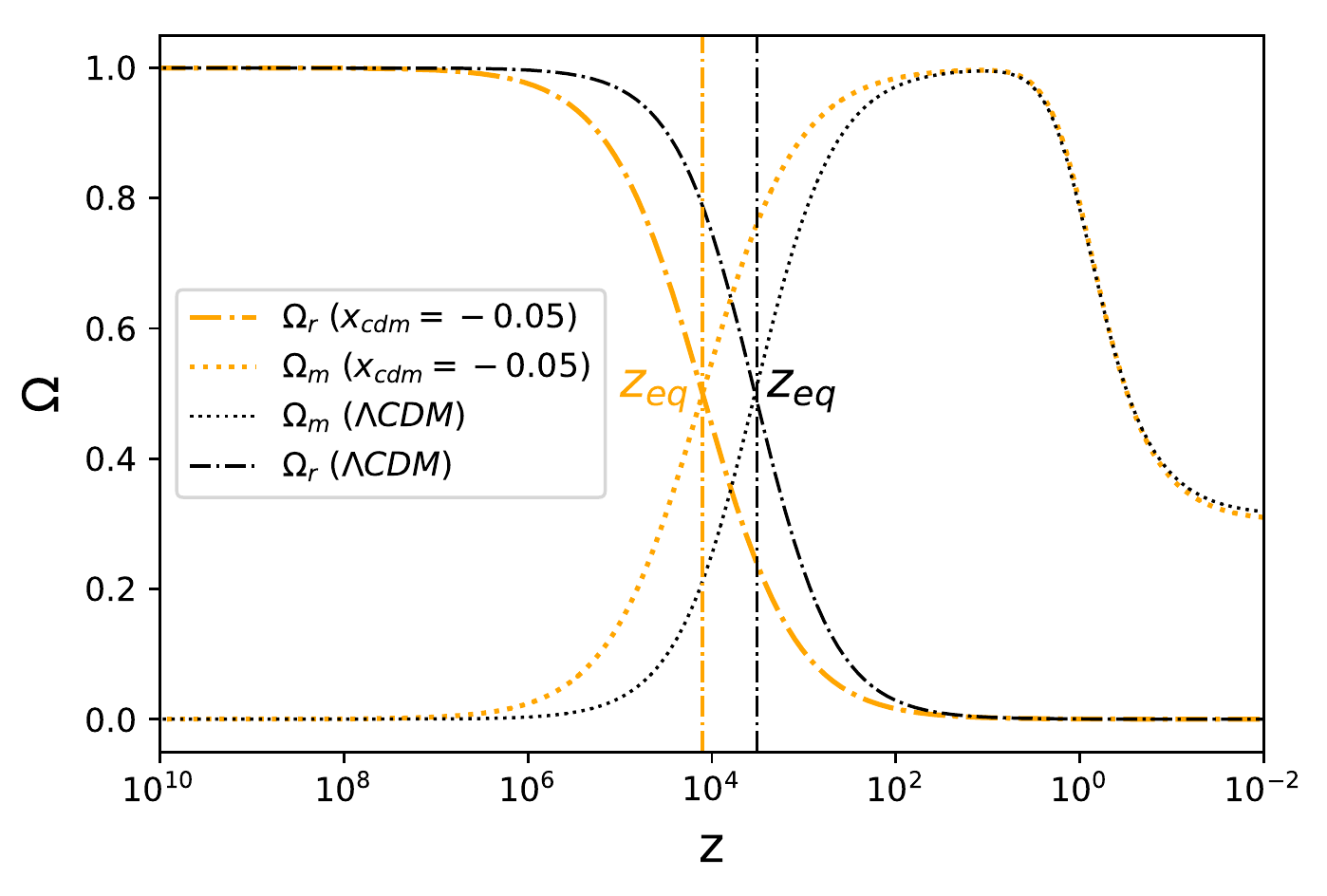}
    \caption{Evolution of $\Omega_{m}$ and $\Omega_{r}$ for the $\Lambda$CDM model (black lines) and the barotropic model (orange lines). The redshift for the matter--radiation equality era $z_{eq}$ depends on the diffusion parameter $x_{cdm}$. For $x_{cdm}=-0.05$, we have $z_{eq} \simeq 12.7\times 10^3$ (vertical dash--dotted orange line), whereas for $\Lambda$CDM we have $z_{eq} \simeq 3.3\times 10^3$ (vertical dash--dotted black line).}
    \label{omegas_diff3_eq}
\end{figure}
\end{center}

We can see that, different from Models 1 and 2, this model presents diffusion during certain period of time of the cosmological evolution ($z\lesssim 10^6$), and not only at a given characteristic redshift. In this sense, this model could have different implications in the cosmic history, as for example in the matter-radiation equality era $z_{eq}$, as is shown in Figure~\ref{omegas_diff3_eq}. In fact, the most notorious effect on the cosmological parameters $\Omega_{cdm}$ and $\Omega_{\Lambda}$ are not at the present day, but approximately from $z\simeq 10
^6$ to $\sim 5$.

\subsubsection{Model 4: Continuous Spontaneous Localization model}
\begin{subequations}
An interesting model also studied in~\cite{Corral:2020lxt} is the \textit{Continuous Spontaneous Localization} (CSL) model~\cite{Pearle:1976ka,Pearle:1988uh,Ghirardi:1989cn,Bassi:2003gd,Perez:2005gh,Lochan:2012di,Martin:2012pea,Canate:2013isa,Josset:2016vrq,Piccirilli:2017mto,Leon:2017yna,Leon:2020sqt}. This model arises as a proposal to explain the spontaneous collapse of the wave function in quantum mechanics, where the stochastic nature of the collapse is encoded in a new correction term in the Schrödinger equation. Such modification consists in a stochastic noise describing a diffusion process of the wave function in Hilbert space. The source of such noise can be of cosmological origin, for instance, due to the dark matter component in the Universe~\cite{Adler_2007,Adler_2008}. In the case we are interested in, the predicted form of the diffusion function $Q$ according to the CSL model is~\cite{Corral:2020lxt}
\begin{equation}
    \dot{Q} = -\xi_{CSL}\ \rho_{cdm}\, ,
\end{equation}
where $\xi_{CSL}$ is the localization rate, which is interpreted as the frequency of the localization events. After integration of the above expression, we have,
\begin{equation}
    Q(t) = Q_i - \xi_{CSL}\int_0^t \rho_{cdm}(t^{\prime})dt^{\prime}\, ,
\end{equation}
where $Q_i=Q(t=0)$ is an integration constant that will contribute to the total dark energy density, as we will show. The CDM energy density is then given by, 
\begin{equation}
    \rho_{cdm}(z) = \rho_{cdm}(1+z)^3e^{\xi_{CSL}t}\, ,
\end{equation}
which lead to the following Friedmann equation,
\begin{equation}\label{ez_4}
    E(z)\equiv \frac{H(z)}{H_0} = \sqrt{ \Omega_{0,r}(1+z)^4 + \Omega_b (1+z)^3 + \Omega_{cdm}\left[ e^{\xi_{CSL}t}(1+z)^3 - \xi_{CSL}\int_0^t e^{\xi_{CSL}t^{\prime}}[1+z(t^{\prime})]^3dt^{\prime} \right] + \Omega_{\Lambda_{eff}}}\, ,
\end{equation}
where $\Omega_{\Lambda_{eff}}\equiv \Omega_{\Lambda} + (\kappa^2Q_i/3H_0^2)\, .$ Once implemented a generalized form of the \textit{lower incomplete Gamma function} to deal with the integral in Eq.~\eqref{ez_4}, the Friedmann equation can be written as follows (see Appendix~\ref{igf})

\begin{equation}
   E(z) =  \left\lbrace\begin{matrix} 
\sqrt{\Omega_{r}(z) + \Omega_{b}(z) + \Omega_{cdm}(z)\left[e^{-u_0(1+z)^{-2}} + 2u_0(1+z)^{-2}\right] + \Omega_{\Lambda_{eff}}}\, , & {\rm{for\ Radiation\ domination}}\, , \\
\\
\sqrt{\Omega_{r}(z) + \Omega_{b}(z) + \Omega_{cdm}(z)\left[e^{-u_0(1+z)^{-3/2}} - u_0(1+z)^{-3/2}\right] + \Omega_{\Lambda_{eff}}}\, , & {\rm{for\ Matter\ domination}}\, . 
\end{matrix}\right.
\label{ez_4_gen}
\end{equation}

The evolution of each $\Omega_i(z)$ is shown in Figure~\ref{omegas_diff4}. Notice that we have introduced a dimensionless parameter $u_0\equiv -\xi_{CSL}t_0$, where $t_0$ is the current age of the Universe. It seems that the condition $\xi_{CSL}\leq 0$ is required in order to have a physically consistent cosmological evolution of this diffusion model during radiation dominated era. We will consider, however, both positive and negative values of $u_0$ in our analysis. For more details, see Appendix~\ref{igf}.
\begin{center}
\begin{figure}[htp!]
    \centering
   \includegraphics[width=0.55\linewidth]{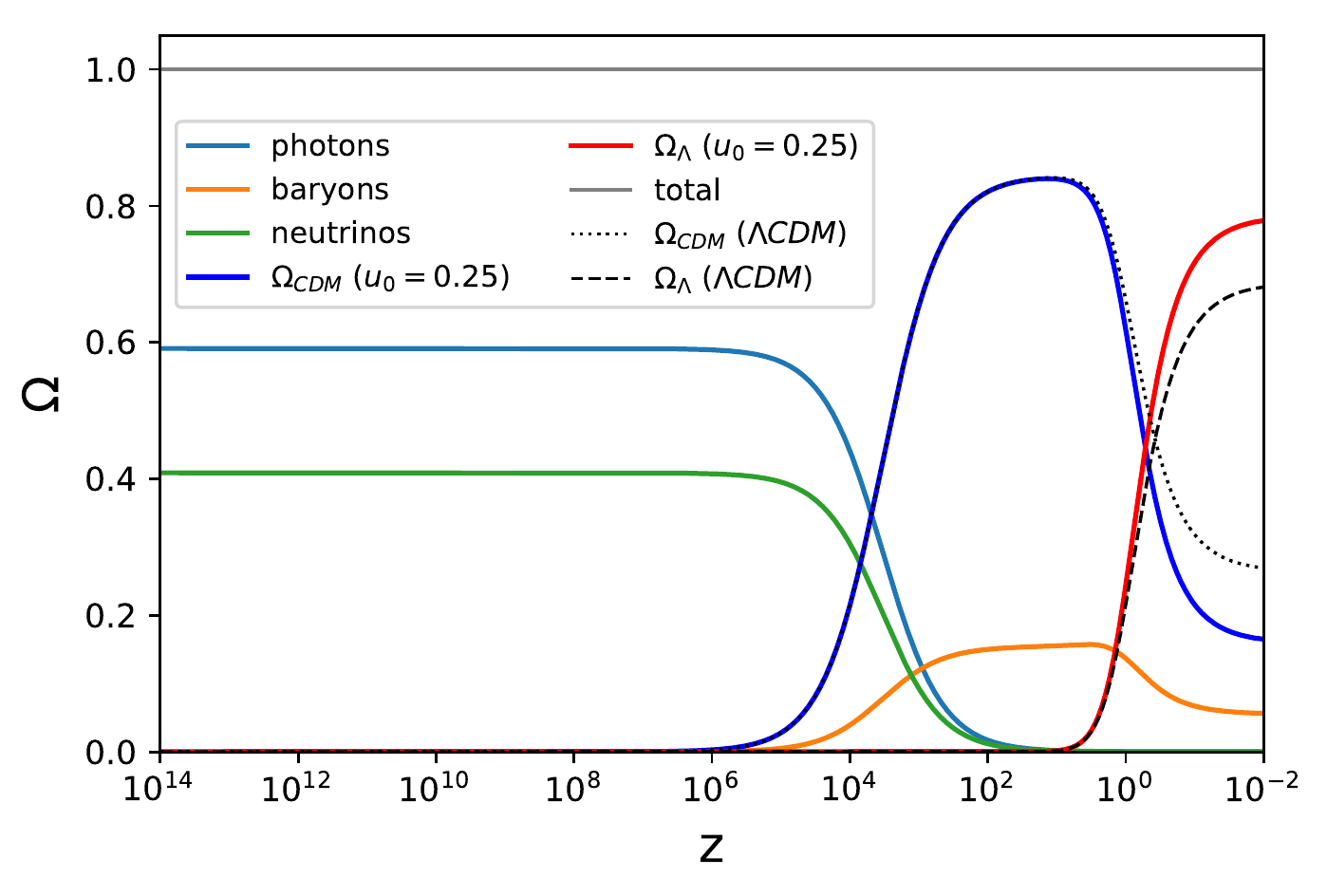}
    \caption{The energy density parameter of each matter component of the Universe as a function of the redshift considering the CSL diffusion process. As in previous Figures, the horizontal gray line stands for $\sum_i\Omega_i(z) = 1\, .$}
    \label{omegas_diff4}
\end{figure}
\end{center}

The sign of the diffusion parameter affects the behavior of both $\Omega_{cdm}$ and $\Omega_{\Lambda}$. This can be seen in Figure~\ref{omegas_diff4_cas}, where $u_0>0$ (blue lines) increases the energy density for the cosmological constant as CDM decreases. The opposite occurs for $u_0<0$ (orange lines).
\begin{center}
\begin{figure}[htp!]
    \centering
   \includegraphics[width=0.55\linewidth]{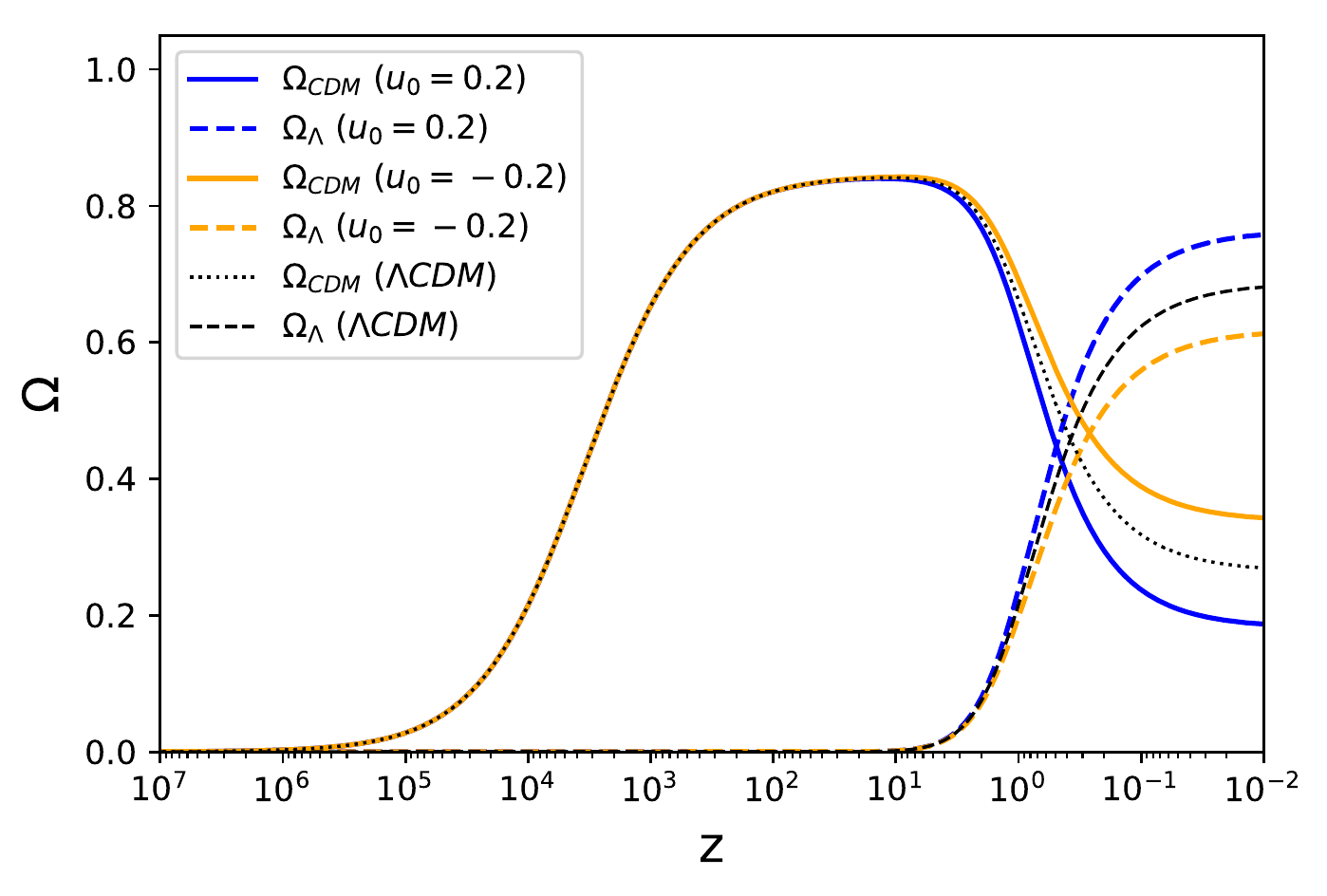}
    \caption{$\Omega_{cdm}(z)$ and $\Omega_{\Lambda}(z)$ for the CSL model for $u_0=-0.2\, , 0.2\, .$ Black lines indicate the standard $\Lambda$CDM result.}
    \label{omegas_diff4_cas}
\end{figure}
\end{center}

\end{subequations}

\begin{center}
\begin{figure}[htp!]
    \centering
  \includegraphics[width=0.7\linewidth]{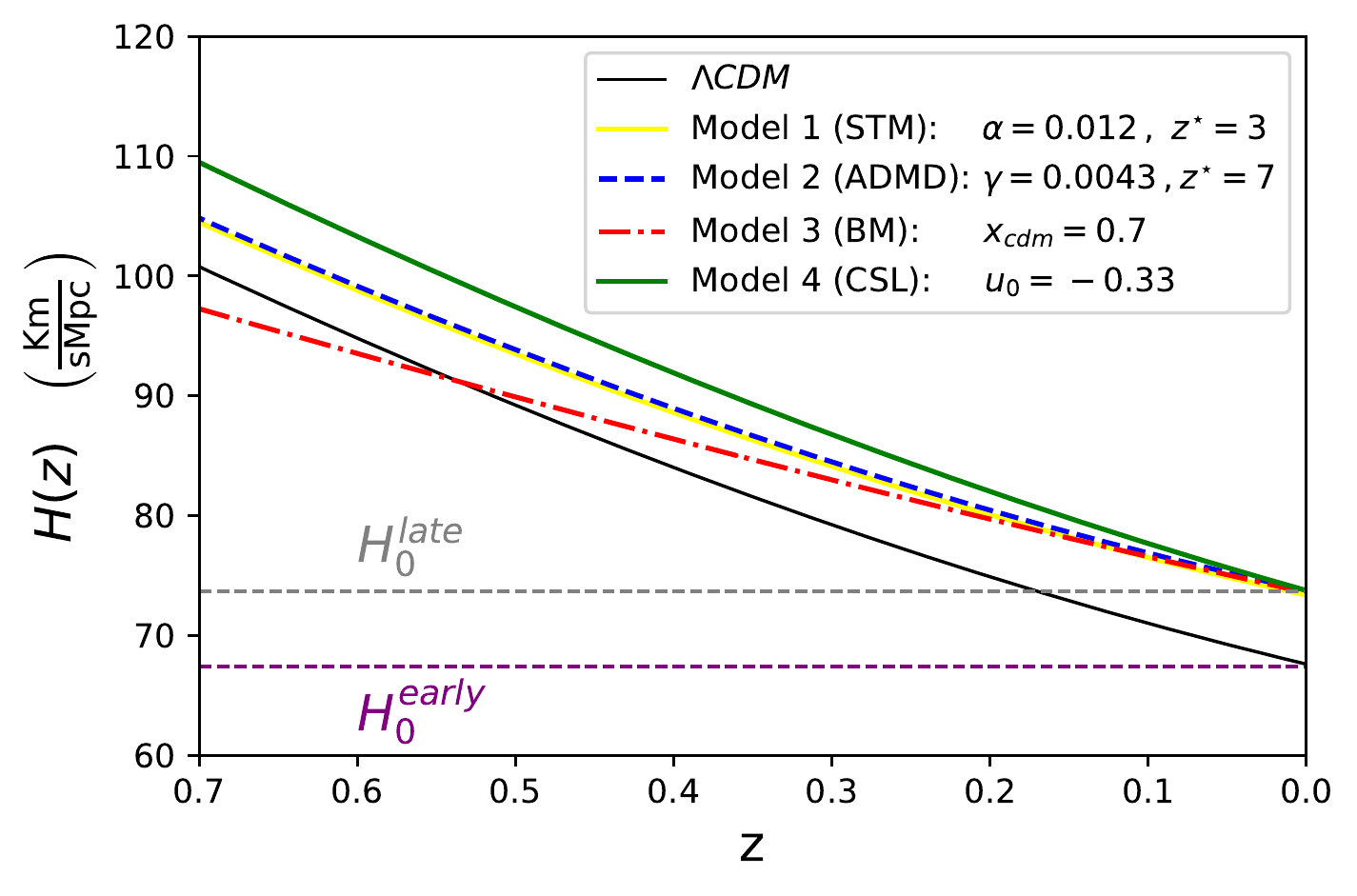}
    \caption{Cosmological evolution of the Hubble parameter $H(z)$ for all the diffusion models: STM (yellow), ADMD (blue), BM (red), and CSL (green). The solid black line shows the standard evolution without diffusion, i.e., the diffusion parameters set to zero. Horizontal dashed lines indicate the value of $H_0$ inferred from the CMB (purple) and local observations (gray), whose values are respectively given by $H_0^{early}\simeq 68$kms$^{-1}$Mpc$^{-1}$ and $H_0^{late}\simeq 73$kms$^{-1}$Mpc$^{-1}$. }
    \label{h_all}
\end{figure}
\end{center}

As we mentioned before, all these changes in $\Omega_{cdm}$ and $\Omega_{\Lambda}$ due to the diffusion processes will affect the current value of the Hubble parameter. In particular, if the current values for each energy density parameter $\Omega_{0,i}$ ($i=b\, ,\gamma\, ,\nu\, , cdm\, , \Lambda$), as well as for $H_0$ are those obtained from CMB observations (and thus $H_0 = H_0^{early}$ for $\Lambda$CDM), it is possible to find values of the diffusion parameters such that $H_0$ shifts to the value inferred from late time observations, i.e., $H_0 = H_0^{late}$. We show the evolution of the Hubble parameter in Figure~\ref{h_all}. Considering particular values for the parameters of each of the diffusion models studied, we show that it is possible to obtain a cosmological solution for $H(z)$ consistent with the reported value of $H_0$ from local observations. Once the diffusion parameters are set to zero, the $\Lambda$CDM case is recovered and $H_0 = H_0^{early}$.

Therefore, from the numerical solutions that we have obtained, the four diffusion models studied seem to be good candidates to solve the tension in the value of $H_0$. Focused on an analysis for the late time Universe, this was already shown for model 1 (STM) and model 2 (ADMD) in~\cite{Perez:2020cwa}, where the authors explore the parameter space in order to find the possible combinations of the diffusion parameters such that $H_0^{early}\rightarrow H_0^{late}\, ,$ although no direct tests with observations are made. On the other hand, the authors in~\cite{Corral:2020lxt} studied models 3 and 4 (BM and CSL respectively), at late times as well. Particularly, in their study, they analyze by parts, each model depending on the contribution of the dark energy component: for the BM, they consider the cases in which 1) $\Omega_{0,\Lambda} = 0\, ,$ 2) $\Omega_{0,\Lambda} = 1\, ,$  3) $0<\Omega_{0,\Lambda} < 1\, ,$ and 4) $\Omega_{0,\Lambda} < 0\, .$ For the CSL model, the cases that the authors analyze are 1) $\Omega_{0,\Lambda}>\Omega_{0,\xi}\, ,$ 2) $\Omega_{0,\Lambda}<\Omega_{0,\xi}\, ,$ and 3) $\Omega_{0,\Lambda}=\Omega_{0,\xi}\, ,$ where $\Omega_{0,\xi} \equiv \xi_{CSL}^2/(9H_0^2)$. The authors include statistical analysis, but only considering local observations, such as Observational Hubble Data (OHD) used in~\cite{Magana:2017nfs}, and supernovae (SNe Ia)~\cite{Scolnic:2017caz}.

We want to emphasise that it is crucial for any model trying to solve the $H_0$ tension to consider the radiation component of the Universe, since it plays an important role in the physics of the last scattering surface, specifically its contribution in the sound speed of the photon--baryon acoustic wave. In particular for models as those studied here, easing the $H_0$ tension should translate in the case in which CMB data are consistent with late time observations. This important part of the analysis is lacking in the previous works mentioned above, and it is the step further that we are giving in this work in order to study the true viability of these models. We do not separate the analysis by cases, or restrict the diffusion models to low redshifts, but rather we contemplate all the cosmological evolution considering all the matter components present in the Universe.

Let us summarize the results obtained in this Section as follows: assuming that the values of the cosmological parameters are those inferred from CBM observations, in Figure~\ref{h_all} it is shown that, with the cosmological field equations obtained from UG, it is possible to obtain a value of $H_0$ consistent with local observations by the means of diffusion processes between cold dark matter and dark energy in the form of a variable cosmological "constant". This is so, of course, for particular choices of the diffusion parameters. A statistical analysis has to be performed in order to infer the most likely values of such parameters in the light of current data taken from several astrophysical and cosmological observations.

\section{Statistical analysis}\label{stat}

In the previous Section, we have shown that diffusion processes in UG can ease the current tension on the $H_0$ parameter. Specifically, at the level of the numerical solutions for the background dynamics, the parameters of each diffusion model allow to have a consistent match between the $H_0$ inferred from CMB with that of local observations. It is crucial then, to analyze the viability of these models in the light of cosmological observations. Particularly, we want to constraint the diffusion parameters for models 1, 2, 3, and 4 with data from CMB observations. If these diffusion models ease the $H_0$ tension, then CMB data should allow values for $\alpha$ and $z_{\star}$ (for model 1), $\gamma$ and $z
^{\star}$ (for model 2), $x_{cdm}$ (for model 3), and $\xi_{CSL}$ (for model 4) such that the value of $H_0$ inferred by CMB coincides with that of local observations. 

\subsection{Constraints with CMB data set}
Since we focus on the background evolution, we will use the \textit{Planck Compressed 2018} (PC2018) data~\cite{Chen:2018dbv}, instead of the full Planck 2018 likelihoods. Such compressed version of the CMB data have been probed to be as useful as the full version to constraint not only the standard $\Lambda$CDM model, but also alternative models of dark energy, such as $\omega$CDM, CPL model, interacting dark energy, early dark energy, and all those models dubbed as \textit{smooth dark energy}, which are models phenomenologically similar to a cosmological constant~\cite{Malekjani:2018qcz,Cedeno:2019cgr,Rivera:2016zzr,Li:2019san,Li:2019ypi,Davari:2019tni,Montiel:2020rnd}. In our case, even when the gravitational theory is Unimodular Gravity, we have shown that the cosmological scenario is basically that of General Relativity with a non-gravitational interaction between the dark sector components. Particularly, we still have a cosmological ``constant'' which only changes its current value. Therefore, we can safely use PC2018 data to constraint the diffusion parameters.

The two main physical quantities to use in order to constraint cosmological models with PC2018 data are the acoustic scale $l_A$, which characterize the CMB temperature power spectrum in the transverse direction (and therefore leading to variations of the peak spacing), and the shift parameter $R$, which affects the CMB temperature spectrum in the line-of-sight direction (and therefore affecting the heights of the peaks),
\begin{equation}\label{la_R}
    l_A = (1+z_{*})\pi \frac{D_A(z_{*})}{r_s(z_*)}\, ,\quad R = (1+z_{*}) \frac{\Omega_{0,m}^{1/2}H_0}{c}D_A(z_{*})\, ,
\end{equation}
where $z_*$ is the value of the redshift when photons decouple from baryons ($z_*\simeq 10^3$), $c$ is the speed of light, and $r_s$ and $D_A$ are the comoving sound horizon and the angular diameter distance respectively,
\begin{subequations}\label{distances}
\begin{eqnarray}
    r_s(z) &=& \frac{1}{H_0}\int_{z}^{\infty}\frac{c_s(z^{\prime})dz^{\prime}}{E(z^{\prime})}\, ,\quad {\rm{with}}\quad c_s(z) = c\left[3\left( 1 + \frac{3\Omega_b}{4\Omega_{r}(1+z)} \right)\right]^{-1/2}\, , \\
    D_A(z) &=& \frac{c}{H_0(1+z)}\int_0^z \frac{dz^{\prime}}{E(z^{\prime})}\, ,\quad {\rm{for}}\quad \Omega_k=0\, ,\label{ang_dist}
\end{eqnarray}
\end{subequations}
where $c_s(z)$ is the sound speed of the photon-baryon acoustic wave. Additionally, the physical baryon energy density parameter $\Omega_{b}h^2$ with $h=H_0/100$, and the scalar spectral index $n_s$ are included in the PC2018 likelihood as well. It can be seen that the diffusion parameters enter through $E(z)$ given by Eq.~\eqref{ez_1},~\eqref{ez_2},~\eqref{ez_3},~\eqref{ez_4_gen}, for model 1, 2, 3, and 4 respectively, in the integrands shown in Eq.~\eqref{distances}. Here it is evident the importance to have the radiation component in the analysis, since we have to evaluate integrals on $z_{*}$ in order to compute $l_A$ and $R$ (see Eq.~\eqref{la_R}).

The parameter space of interest for each diffusion model will be spanned by
\begin{subequations}
\begin{eqnarray}
    \Theta_1 &=& \left\lbrace \Omega_b h^2\, , \Omega_{cdm} h^2\, , H_0\, ,\alpha\, ,z^{\star}_1 \right\rbrace\, , \\
    \Theta_2 &=& \left\lbrace \Omega_b h^2\, , \Omega_{cdm} h^2\, , H_0\, ,\gamma\, ,z^{\star}_2 \right\rbrace\, , \\
    \Theta_3 &=& \left\lbrace \Omega_b h^2\, , \Omega_{cdm} h^2\, , H_0\, ,x_{cdm} \right\rbrace\, , \\
    \Theta_4 &=& \left\lbrace \Omega_b h^2\, , \Omega_{cdm} h^2\, , H_0\, ,\xi_{CSL} \right\rbrace\, ,
\end{eqnarray}
\end{subequations}

This will explicitly show whether the diffusion parameters can take non--null values consistent with CMB data, and simultaneously solving the $H_0$ tension. The (logarithmic) likelihood function $\log\mathcal{L}(\Theta)$ is given by
\begin{equation}
    \log\mathcal{L}_{{\rm{CMB}}}(\Theta) = -\frac{1}{2}\left[ \vec{\mu}_{{\rm{CMB}}} - \vec{\mu}(\Theta) \right]^{T} \mathbf{C}^{-1}_{{\rm{CMB}}} \left[ \vec{\mu}_{{\rm{CMB}}} - \vec{\mu}(\Theta) \right]\, ,
\end{equation}
where $\mathbf{C}^{-1}_{{\rm{CMB}}}$ is the inverse covariance matrix of the CMB observation, $\vec{\mu}_{{\rm{CMB}}} = (R^{{\rm{CMB}}}\, ,l_A^{{\rm{CMB}}}\, ,(\Omega_b h^2)^{{\rm{CMB}}}\, ,n_s^{{\rm{CMB}}})$ is the vector of CMB observations, and $\vec{\mu}(\Theta_i)= (R(\Theta_i)\, ,l_A(\Theta_i)\, ,\Omega_b h^2\, ,n_s)$ is the vector of their corresponding theoretical values according to the diffusion models $i=1,2,3,4$.

To compute the posterior probabilities for each of the diffusion models, we use the software \textsc{monte python}~\cite{Audren:2012wb}. We have considered flat priors, since they are the most conservatives to use when there is not previous knowledge about the parameters to analyze, as is the case for the diffusion models. As can be seen in Table~\ref{paramfile}, we have chosen the mean values of the diffusion parameters to be those agreeing with $\Lambda$CDM, but with broad priors. For instance, the characteristic redshift $z^{\star}_1$ ($z^{\star}_2$) for Model 1 (Model 2) is such that, the diffusion process can take place at any moment between the last scattering surface ($z\simeq 1100$) and the present day ($z=0$). The prior for $H_0$ is such that it contains the two preferred and different mean values reported by cosmological and local observations. On the other hand, the prior for the physical baryon density parameter $\Omega_{b}h^2$ was established such that $0\leq \Omega_{b}\leq 1$, and according to the prior for $H_0$ mentioned above $0.65\leq h\leq 0.75$. The prior for the CDM parameter $\Omega_{cdm}h^2$ was established in the same way.
\begin{table}[h!]
\centering
\begin{tabular}{cccccc}
\hline
\hline
model  & parameter    & mean & min prior & max prior & Std. Dev. \\
\hline
\hline
& $\Omega_b h^2$ & 0.0224  & 0 & 0.5625 & 0.015 \\
$\Lambda$CDM & $\Omega_{cdm}h^2$ & 0.120  & 0 & 0.5625 & 0.0013 \\
& $H_0$ & 70  & 65 & 75 & 0.01 \\
\hline
Model 1 & $\alpha$ & 0 & -1 & 1 & 0.005 \\
& $z^{\star}_1$ & 0 & 0 & 1100 & 0.005 \\
\hline
Model 2 & $\gamma$ & 0 & -10 & 10 & 0.005 \\
& $z^{\star}_2$ & 0 & 0 & 1100 & 0.005 \\
\hline
Model 3 & $x_{cdm}$ & 0 & -0.01 & 0.05 & 0.001 \\
\hline
Model 4 & $u_0$ & 0 & -10 & 10 & 0.005 \\
\hline
\end{tabular}
\caption{Input of the parameters of each diffusion model to generate the MCMC. Each diffusion model also has the same input as those of the $\Lambda$CDM parameters.}
\label{paramfile}
\end{table}

\begin{figure}[h!]
\centering
  \includegraphics[width=0.4\linewidth]{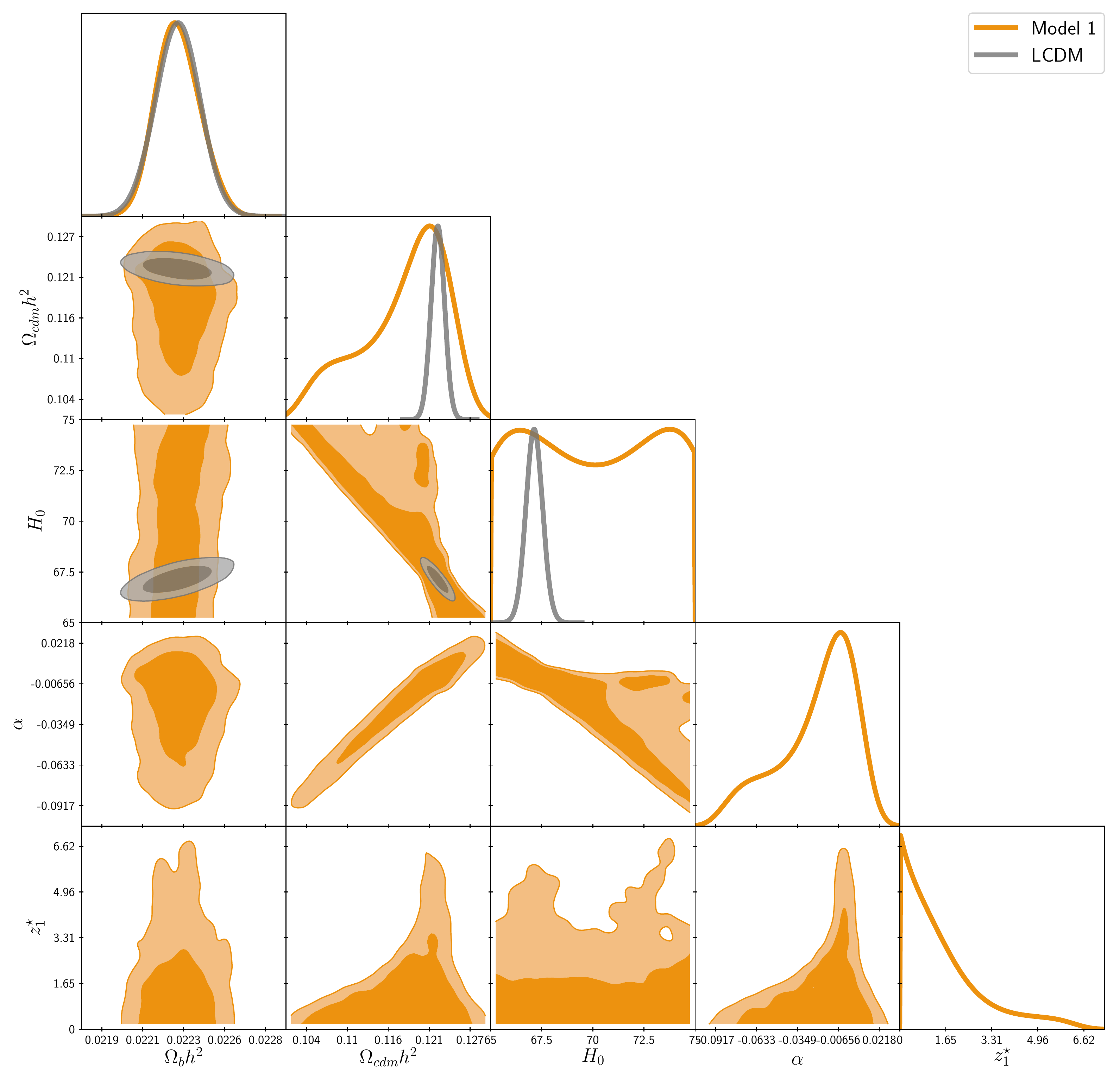}
\qquad
  \includegraphics[width=0.4\linewidth]{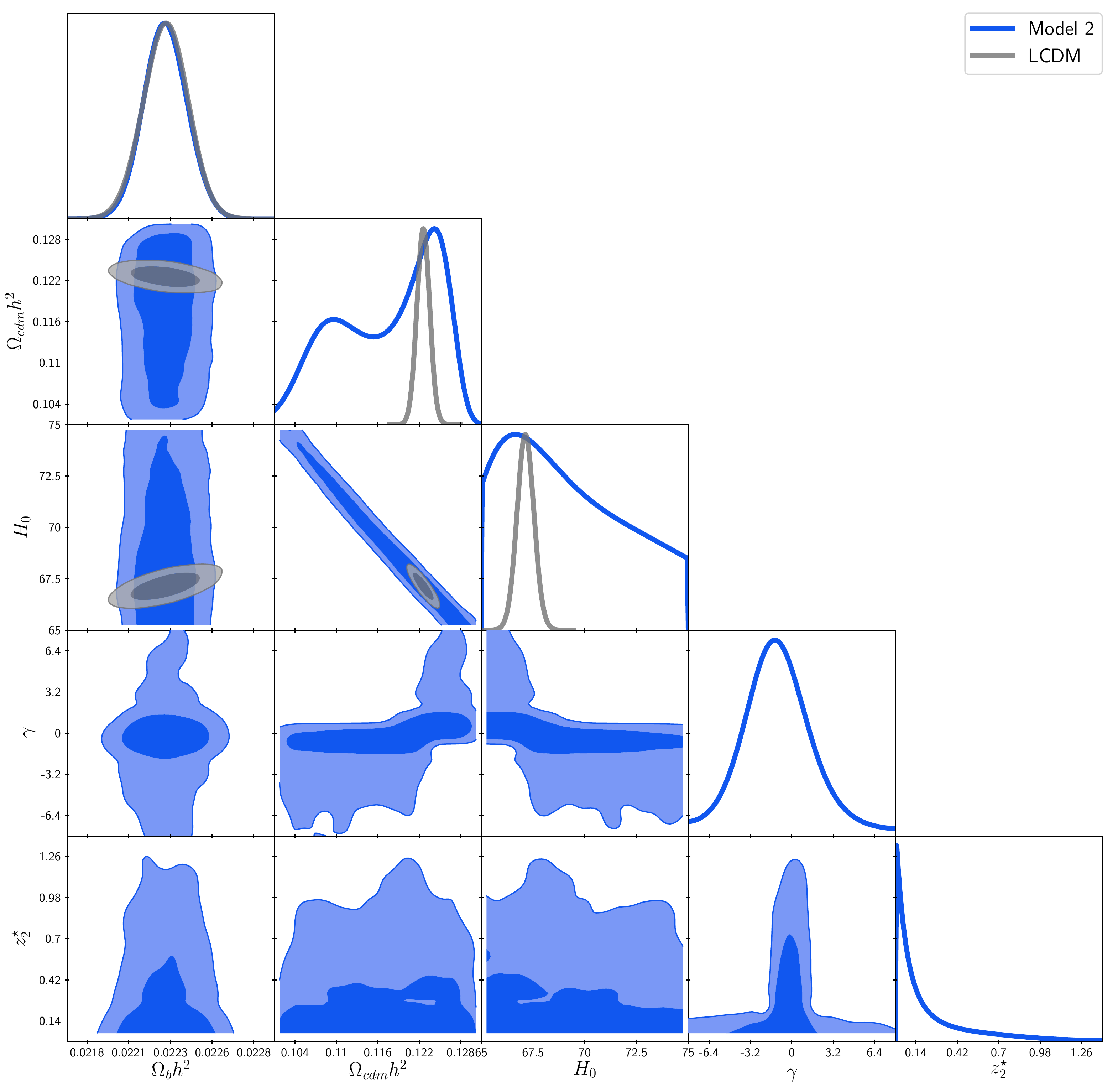}
\qquad
  \includegraphics[width=0.4\linewidth]{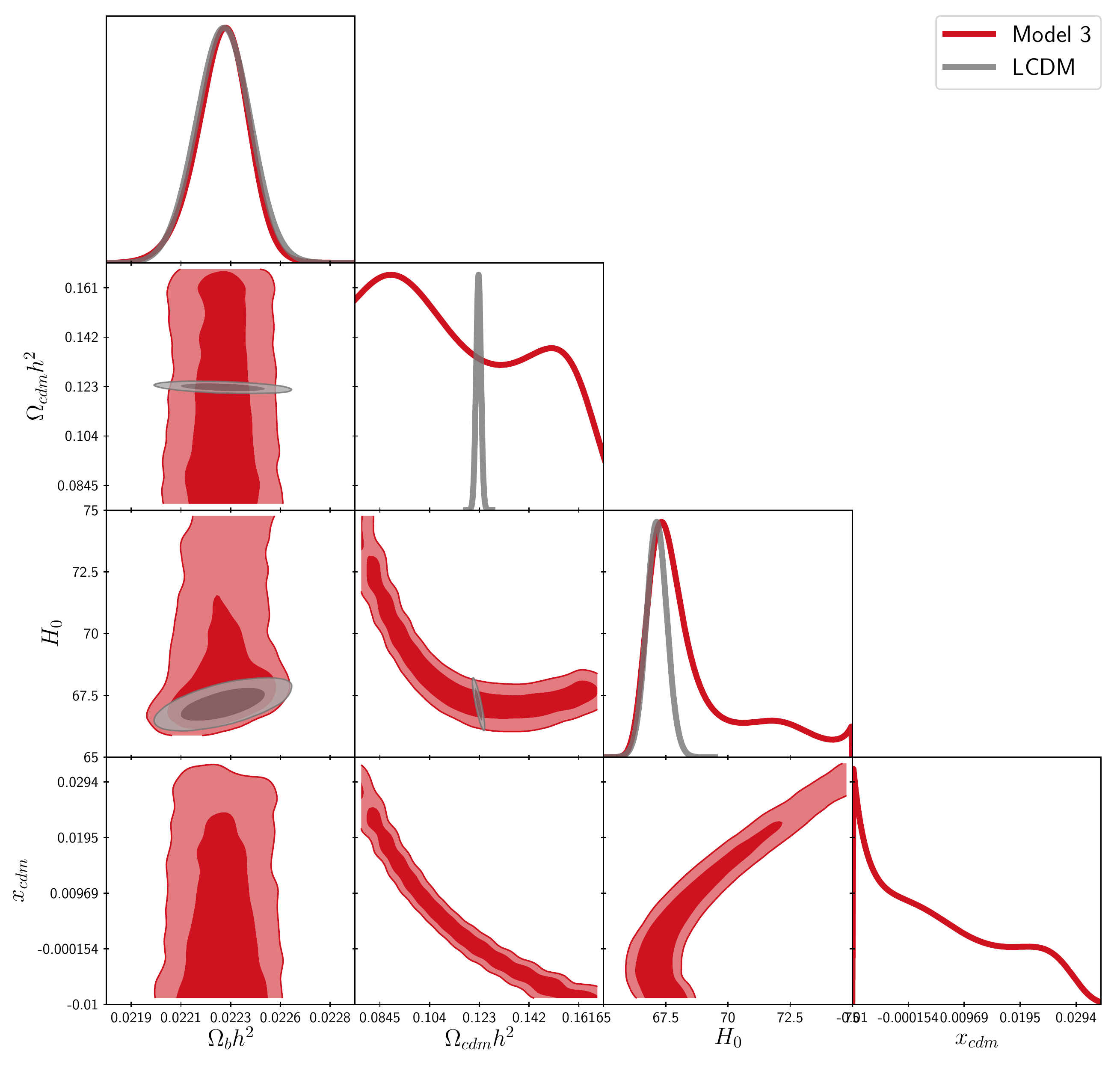}
\qquad
  \includegraphics[width=0.4\linewidth]{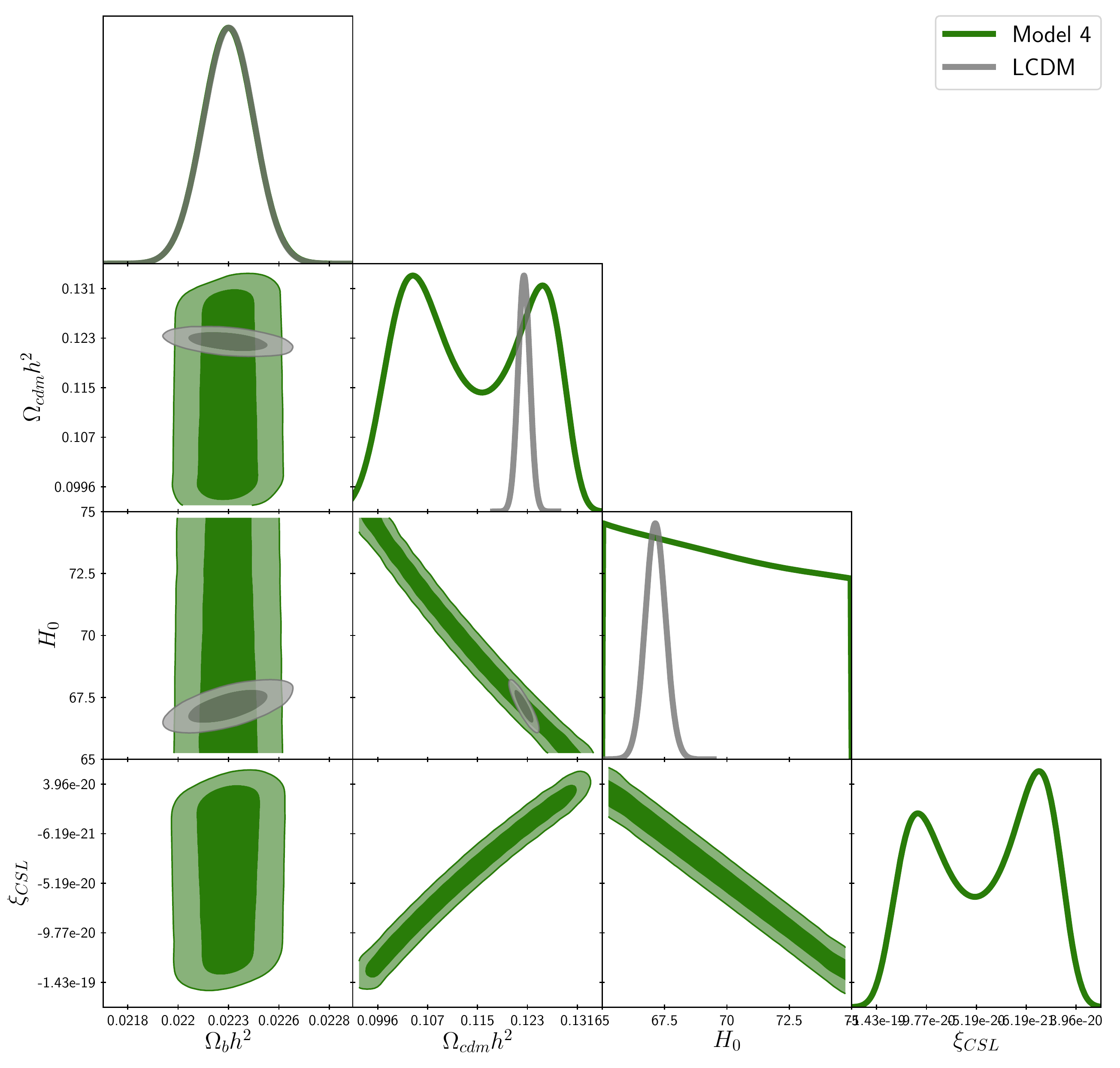}
\caption{Posterior probability distributions considering CMB data only, for all diffusion models under study: Sudden Transfer model (top right, orange), Anomalous Decay of the Matter Density model (top right, blue), Barotropic model (bottom left, red), and Continuous Spontaneous Localization model (bottom right, green). We have included the posteriors for the $\Lambda$CDM parameters as well (gray).}
\label{post_cmb}
\end{figure}

When running the chains, we have monitored the convergence with the Gelman--Rubin criterion~\cite{Gelman:1992zz}, by considering $R-1<0.05$. Figure~\ref{post_cmb} shows the posteriors for the parameters of Model 1 (top left, orange), 2 (top right,blue), 3 (bottom left, red), and 4 (bottom right, green). The posteriors for the $\Lambda$CDM parameters (gray) are shown for comparison.

As general features shared by all the diffusion models, we observe from the posteriors of the standard $\Lambda$CDM parameters that 1) the amount of baryonic matter $\Omega_b h^2$ remains unchanged, 2) the amount of cold dark matter $\Omega_{cdm} h^2$ gets a broader range of values, and since it is anticorrelated with the Hubble parameter\footnote{While this anticorrelation is true in the $\Lambda$CDM case, strictly speaking, the anticorrelation between $\Omega_{cdm}h^2$ and $H_0$ occurs for the diffusion models 1, 2, and 4. In the case of model 3 the $\left\lbrace \Omega_{cdm}h^2\, ,H_0\right\rbrace$--plane present a banana--shaped posterior, and the anticorrelation occurs only for the region corresponding to positive values of $x_{cdm}$.}, 3) $H_0$ is weakly constrained, and its posterior gets ``stretched''. This is not the ideal way to ease the $H_0$ tension, in the sense that it would be desirable CMB data to be consistent with local observations by shifting the mean value of $H_0$ from $H_0^{early}$ to $H_0^{late}$.

In the case of the diffusion parameters from models 1 and 2 we have,
\begin{subequations}
\begin{eqnarray}
    {\rm{Model\ 1\ (STM)}}&:& \alpha = -0.0202^{+0.0317}_{-0.0152}\, ,\quad z^{\star}_1 = 1.58^{+0.256}_{-1.58}\, . \\
    {\rm{Model\ 2\ (ADMD)}}&:& \gamma = -0.633^{+1.67}_{-0.728}\, ,\quad z^{\star}_2 = 0.247^{+0.0269}_{-0.247}\, .
\end{eqnarray}
\end{subequations}

Whereas in~\cite{Perez:2020cwa} the diffusion parameters $\alpha$ and $\gamma$ are defined strictly positives, we are regarding their values to be constrained by observations, and in principle there is no restriction on the sign of these parameters. Moreover, if they have to be consistent with $\Lambda$CDM, it is convenient to consider a symmetric range around zero. This allows us to see that the most likely values for those parameters have negative mean values, with a standard deviation including $\alpha\, , \gamma$ equal to zero, where the $\Lambda$CDM model is recovered. On the other hand, the characteristic redshift is $z^{\star}\geq 0$, since the energy transfer between CDM and $\Lambda$ occurs at some time between the last scattering surface and the present day. Both models present a transfer of energy taking place at $z^{\star}_1 = 1.58$ for model 1, and $z^{\star}_2 = 0.247$ for model 2, which is consistent with a late time diffusion process. The novelty here is that we were able to infer such values through CMB observations, which requires to calculate the angular diameter distance $D_A$ at $z_*\simeq 1100$ in order to compute the observables $l_A$ and $R$ (see Eq.~\eqref{la_R} and Eq.~\eqref{ang_dist}),
\begin{equation}
    D_A(z_*) = \frac{c}{H_0(1+z_*)}\int_0^{z_*} \frac{dz^{\prime}}{E(z^{\prime})} = \frac{c}{H_0(1+z_*)}\left[\int_0^{z^{\star}_i} \frac{dz^{\prime}}{E_{i}(z^{\prime})}+\int^{z_*}_{z^{\star}_i} \frac{dz^{\prime}}{E_{\Lambda CDM}(z^{\prime})}\right]\, ,
    \label{dist_ang_diff}
\end{equation}
where $E_{i}(z)$ stands for the normalized Friedmann equations~\eqref{ez_1},~\eqref{ez_2} according to the models $i=1\, , 2$ respectively, with $z_i^{\star}$ the corresponding characteristic redshift, and $E_{\Lambda CDM}(z)$ for the standard $\Lambda$CDM case. The first integral within the square brackets quantify the modification induced by the diffusion process.

In the case of model 3, the diffusion parameter presents an upper bound at $x_{cdm}\simeq 0.03$. This allows to the CDM energy density parameter to have a broader range of values. However, the constraint on $H_0$ is not as weak as that imposed on such parameter by the other diffusion models. The parameter $\xi_{CSL}$ of model 4 presents a posterior with two peaks located at $\xi_{CSL} \simeq \left(-1.09\times 10^{-19}\, , 5.26\times 10^{-21}\right){\rm{s}}^{-1}\, ,$ which constitute the most likely values for such parameter. These peaks on $\xi_{CSL}$ induce a bimodal posterior in the CDM density parameter whose peaks are located at $\Omega_{cdm}h^2 \simeq \left( 0.105\, , 0.125 \right)\, .$ While one of the peaks is approximately consistent with the amount of CDM according to the $\Lambda$CDM model, the second peak is located at less CDM contribution, which lead to higher values of $H_0\, .$

Therefore, whereas for $\Lambda$CDM we obtained the expected values inferred from CMB for $\Omega_b h^2\, , \Omega_{cdm}h^2\, ,$ and $H_0$ given by
\begin{equation}
    \Omega_b h^2 = 0.0223^{+0.000136}_{-0.000137}\, ,\quad \Omega_{cdm} h^2 = 0.122^{+0.000968}_{-0.00095}\, ,\quad H_0 = 67.1^{+0.427}_{-0.438}\ {\rm{km}}\ {\rm{s}}^{-1}{\rm{Mpc}}^{-1}\, ,
    \label{lcdm_cmb}
\end{equation}
the diffusion parameters allow to increase the range of values that CDM energy density and $H_0$ usually have within the $\Lambda$CDM model according to CMB data. In particular, $H_0$ is weakly constrained and it can take values from $H_0^{early}$ to $H_0^{late}$ while being in agreement with CMB observations. However, and as we mentioned above, this is not the ideal way to ease the $H_0$ tension, in the sense that the mean value of $H_0$ should shift from $H_0^{early}$ to $H_0^{late}$, if it is the case that these models solve the tension successfully. The latter can be explored by adding new information to the analysis, new data sets that help to break some degeneracies between the parameters.

\subsection{Constraints with CMB + Late time data set}
So far we have obtained that the diffusion models presented here seem to ease the $H_0$ tension, but in such a way that, instead of an effective shift of the mean value, the constraint on $H_0$ is weaker than in the $\Lambda$CDM case due to the presence of the diffusion parameters. Therefore, let us now include observations from the late Universe to study whether the diffusion models are in agreement with $H_0 = H_0^{{\rm{late}}}$.

Besides the cosmological observation from the CMB ($z_*\sim 1100$), we now include some local observations in the analysis: \textbf{Pantheon}, light-curve from 1048 supernovae (SNe Ia) within the redshift range $0.01<z<2.26$~\cite{Scolnic:2017caz}. \textbf{H0LICOW}\footnote{For the \textsc{monte python} implementation of the H0LICOW likelihoods see \href{https://zenodo.org/record/3632967\#.XjrsmRd7k0o}{here}~\cite{stefan_taubenberger_2020_3632967}.}, time-delay distances of 6 lensed quasars at redshifts $z = 0.654\, , 1.394\, , 1.662\, , 1.693\, , 1.722\, , 1.789\,$~\cite{Suyu:2009by,Suyu:2013kha,Wong:2016dpo,Birrer:2018vtm,Jee:2019hah,Chen:2019ejq,Rusu:2019xrq,Wong:2019kwg}. \textbf{SH0ES}, $H_0$ measurement from Cepheids in the Large Magellanic Cloud ($d\sim 50$kpc)~\cite{Riess:2019cxk}.

\begin{center}
\begin{figure}[htp!]
    \centering
   \includegraphics[width=0.7\linewidth]{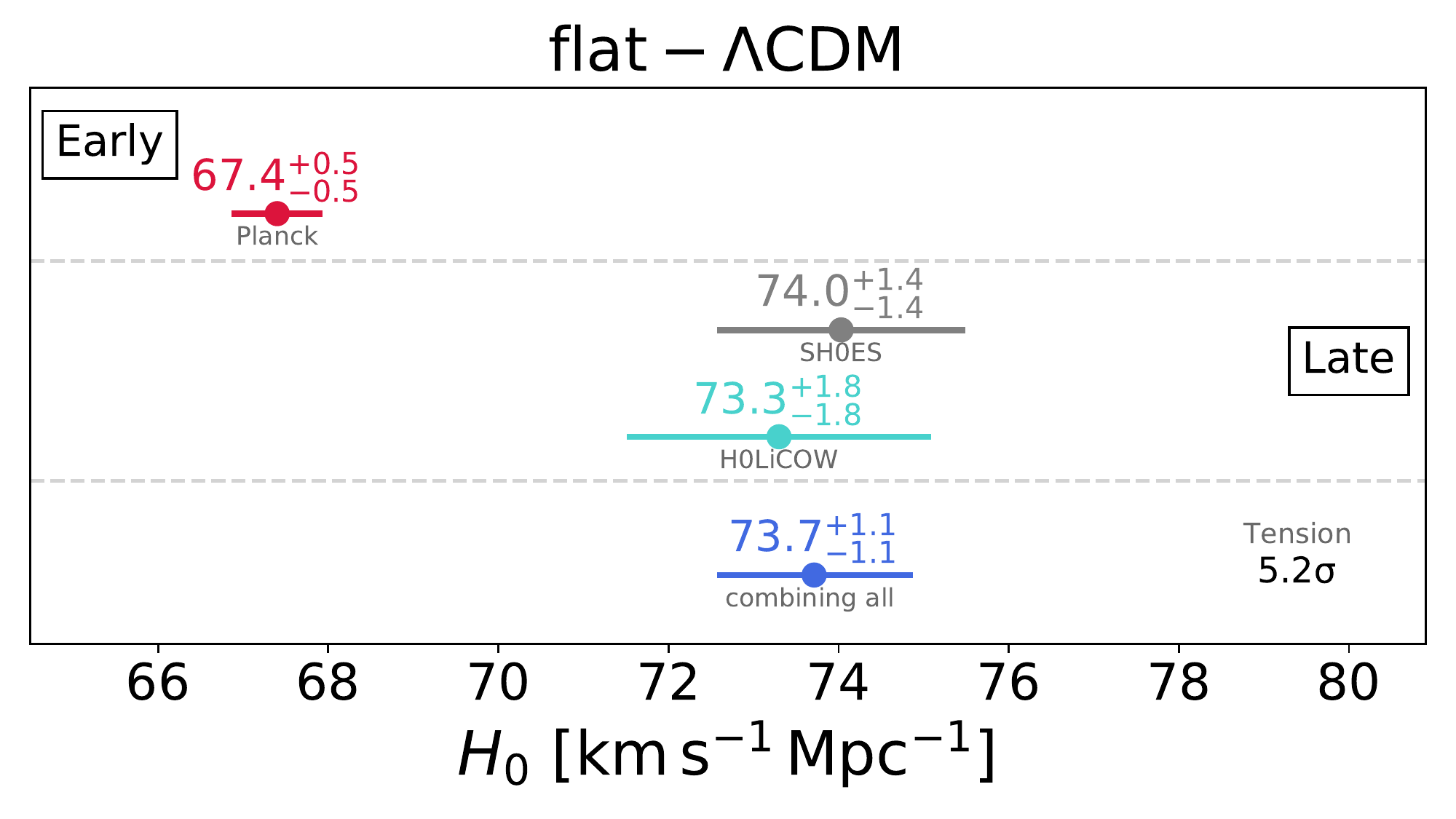}
    \caption{Reduced version of Figure~\ref{h0_tension_all} considering the observations we will use. Constraints on $H_0$ from early (top) and late (middle) time observations. The value of $H_0$ inferred from the combination of the two late Universe observations considered in this work (bottom) is also shown. There is a tension of $5.2\sigma$ between the value of $H_0$ inferred from the early Universe (Planck) and that obtained from late time observations (SH0ES+H0LICOW). Pantheon is not shown since it alone does not constraint $H_0$. See text for more details.}
    \label{h0_tension}
\end{figure}
\end{center}

The current constraints on $H_0$ from Planck, SH0ES, and H0LICOW can be seen in Figure~\ref{h0_tension}\footnote{For an editable version of Figure~\ref{h0_tension_all}, like the one presented in Figure~\ref{h0_tension} see \href{https://zenodo.org/record/3635517\#.Xr4Xwy_mFQJ}{here}~\cite{vivien_bonvin_2020_3635517}.}, where it is also included the $H_0$ value for the combination of these two late Universe observations, as well as the corresponding tension $T_{H_0}$ calculated according to the estimator~\cite{Camarena:2018nbr}
\begin{equation}
    T_{H_0} = \frac{\mid \mu_{early} - \mu_{late} \mid}{\sqrt{\sigma_{early}^2 + \sigma_{late}^2}}\, ,
    \label{tension_eq}
\end{equation}
where $\mu_{early}$ ($\mu_{late}$) is the mean value of $H_0^{early}$ ($H_0^{late}$), and $\sigma$ their corresponding standard deviation.

\subsection*{Supernovae (SNe Ia)}\label{sneia}
\begin{subequations}
Observations from the luminosity of SNe Ia allow to estimate their distances. The model for the observed distance modulus $\mu_{SNe}$ is the following~\cite{Scolnic:2017caz},
\begin{equation}\label{sne_obs}
    \mu_{SNe} = m_{B}^{*} - M\, ,\quad {\rm{with}}\quad m_{B}^{*} = m_B+ \tilde{\alpha} X_1 - \beta C + \Delta_M + \Delta_B\, ,
\end{equation}
where $m_B^{*}$ corresponds to the corrected apparent peak magnitude, $X_1$ describes the time stretching of the light-curve, $C$ stands for the supernova color at maximum brightness, and $\tilde{\alpha}\, , \beta\, ,M\, \Delta_M\, ,\Delta_B$ are nuisance parameters\footnote{We have used a tilde on the nuisance parameter $\tilde{\alpha}$ to avoid confusions with the diffusion parameter $\alpha$ from Model 1 STM. }
~\cite{Betoule:2014frx,Kessler:2016uwi,Jones:2018vbn}. On the other hand, the distance modulus which relies on the cosmological model is
\begin{equation}\label{sne_theo}
    \mu(z) = 5\log\left[ \frac{d_L(z)}{10{\rm{pc}}} \right]\, ,\quad {\rm{with}} \quad d_L(z) = \frac{c(1+z)}{H_0}\int_0^z \frac{dz^{\prime}}{E(z^{\prime})}\, ,
\end{equation}
where $d_L(z)$ is the luminosity distance. Then, the likelihood function will be,
\begin{equation}
    \log\mathcal{L}_{{\rm{SNe}}}(\Theta) = -\frac{1}{2}\left[ \vec{\mu}_{{\rm{SNe}}} - \vec{\mu}(\Theta) \right]^{T} \mathbf{C}^{-1}_{{\rm{SNe}}} \left[ \vec{\mu}_{{\rm{SNe}}} - \vec{\mu}(\Theta) \right]\, ,
\end{equation}
where $\vec{\mu}_{{\rm{SNe}}}$ and $\vec{\mu}(\Theta)$ are respectively given by Eq.~\eqref{sne_obs} and~\eqref{sne_theo} for each supernova, and $\mathbf{C}_{{\rm{SNe}}}$ is the covariance matrix. 

It is important to mention that $H_0$ can not be constrained when using only SNe Ia because $H_0$ and the nuisance parameter $M$ are strongly degenerated~\cite{Scolnic:2017caz}. Thus, data from SNe Ia should be combined with other observations in order to constraint $H_0$ properly.
\end{subequations}
\subsection*{Quasars (QSR)}
\begin{subequations}
Strong gravitational lenses allow to measure distances through the time delay between the multiple images of the lensed source. The $H_0$ Lenses in COSMOGRAIL'S Wellspring (H0LICOW) program have measured the current value of the Hubble parameter from a joint analysis of six gravitationally lensed quasars with measured time delay.

The time delay of an image $i$ in comparison with the no lensing case is~\cite{Suyu:2018vqs},
\begin{equation}
    t(\theta_i\, ,\beta) = \frac{D_{\Delta t}}{c}\phi(\theta_i\, ,\beta)\, ,
\end{equation}
where $\theta_i$ is the position of the lensed image of $i$, $\beta$ is the source position, $D_{\Delta t}$ is the so-called \textit{time-delay distance}, $\phi$ is the \textit{Fermat potential}, and $c$ is the speed of light. For a lens at redshift $z_d$ and a source at redshift $z_s$, the time-delay distance is given by,
\begin{equation}
    D_{\Delta t} = (1+z_d)\frac{D_dD_s}{D_{ds}}\, ,
    \label{qsr_theo}
\end{equation}
where $D_d$ and $D_s$ are the angular diameter distances to the lens and to the source respectively, whereas $D_{ds}$ is the angular diameter distance between the lens and the source (see Eq.~\eqref{ang_dist}). By measuring the time delay between two images $i$ and $j$, we have
\begin{equation}
    \Delta t_{ij} = t(\theta_i\, ,\beta) - t(\theta_j\, ,\beta) = \frac{D_{\Delta t}}{c}\Delta \phi_{ij}\, .
    \label{qsr_obs}
\end{equation}

Once determined the Fermat potential (by modeling the lens mass distribution), and with $\Delta t$ measured, it is possible to infer the value of the time-delay distance $D_{\Delta t}$. Thus, the likelihood function is,
\begin{equation}
    \log\mathcal{L}_{{\rm{QSR}}}(\Theta) = -\frac{1}{2}\left[ \vec{\mu}_{{\rm{QSR}}} - \vec{\mu}(\Theta) \right]^{T} \mathbf{C}^{-1}_{{\rm{QSR}}} \left[ \vec{\mu}_{{\rm{QSR}}} - \vec{\mu}(\Theta) \right]\, ,
\end{equation}
where $\vec{\mu}(\Theta)$ and $\vec{\mu}_{{\rm{QSR}}}$ are respectively given by Eq.~\eqref{qsr_theo} and~\eqref{qsr_obs} for each lensed quasar, and $\mathbf{C}_{{\rm{QSR}}}$ is the covariance matrix. 
\end{subequations}

\subsection*{Cepheids (CPH)}
Cepheids provide a way to measure distances through the period-luminosity relation characterizing them. Particularly, from 70 long-period Cepheids observed by the Hubble Space Telescope (HST) in the Large Magellanic Cloud, the Supernova $H_0$ for the Equation of State (SH0ES) collaboration  have determined the current value of the Hubble parameter $H_0^{CPH} = 74.03\pm 1.42$km/s/Mpc~\cite{Riess:2019cxk}. The likelihood function is then given by,
\begin{equation}
    \log\mathcal{L}_{{\rm{CPH}}}(H_0) = -\frac{1}{2}\frac{\left[ H_0^{{\rm{CPH}}} - H_0 \right]^2}{\sigma^2_{CPH}} \, .
\end{equation}

The likelihood function associated with these late Universe observations is given by,
\begin{equation}
    \log\mathcal{L}_{{\rm{Late}}}(\Theta) = \log\mathcal{L}_{{\rm{SNe}}}(\Theta) + \log\mathcal{L}_{{\rm{QSR}}}(\Theta) + \log\mathcal{L}_{{\rm{CPH}}}(H_0)\, ,
\end{equation}
and finally, the total likelihood function will be the following,
\begin{equation}
    \log\mathcal{L}_{tot}(\Theta) = \log\mathcal{L}_{{\rm{CMB}}}(\Theta) + \log\mathcal{L}_{{\rm{Late}}}(\Theta)\, .
\end{equation}

\begin{figure}[h!]
\centering
  \includegraphics[width=0.37\linewidth]{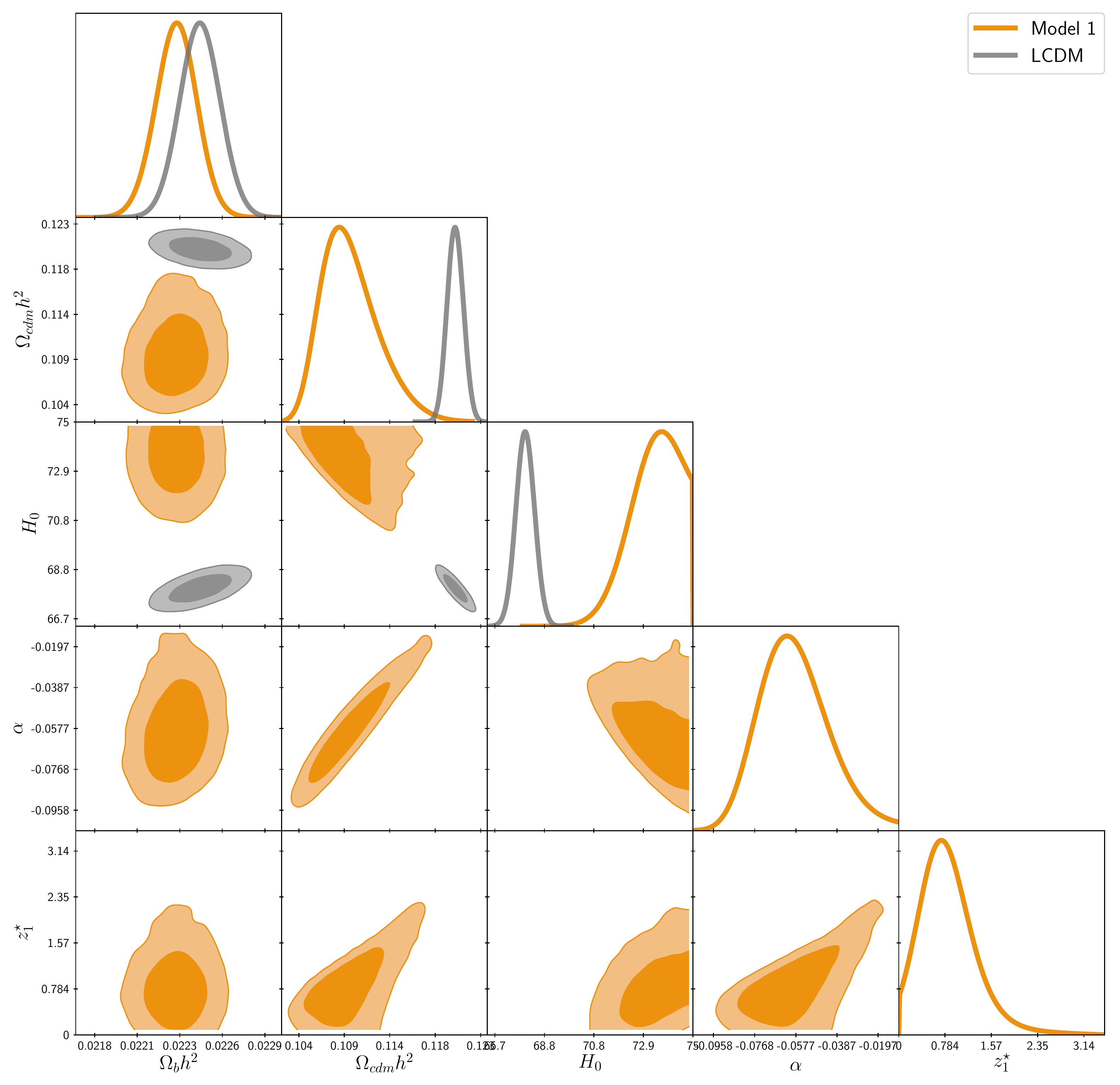}
\qquad
  \includegraphics[width=0.37\linewidth]{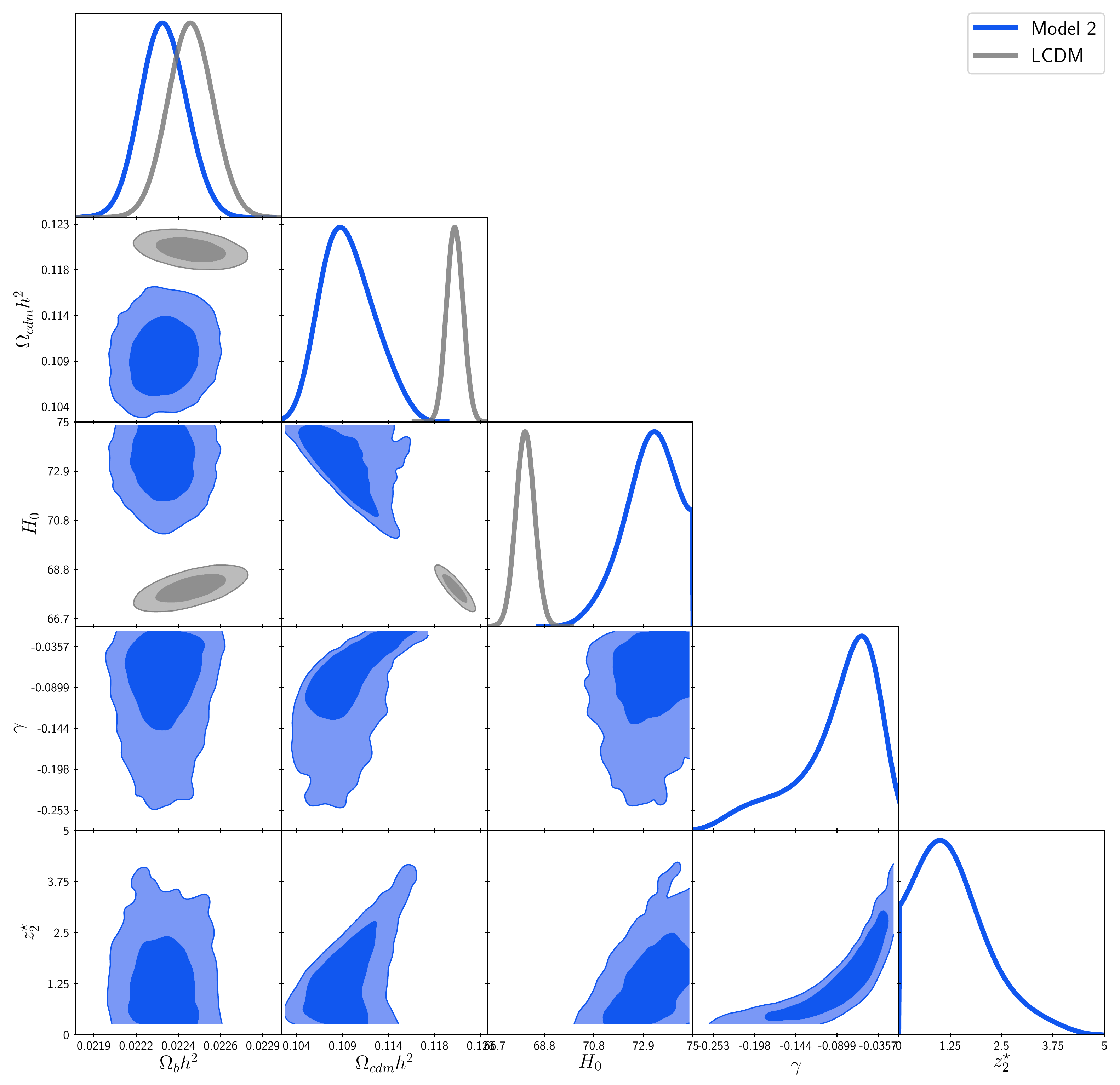}
\qquad
  \includegraphics[width=0.37\linewidth]{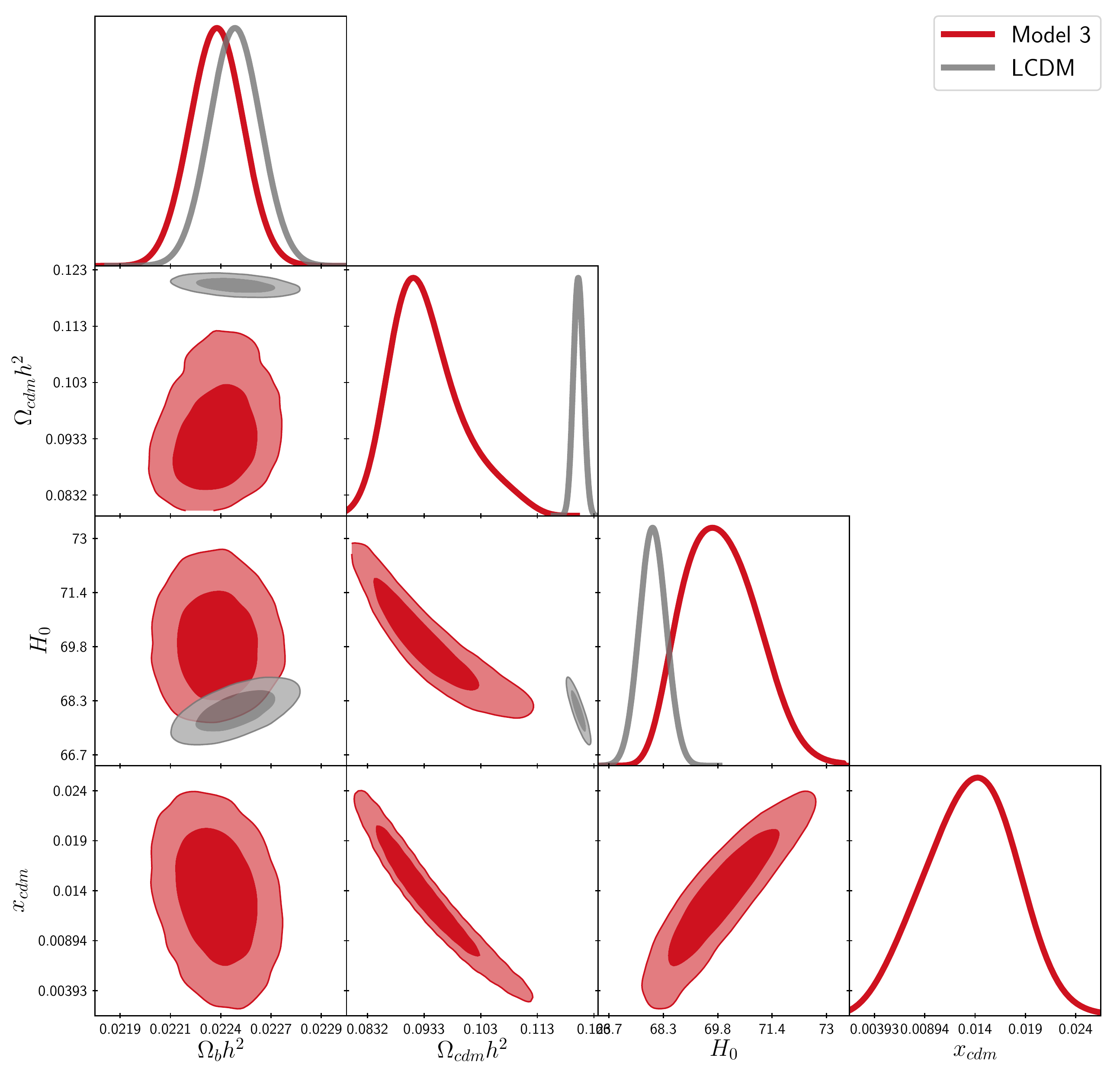}
\qquad
  \includegraphics[width=0.37\linewidth]{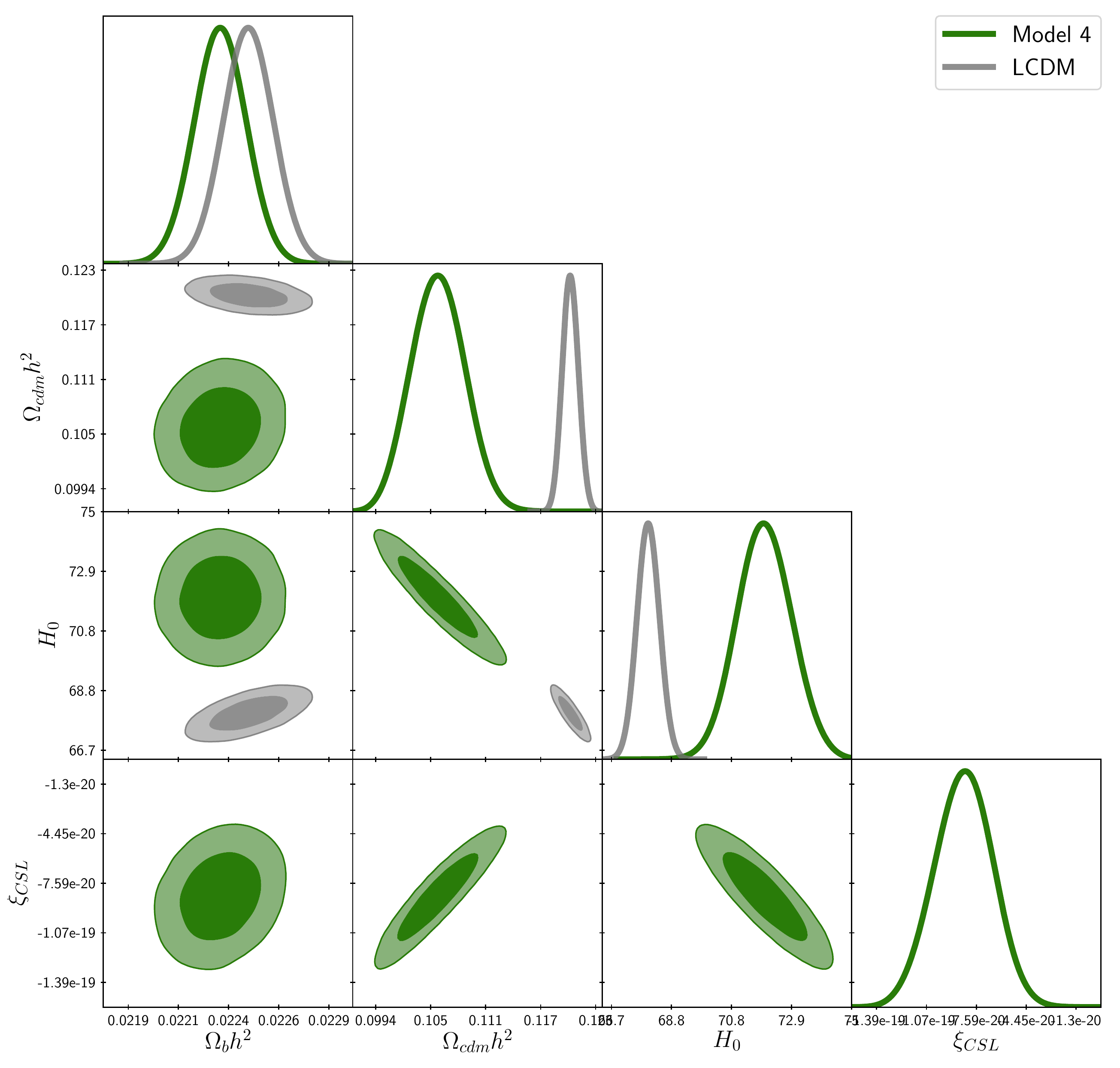}
\caption{Posterior probability distributions considering CMB+SNe+QSR+CPH data for all diffusion models under study: Sudden Transfer model (top right, orange), Anomalous Decay of the Matter Density model (top right, blue), Barotropic model (bottom left, red), and Continuous Spontaneous Localization model (bottom right, green). We have included the posteriors for the $\Lambda$CDM parameters as well (gray).}
\label{post_comb}
\end{figure}

Considering the same initial mean values, priors, and standard deviations shown in Table~\ref{paramfile}, Figure~\ref{post_comb} shows the posteriors for each diffusion model as well as those for the $\Lambda$CDM model. It can be observed that the posteriors for the $\Lambda$CDM parameters (gray) do not change considerably with respect to the posteriors obtained in the previous analysis when only CMB data were considered (see Eq.~\eqref{lcdm_cmb}). This time we have,
\begin{equation}
    \Omega_b h^2 = 0.0225^{+0.000134}_{-0.000132}\, ,\quad \Omega_{cdm} h^2 = 0.121^{+0.000877}_{-0.000887}\, ,\quad H_0 = 67.9^{+0.399}_{-0.403}\ {\rm{km}}\ {\rm{s}}^{-1}{\rm{Mpc}}^{-1}\, .
\end{equation}

Particularly, the mean value of $H_0$ slightly increases its value from $67.1$ to $67.9$ km s$^{-1}$ Mpc$^{-1}$ and, as expected from the $\Lambda$CDM model, the tension persists when this result is compared with the mean value of $H_0$ inferred from late time observations.

With the aim of quantifying how much the tension is eased in the case of the diffusion models, let us propose a similar estimator to that of Eq.~\eqref{tension_eq},
\begin{equation}
    \mathcal{T}_{H_0} \equiv \frac{\mid \mu_{comb} - \mu_{late} \mid}{\sqrt{\sigma_{comb}^{2} + \sigma_{late}^2}}\, ,
    \label{tension_eq_mod}
\end{equation}
where $\mu_{comb}$ and $\sigma_{comb}$ are respectively the mean and standard deviation of $H_0$ according to the combined analysis CMB+SNe+QSR+CPH, whereas $\mu_{late}$ and $\sigma_{late}$ are the mean and standard deviation of $H_0$ from the SH0ES and H0LICOW combined observations given by $H_0=73.7^{+1.1}_{-1.1}$km s$^{-1}$Mpc$^{-1}$ (see bottom panel of Figure~\ref{h0_tension})\footnote{The presence of SNe Ia in the combined analysis we have made does not affect the results obtained from the estimator~\eqref{tension_eq_mod}. Since SH0ES and H0LICOW impose the late time constraint on $H_0$, once combined, SNe Ia data are in agreement with such constraints. Thus, we do not expect a substantial change in the combined SH0ES+H0LICOW value $H_0^{late} = 73.7^{+1.1}_{-1.1}$km s$^{-1}$ Mpc$^{-1}$ due to SNe Ia. }. In particular, we define $\sigma_{comb}$ as the following mean value $\sigma_{comb}\equiv (\sigma_{comb}^+ + \sigma_{comb}^{-})/2$ to incorporate the case of an asymmetric standard deviation.

It is important to recall that when only CMB data are considered, the posteriors of $H_0$ for the diffusion models are weakly constrained, and there is not a mean value with an associated standard deviation that can be used to compare with local observations. As discussed in the previous Section, the presence of new parameters allows a broad range of values for the current Hubble parameter. On the other hand, since the value of $H_0$ does not change considerably between the CMB analysis and that of CMB+SNe+QSR+CPH, the results from the latter give approximately the same tension as that obtained from the former. This motivates us to propose Eq.~\eqref{tension_eq_mod}. Table~\ref{output_info} shows the mean and the 1--$\sigma$ confidence level for the $\Lambda$CDM and diffusion parameters. The tension $\mathcal{T}_{H_0}$ for each model is shown as well.
\begin{table}[h!]
\centering
\begin{tabular}{cccccc}
\hline
\hline
model  & parameter    & mean$^{+1\sigma}_{-1\sigma}$ & $\mathcal{T}_{H_0}$  \\
\hline
\hline
& $\Omega_b h^2$ & $0.0225^{+0.000134}_{-0.000132}$  & $\cdots$  \\
$\Lambda$CDM & $\Omega_{cdm}h^2$ & $0.121^{+0.000877}_{-0.000887}$  & $\cdots$  \\
& $H_0$ & $67.9^{+0.399}_{-0.403}$  & $5.0\sigma$  \\
\hline
& $\Omega_b h^2$ & $0.0223^{+0.000135}_{-0.000135}$  & $\cdots$  \\
& $\Omega_{cdm}h^2$ & $0.109^{+0.00233}_{-0.00356}$  & $\cdots$  \\
Model 1 & $H_0$ & $73.4^{+1.46}_{-0.588}$  & $0.2\sigma$  \\
& $\alpha$ & $-0.0582^{+0.0134}_{-0.0182}$ & $\cdots$  \\
& $z^{\star}_1$ & $0.842^{+0.318}_{-0.547}$ & $\cdots$  \\
\hline
& $\Omega_b h^2$ & $0.0223^{+0.000134}_{-0.000139}$  & $\cdots$  \\
& $\Omega_{cdm}h^2$ &  $0.109^{+0.00254}_{-0.0032}$ & $\cdots$  \\
Model 2 & $H_0$ & $73.2^{+1.38}_{-0.86}$  & $0.3\sigma$  \\
& $\gamma$ & $-0.089^{+0.0672}_{-0.0212}$ & $\cdots$  \\
& $z^{\star}_2$ & $1.41^{+0.414}_{-1.13}$ & $\cdots$  \\
\hline
& $\Omega_b h^2$ & $0.0224^{+0.000138}_{-0.000139}$  & $\cdots$  \\
& $\Omega_{cdm}h^2$ & $0.094^{+0.00467}_{-0.00784}$  & $\cdots$  \\
Model 3 & $H_0$ & $70^{+0.949}_{-1.17}$  & $2.4\sigma$  \\
& $x_{cdm}$ & $0.0135^{+0.00475}_{-0.00436}$ & $\cdots$  \\
\hline
& $\Omega_b h^2$ & $0.0223^{+0.000136}_{-0.000138}$  & $\cdots$  \\
& $\Omega_{cdm}h^2$ & $0.106^{+0.00284}_{-0.00301}$  & $\cdots$  \\
Model 4 & $H_0$ & $72^{+0.951}_{-0.969}$  & $1.1\sigma$  \\
& $\xi_{CSL}$ & $\left(-8.4^{+1.86}_{-1.82}\right)\times 10^{-20}$ & $\cdots$ \\
\hline
\end{tabular}
\caption{Mean values and their corresponding $1\sigma$ confidence level for the parameters of each diffusion model, as well as for the $\Lambda$CDM parameters. The units of $H_0$ and $\xi_{CSL}$ are given respectively by km s$^{-1}$ Mpc$^{-1}$ and s$^{-1}$. The last column indicates the tension $\mathcal{T}_{H_0}$ according to Eq.~\eqref{tension_eq_mod}. See the text for more details.}
\label{output_info}
\end{table}

We want to end this Section by highlighting the following: having included local observations in the analysis led to specific and well--defined values of the diffusion parameters. In particular, the amount of CDM and $H_0$ get precise values, opposite of the case in which only CMB data were used. The amount of total matter $\left(\Omega_b + \Omega_{cdm}\right)h^2$ predicted by the diffusion models is less than the $\Lambda$CDM case, which translates in a higher value for $H_0$. This indicates that the diffusion models are, effectively, good candidates alleviating the $H_0$ tension. In this sense, our statistical analysis verifies what we have obtained numerically in the previous Section: within the framework of Unimodular Gravity, the non--gravitational interaction between the dark sector components through diffusion processes, leads to an inferred value of $H_0$ consistent with local observation when constraining the diffusion parameters with CMB data and CMB+local observations.

\section{Discussion and final remarks}\label{final}

Whereas the main ingredient responsible of the dynamics of the late Universe remains unknown, the best model to explain the current accelerated expansion is given by the cosmological constant in the Einstein field equations. Nonetheless, the origin of such constant is still unknown, although there are several theoretical proposal trying to understand its nature. Unimodular Gravity offers an explanation of the origin of the cosmological constant: it arises as an integration constant once volume--preserving diffeomorphisms are considered in the Einstein--Hilbert action. The dynamics of the spacetime metric is governed by the trace--free version of the Einstein field equations and, as a direct consequence, the energy--momentum tensor is not conserved once the Bianchi identities are applied. The non--conservation of the energy--momentum tensor allows matter components to interact with each other non--gravitationally. Since the physics of the standard model of particles is very well understood, we have assumed the non--gravitational interaction to be possible only between the dark sector components, which are given by a cold dark matter fluid $\rho_{cdm}$ and an effective cosmological constant with dependence on the cosmic time $\Lambda(t)$. To describe the interaction between $\rho_{cdm}$ and $\Lambda(t)$, we have considered phenomenological models previously studied in~\cite{Perez:2020cwa,Corral:2020lxt}, where the authors propose diffusion processes as the mechanism for the energy transfer from one dark component to the other.

In general, models with non--gravitational interactions in the dark sector have been studied with emphasis in the $H_0$ tension, this is, the current discrepancy between the value of the Hubble parameter at present day inferred from the early Universe, and that inferred from local observations. The authors in~\cite{Perez:2020cwa,Corral:2020lxt} have also analyzed the possibility of easing such tension through diffusion processes between $\rho_{cdm}$ and $\Lambda(t)$ in UG. However, their analysis has been restricted to the dynamics of the late Universe, which, although some hints about the viability of these models can be obtained, a more suitable analysis must involve physics of the early Universe, in particular the radiation component has to be included in the cosmological evolution. This is important because any model addressing the $H_0$ tension must be tested not only with late time observations, but with CMB data as well. For the latter, it is mandatory to evaluate the model at high redshifts, in particular at the moment of photon--baryon decoupling occurring at $z_*\simeq 1100$. Thus, we have implemented each of the diffusion models studied in the works mentioned before, in a more realistic cosmological scenario, where the background evolution includes the presence of photons and ultra--relativistic neutrinos as the components of radiation in the Universe.

Considering a spatially--flat FRW line element, we have solved the UG field equations, from very deep in the radiation dominated era (initial redshift $z(t_i) = 10^{14}$) to the present day ($z(t_0) = 0$), with special interest in the cosmological evolution of the energy density parameters $\Omega_i$ where $i=b\, , \gamma\, , \nu\, , cdm\, , \Lambda$ stands for baryons, photons, neutrinos, cold dark matter, and cosmological constant respectively. We explicitly showed how the diffusion process takes place between $\Omega_{cdm}$ and $\Omega_{\Lambda}$ for each of the four diffusion models studied: Sudden Transfer Model (STM) and Anomalous Decay of the Matter Density (ADMD) from~\cite{Perez:2020cwa}, and Barotropic Model (BM) and Continuous Spontaneous Localization (CSL) from~\cite{Corral:2020lxt}, for which we have obtained the corresponding modified Friedmann equation~\eqref{ez_1},~\eqref{ez_2},~\eqref{ez_3}, and~\eqref{ez_4_gen}. For the BM and CSL model, we have implemented a different approach to that of the authors in~\cite{Corral:2020lxt}: instead of separating the analysis by cases depending on the contribution of the cosmological constant energy density, we have solved the full set of background equations for all the cosmological evolution considering the modification on $H(z)$ due to the diffusion models. In particular, this was not direct to do for the CSL model, whose modification involves an integro--differential equation for $z(t)$. In appendix~\ref{igf} we show the method we have used to address this difficulty, which allowed us to write in a closed form the modification in $H(z)$ due to the CSL diffusion model. It was crucial to have the expression of the modified Friedmann equation for each diffusion model, since later in the statistical analysis, $H(z)$ enters in the angular diameter distance $D_A$ that is needed to compute the observables from CMB (Eq.~\eqref{la_R}) and Quasars (Eq.~\eqref{qsr_theo}), as well as in the distance modulus $\mu$ through the luminosity distance $d_L$ for SNe Ia (Eq.~\eqref{sne_theo}).

We explored the influence of the diffusion parameters on the dynamics of the dark sector, as well as how these parameters modify the current amount of the dark components energy densities. The magnitude and sign of the diffusion parameters set the contribution of both, $\Omega_{cdm}$ and $\Omega_{\Lambda}$, and thus, the Hubble parameter at present day change its value as well. In particular, when considering that the values of each $\Omega_i$ and $H_0$ are those reported by CMB observations (and then $H_0 = H_0^{early} \simeq 67$ km s$^{-1}$ Mpc$^{-1}$), we have shown that it is possible to find values of the diffusion parameters such that the diffusion models lead to a Hubble parameter at $z=0$ consistent with the value obtained from data of the local Universe (i.e., $H_0 = H_0^{late} \simeq 73$ km s$^{-1}$ Mpc$^{-1}$). Therefore, at the level of the numerical solutions, all the diffusion models ease the $H_0$ tension. Nonetheless, to truly test these models and to analyze properly their viability as compelling candidates to solve the discrepancy on the measurement of the current value of the Hubble parameter, it is mandatory a statistical analysis. While such analysis was not performed in~\cite{Perez:2020cwa}, the authors in~\cite{Corral:2020lxt} constraint the BM and CSL model by using data from late time observations (Observational Hubble Data (OHD) and supernovae (SNe Ia)). As we mentioned before, to establish a proper comparison between the predictions of the diffusion models in what the $H_0$ tension is concerned, not only data set of local observations have to be considered, but from the early Universe as well. We have done this by using the Planck Compressed 2018 data. 

Since the diffusion models predict less amount of matter in order to raise the value of $H_0$, we have let to vary both the physical energy density for baryons $\Omega_{b}h^2$ and CDM $\Omega_{cdm}h^2$, considering the broadest priors possible. Besides, the prior in $H_0$ included the observed mean values from both, early and late time Universe. In the case of the diffusion parameters, we also considered the broadest priors possible. In fact, by setting the more broadest priors have costed to expend more time for the chains to converge when exploring the parameter space. This is so because the new parameters from the diffusion models induce a longer anticorrelation between $\Omega_{cdm}h^2$ and $H_0$, and more possible combinations of such parameters are allowed by the observations in comparison with the $\Lambda$CDM case. When considering only CMB observations, the diffusion parameters allow to ease the tension on $H_0$, for which weaker constraints are imposed. In fact, the CDM energy density parameter $\Omega_{cdm}$ is weakly constrained as well, and given that the diffusion parameters lead to lower values of $\Omega_{cdm}$, this translates to higher values of $H_0$. Thus, the posteriors for $H_0$ get a broader range of values, which includes those inferred from early and late time Universe (see Figure~\ref{post_cmb}). It is when local observations are included in the analysis that an actual shift in the mean value of $H_0$ occurs (see Figure~\ref{post_comb}).

In the case of the model 1 (STM), negative values of the diffusion parameter $\alpha$ are preferred by observations. This is not the case in~\cite{Perez:2020cwa} where such parameter is defined such that $0<\alpha<1$, and the parameter space is restricted for positive $\alpha$ only. However, our result is not in contradiction with the prediction of the model: the contribution of matter today will be less than the predicted by $\Lambda$CDM. In fact, this is a prediction of all the diffusion models studied, as can be seen in Figure~\ref{post_comb} where both $\Omega_b h^2$ and $\Omega_{cdm}h^2$ for the STM (red), ADMD (blue), BM (red), and CSL (green) are shifted to the left in comparison with the $\Lambda$CDM case (gray). Since the amount of baryons is a well--measured quantity, the shift is less than $1\%$, whereas for CDM the differences are approximately $9.9\%$ for STM and ADMD, $22.3\%$ for BM, and $12.4\%$ for CSL. For the diffusion parameter $\gamma$ from model 2 (ADMD), we observe that a negative value is preferred as well. The characteristic redshift $z^{\star}$ for models 1 and 2 are constrained to low redshifts, this is, the diffusion processes described by these models occur in the late Universe. Such constraints were obtained under the assumption that the diffusion process could take place at any time between the last scattering surface and the present day, which was important to consider since, according to the analysis of the parameter space carried out in~\cite{Perez:2020cwa}, larger values of $z^{\star}$ lead to values of $\alpha$ and $\gamma$ such that the $\Lambda$CDM case is recovered, i.e., for $z^{\star}>>1\, ,$ $\alpha\, , \gamma \rightarrow 0$ (see Figures 3 and 6 from~\cite{Perez:2020cwa}).

The constraint we have obtained on the diffusion parameter $x_{cdm}$ from model 3 (BM) is in agreement with that reported in~\cite{Corral:2020lxt} for the case in which the authors infer the value of $x_{cdm}$ from OHD and SNe Ia independently ($x_{cdm}=0.027^{+0.046}_{-0.045}\, , 0.041^{+0.310}_{-0.194}$ for OHD and SNe Ia respectively). In the joint analysis OHD--SNe Ia, the constraint obtained in~\cite{Corral:2020lxt} is given by $x_{cdm}=0.054^{+0.035}_{-0.032}$, which is larger than the inferred value in our analysis. Nonetheless, as can be compared with the reported values in~\cite{Corral:2020lxt} shown above and our results, our uncertainties for $x_{cdm}$ are smaller (see Model 3 in Table~\ref{output_info}). In the case of $H_0$, our result is consistent with that of the authors, at least when compared with their analysis for OHD and OHD+SNe Ia. As we mentioned, the Pantheon data set alone cannot be used to constraint the $H_0$ since it is degenerated with $M$~\cite{Scolnic:2017caz}. In order to have a $H_0$ measurement only from SNe Ia, we have noticed that in~\cite{Corral:2020lxt} the authors have rewritten the supernovae likelihood only as a function of the redshift and the cosmological parameters, and no nuisance parameters are present (see Eq.55 in~\cite{Corral:2020lxt}). This gives a value of $H_0=73.2^{+1.7}_{-1.7}$, which is larger than the inferred value of $H_0$ in our analysis for model 3. In the case of model 4 (CSL), the inferred value for $\xi_{CSL}$ is incompatible with the constraints found in~\cite{Josset:2016vrq,Martin:2019jye,Corral:2020lxt}, where positive values for such parameter are inferred, whereas from our analysis we have obtained that $\xi_{CSL}<0$. Nonetheless, let us to pinpoint out the following aspects to be regarded: in the works mentioned above, the non--gravitational interaction includes ordinary matter, since the authors consider energy transfer between the cosmological constant and the total matter content $\rho_m$. It has to be recalled that in our case, the violation of the energy--momentum tensor is produced only through the components of the dark sector. On the other hand, as discussed in~\cite{Josset:2016vrq} the sign of $\xi_{CSL}$ can be negative, which implies an endothermic evolution. The latter scenario can be obtained, for instance, from approaches to quantum gravity such as \textit{causal set}~\cite{Dowker:2003hb,Philpott:2008vd}.

With the aim of quantifying how much these diffusion models ease the tension on $H_0$, we proposed an estimator $\mathcal{T}_{H_0}$ to compute the differences between our combined analysis CMB+SNe+QSR+CPH and local observations from SH0ES and H0LICOW (see Eq.~\eqref{tension_eq_mod}). The model 1 (STM) is the one reducing more the tension with a difference of $0.2\sigma$, whereas model 3 (BM) gives the larger difference by easing the tension at $2.4\sigma$. These results indicate that diffusion models in UG are viable theoretical proposals to solve the $H_0$ tension.

It will be interesting to explore the cosmological perturbations for the diffusion models studied. Previous works have made some contributions in this direction, although imposing by hand the conservation of the energy--momentum tensor, which as we have shown it is not the most general way to solve the UG field equations. Even so, in~\cite{Gao:2014nia} the authors show that temperature fluctuations of the Cosmic Microwave Background (CMB) radiation, and specifically the Sachs--Wolfe effect~\cite{Sachs:1967er} has a correction term given by a scalar metric perturbation that is demanded to be non--vanishing in UG. Assuming adiabatic fluctuations, the authors of the mentioned work found that the main difference between the GR and UG prediction is only a dipole--like term which is suppressed at large scale, and thus, the effect induced by UG is negligible. Moreover, in terms of gauge--invariant quantities, it was shown in~\cite{Basak:2015swx} that the GR and UG cosmological perturbations are identical, and then, CMB photons will not distinguish between the two theories. An analysis including the non--conservation of the energy--momentum tensor has to be made in order to explore possible deviations of GR in cosmological observables such as CMB anisotropies and large scale structures.

While the discrepancies between early and late Universe in the $H_0$ measurements may be due to still unaccounted systematic errors in the observations, there exists the possibility that new physics is needed in order to unravel this cosmological conundrum. A deeper comprehension of the physical mechanisms driving the diffusion processes is needed in order to have a complete description of the non--gravitational interaction between the dark sector components. A recent theoretical proposal for such diffusion processes, is that of considering the effect of spacetime granularity at Planck scale on rotating black holes, where it is found that the effect would be mainly due to the energy associated to the slowing down of the rotation of Kerr-type black holes~\cite{Perez:2019gyd}. Therefore, the phenomenological models studied here might constitute a compelling variation to the $\Lambda$CDM paradigm. In this sense,  Unimodular Gravity offers not only an explanation to the origin of the cosmological constant, but also the non--conservation of the energy--momentum tensor that naturally arises in this gravitational framework allows to address the $H_0$ tension successfully.

\section*{Acknowledgements}

We thank Cristóbal Corral, Daniel Sudarsky, and Yuri Bonder for useful comments and suggestions. Francisco X. Linares Cedeño acknowledges the receipt of the grant from the Abdus Salam International Centre for Theoretical Physics, Trieste, Italy. FXLC and UNG acknowledge Consejo Nacional de Ciencia y Tecnolog\'ia (CONACYT), Sistema Nacional de Investigadores (SNI) and the Programa para el Desarrollo Profesional Docente (PRODEP) for financial support. UNG acknowlegdes Coordinación de la Investigación Científica de la Universidad Michoacana de San Nicolás de Hidalgo (UMSNH). FXLC thanks the support from CONACyT Mexico under Grant No. A1-S-17899; and the Instituto Avanzado de Cosmologia Collaboration.

\appendix
\section{The Incomplete Gamma Function and the CSL model}\label{igf}

The Continuous Spontaneous Localization (CSL) model leads to an expression for the CDM energy density that involves the following integral
\begin{equation}
    I(t)\equiv \int_0^te^{\xi_{CSL}t^{\prime}}[1 + z(t^{\prime})]^3dt^{\prime}\, .
    \label{int_orig}
\end{equation}

We need to solve the background evolution to be able to give the explicit function $z(t)$, and thus, the Friedmann equation~\eqref{ez_4} is an integro--differential equation,
\begin{equation}
    \frac{1}{H_0^2[1 + z(t)]^2}\left[\frac{dz(t)}{dt}\right]^2 = \Omega_{r}[z(t)] + \Omega_{b}[z(t)] + \Omega_{cdm}[z(t)]e^{\xi_{CSL}t} - \Omega_{cdm}\xi_{CSL}\int_0^t e^{\xi_{CSL}t^{\prime}}[1+z(t^{\prime})]^3dt^{\prime}  + \Omega_{\Lambda_{eff}}\, .
    \label{int_diff_fried}
\end{equation}

This will be the case even if we try to integrate $I(t)$ by parts: let be $f(t)\equiv[1+z(t)]^3$, then
\begin{eqnarray}
    I(t) &=& \left.\frac{f(t^{\prime})e^{\xi_{CSL}t^{\prime}}}{\xi_{CSL}}\right|_0^t - \int_0^t \frac{e^{\xi_{CSL}t^{\prime}}}{\xi_{CSL}}\dot{f}(t^{\prime})dt^{\prime}\nonumber \\
    &=& \frac{[1+z(t)]^3e^{\xi_{CSL}t}}{\xi_{CSL}} - \frac{1}{\xi_{CSL}} + \frac{3}{\xi_{CSL}}\int_0^t e^{\xi_{CSL}t^{\prime}}[1 + z(t^{\prime})]^3H(t^{\prime})dt^{\prime}\, .
\end{eqnarray}

A way to handle this situation, is by proposing an ansatz on the temporal dependence of the scale factor $a(t)$, or equivalently, the redshift $z(t)$. This is in fact usually done in the framework of $\Lambda$CDM model, where it is well--known that a power law relates the scale factor with the cosmic time as $a(t) = (t/t_0)^p$, with $p=1/2\ (p=2/3)$ for radiation (matter) domination era. Thus, considering that the new term arising from the CSL diffusion process will not change drastically such power law relation, let us propose
\begin{equation}
    a(t) = \frac{1}{1 + z(t)} \equiv \left(\frac{t}{t_0}\right)^p\, ,
    \label{ansatz}
\end{equation}
and then the integral~\eqref{int_orig} is written as
\begin{equation}
    I(t)\equiv \int_0^te^{\xi_{CSL}t^{\prime}}\left( \frac{t^{\prime}}{t_0} \right)^{-3p}dt^{\prime}\, .
    \label{int_ansatz}
\end{equation}

If we now introduce the following change of variable $u\equiv -\xi_{CSL}t$, we have
\begin{equation}
    I_s(u)\equiv \frac{t_0}{u_0^s}\int_0^ue^{-u^{\prime}}u^{\prime s-1}du^{\prime}\, ,
    \label{int_u}
\end{equation}
where we have defined $-3p\equiv s-1$, and $u_0\equiv-\xi_{CSL}t_0$. The integral in the last expression looks like the \textit{lower incomplete Gamma function} $\gamma(s,u)$~\cite{abramowitz1965handbook,Gautschi98theincomplete,olver2010nist,jameson_2016}, which has the following series representation~\cite{adams1922smithsonian,1953hft1.book...59E,gradshteyn2014table}
\begin{equation}
    \gamma(s,u) = \int_0^ue^{-u^{\prime}}u^{\prime s-1}du^{\prime} = \sum_{k=0}^{\infty}\frac{(-1)^ku^{k+s}}{k!(k+s)}\, ,\quad {\rm{for}}\ s > 0\ {\rm{and}}\ u\geq 0\, .
\end{equation}

Thus, it is mandatory to verify whether $\left\lbrace u\, ,s\right\rbrace$ satisfy such conditions. Given the change of variable $u\equiv-\xi_{CSL}t$, we observe that $\xi_{CSL}\leq0$ is needed to have $u\geq 0$. On the other hand, $s > 0$ implies $1-3p > 0$. Considering values of the diffusion constant such that $\mid\xi_{CSL}\mid<<1$, we expect that the power law given by the ansatz
~\eqref{ansatz} does not differ too much from the $\Lambda$CDM result, and then $s = -1/2\ (s=-1)$ for radiation (matter) domination era. Moreover, it can be seen that the constant term out of the integral~\eqref{int_u} will be imaginary in radiation domination if $\xi_{CSL} > 0$. This indicates that even when $u$ will be well--defined for $\xi_{CSL} < 0$, the value of $s$ in the cosmological context we are studying lies within the set of negative integers and negative semi--integers. Therefore, we cannot identify our integral $I_s(u)$ in terms of $\gamma(s,u)$, at least in its standard form.

Before going further, let us summarize the above discussion as follows: the only case that will lead to an integral with physical meaning ($I_s(u) \in \mathbb{R}$), is when $\xi_{CSL}\leq0\Rightarrow u\geq 0\, , $ with $s<0\, .$ Since the lower incomplete Gamma function $\gamma(s,u)$ does not admit negative values of the so-called \textit{shape parameter} $s$, some sort of extension or generalization has to be implemented. This has been possible by using the tools of \textit{Neutrix Calculus}~\cite{hadamard1923lectures,van1959introduction,fisher1976neutrices}, when studying the asymptotic behavior of divergent integrals. The main use of neutrix calculus and neutrix limit, is to extract the finite part of divergent quantities. This has been used, for example, in Quantum Field Theory to obtain finite renormalizations in loop calculations~\cite{Ng:2004tk,Ng:2005er}. Within this context, different generalizations of the standard and well--known lower incomplete Gamma function have been developed~\cite{CHAUDHRY199499,fisher2003defining,fisher2004defining,ozccaug2007some,fisher2012some,thompson2013algorithm,ozcaug2016remarks,lin2019incomplete}. Particularly in~\cite{fisher2003defining,ozccaug2007some,ozcaug2016remarks} have developed an extension to consider negative integers given by
\begin{equation}
    \gamma(-s,u) = \frac{(-1)^s}{s!}\ln u + \sum_{\substack{k=0\\ k\neq s}}^\infty \frac{(-1)^ku^{k-s}}{k!(k-s)}\, ,\quad {\rm{for}}\quad s\in \mathbb{N}\ {\rm{and}}\ u>0\, .
    \label{gamma_for_md}
\end{equation}

Since $s\in \mathbb{N}$, the above equation will be useful for the matter domination era, where we are interested in $\gamma(-1,u)$.

On the other hand, another way to write the lower incomplete Gamma function is given by its suitably normalized form~\cite{bohmer1939differenzengleichung,tricomi1950asymptotische}
\begin{equation}
    \gamma^*(s,u) = \frac{u^{-s}}{\Gamma(s)}\gamma(s,u)\, .
    \label{ligf_norm}
\end{equation}

Written like this, $\gamma^*(s,u)$ is a real--valued function for $s,u \in {\rm I\!R}$ and such that $s,u>0$. An extension of this formula was given in~\cite{gautschi1977evaluation,gautschi1979computational} to consider negative and real values of $s$ with $u>0$, where particularly for $s\geq-1/2$ we have
\begin{equation}
    \gamma^*(s,u) = \frac{1}{\Gamma(s+1)} \sum_{k=0}^\infty \frac{s(-u)^k}{k!(k+s)}\, .
\end{equation}

Therefore, using Eq.~\eqref{ligf_norm} we have that
\begin{equation}
    \gamma(s,u) = \frac{\Gamma(s)u^s}{\Gamma(s+1)} \sum_{k=0}^\infty \frac{s(-u)^k}{k!(k+s)}\, ,\quad {\rm{for}}\quad s\geq-1/2\ {\rm{and}}\ u>0\, .
    \label{gamma_for_rd}
\end{equation}

The last equation can then be used during the radiation dominated era, where $s=-1/2$. Thus, the integral in Eq.~\eqref{int_u} can be now formally identified with the \textit{Generalized lower incomplete Gamma function} $\gamma^G(s,u)$ as follows,
\begin{equation}
    I_s(u) = \frac{t_0}{u_0^s}\gamma^G(s,u)\, ,\quad {\rm{for}}\ u>0\, ,
    \label{I_complete}
\end{equation}
where, using the results of Eq.~\eqref{gamma_for_md} and Eq.~\eqref{gamma_for_rd}, we have
\begin{equation}
   \gamma^G(s,u) =  \left\lbrace\begin{matrix} 
\gamma(-1,u) = -\ln u + \sum_{\substack{k=0\\ k\neq 1}}^\infty \frac{(-1)^ku^{k-1}}{k!(k-1)}\, , & {\rm{for}}\quad s=-1\, , \\
\gamma(-1/2,u) = -\frac{1}{u^{1/2}}\sum_{k=0}^\infty \frac{(-u)^k}{k!(k-1/2)}\, , & {\rm{for}}\quad s=-1/2\, , 
\end{matrix}\right.
\label{gen_gamma}
\end{equation}
where we have used that $\Gamma(1/2) = \sqrt{\pi}$ and $\Gamma(-1/2) = -2\sqrt{\pi}$. The Gamma function $\Gamma(s)$ is classically defined for positive integers, but it can be extended to both, real and imaginary numbers as well (see for instance~\cite{thukral2014factorials}). Now, let us work on the series shown in Eq.~\eqref{gen_gamma}.

\subsection{$\mathbf{s=-1}:$}
In matter domination era we have
\begin{equation}
    \gamma^G(-1,u) = -\ln u +\sum_{\substack{k=0\\ k\neq 1}}^\infty \frac{(-1)^ku^{k-1}}{k!(k-1)} \simeq -\ln u -\frac{1}{u} + \frac{u}{2} - \frac{u^2}{12} + \dots\, ,
\end{equation}
which, as function of the redshift is written as
\begin{equation}
    \gamma^G(-1,z) \simeq -\ln\left[ \frac{u_0}{(1 + z)^{3/2}}\right] - \frac{(1+z)^{3/2}}{u_0} + \frac{u_0}{2(1 +z)^{3/2}} - \frac{u_0^2}{12(1+z)^3} + \dots\, .
\end{equation}

Then, Eq.~\eqref{I_complete} is given by
\begin{equation}
    I_{-1}(z) \simeq -t_0u_0\ln\left[ \frac{u_0}{(1 + z)^{3/2}}\right] - t_0(1+z)^{3/2} + \frac{t_0u_0^2}{2(1 +z)^{3/2}} - \frac{t_0u_0^3}{12(1+z)^3} + \dots\, .
\end{equation}

Recalling that this integral has a multiplicative factor of $\xi_{CSL}$ (see Eq.~\eqref{int_diff_fried}), and considering the dimensionless variable $u_0\equiv -\xi_{CSL}t_0$ as the parameter with which the expansion is developed,
\begin{equation}
   \xi_{CSL} I_{-1}(z) \simeq  u_0(1+z)^{3/2} + u_0^2\ln\left[ \frac{u_0}{(1 + z)^{3/2}}\right] - \frac{u_0^3}{2(1 +z)^{3/2}} + \frac{u_0^4}{12(1+z)^3} + \dots\, ,\quad {\rm{for}}\ u_0<<1\, .
\end{equation}

Therefore, if we neglect terms of order equal and higher than $\mathcal{O}(u_0^2)$, we have that in the matter domination era, the modified Friedmann equation is written as
\begin{equation}
    E(z) = \sqrt{\Omega_{r}(z) + \Omega_{b}(z) + \Omega_{cdm}(z)\left[e^{-u_0(1+z)^{-3/2}} - u_0(1+z)^{-3/2}\right] + \Omega_{\Lambda_{eff}}}\, .
    \label{fried_MD}
\end{equation}

\subsection{$\mathbf{s=-1/2}:$}
In the case of the radiation dominated era, Eq.~\eqref{gen_gamma} reads as follows,
\begin{equation}
    \gamma^G(-1/2,u) = -\frac{1}{u^{1/2}}\sum_{k=0}^\infty \frac{(-u)^k}{k!(k-1/2)}\simeq  \frac{2}{u^{1/2}} +2u^{1/2} - \frac{u^{3/2}}{3} + \frac{u^{5/2}}{15} +\dots\, .
\end{equation}

When the above expression is written in terms of the redshift we have
\begin{equation}
    \gamma^G(-1/2,z) \simeq \frac{2(1+z)}{u_0^{1/2}} + \frac{2u_0^{1/2}}{1+z} - \frac{u_0^{3/2}}{3(1+z)^3} + \frac{u_0^{5/2}}{15(1+z)^5} + \dots\, ,
\end{equation}
which leads to the integral~\eqref{I_complete}
\begin{equation}
    \xi_{CSL}I_{-1/2}(z) \simeq -2u_0(1+z) - \frac{2u_0^2}{1+z} + \frac{u_0^3}{3(1+z)^3} - \frac{u_0^4}{15(1+z)^5}+\dots\, .
\end{equation}
where we have already included the multiplicative factor $\xi_{CSL}$. Again, neglecting terms of order $\mathcal{O}(u_0^2)$ and higher, the Friedmann equation is written as
\begin{equation}
    E(z) = \sqrt{\Omega_{r}(z) + \Omega_{b}(z) + \Omega_{cdm}(z)\left[e^{-u_0(1+z)^{-2}} + 2u_0(1+z)^{-2}\right] + \Omega_{\Lambda_{eff}}}\, .
    \label{fried_RD}
\end{equation}

Therefore, the Friedmann equation for the CSL model, can be written as Eq.~\eqref{fried_MD} and~\eqref{fried_RD} for matter domination and radiation domination respectively, as is shown in Eq.~\eqref{ez_4_gen}.

\section{On the very far future cosmological evolution}\label{appB}

The current dynamics of the Universe is driven by the dark energy component. According to the $\Lambda$CDM model, such component is described by the Cosmological Constant, $\Lambda$. In such scenario, the fate of the Universe is such that the expansion will continue exponentially, $a(t)\propto e^{\sqrt{\Lambda/3}t}$, leading to an ever-expanding Universe. Here, we explore the cosmological scenario at the very far future in the context of Unimodular Gravity. In particular, we want to determine whether the diffusion models we have studied lead to a different fate of the Universe as that predicted by $\Lambda$CDM.

To study the properties of spacetime in a rigorous way, a set of \textit{singularity theorems} have been established in General Relativity~\cite{Hawking:1973uf,Wald:1984rg}. Whereas one may think about singularities when some geometric quantity, as the curvature tensor, blows up, a more dynamical criterion is that of tracking observers along their trajectories. The concept behind such criterion is that of \textit{geodesic completeness} ($g-$completeness): every geodesic can be extended to arbitrary values of its affine parameter. If this is not the case, we say that the spacetime is geodesically incomplete, and then such spacetime has a singularity. Analysis in this line of ideas have been made for FRW universe in General Relativity and Modified Gravity theories~\cite{FernandezJambrina:2004yy,FernandezJambrina:2006hj,FernandezJambrina:2008dt}.

We will focus on timelike geodesics described by non-relativistic matter, such as baryons and Cold Dark Matter. This will make us to know the behavior of the cosmological evolution at the very far future for the diffusion models.


\subsection{Geodesics in Unimodular Gravity}

The Bianchi identities applied to the Einstein field equations of General Relativity, directly imply $\nabla_{\mu}T^{\mu}_{\ \ \nu}=0$. In the case of dust, this expression tell us that dust particles move on geodesics~\cite{Wald:1984rg}. The scenario is different in Unimodular Gravity, as can be seen in Eq.~\eqref{eff_cc}, where we have,
\begin{equation}
    \nabla^{\nu}T_{\mu \nu}=\frac{1}{\kappa^2}J_{\mu}\, , \quad {\rm{with}}\quad J_{\mu} \equiv \frac{1}{4}\nabla_{\mu}\left( R+\kappa^2 T  \right)\, .
    \label{f_nu}
\end{equation}

At this point, the question is: what are the geodesic equations in Unimodular Gravity, and how do they compare with those of General Relativity? Answering this will lead us to understand the possible future singularities for the diffusion models we have studied.

In what follows, we redefine $J_{\mu}/\kappa^2 \rightarrow J_{\mu}$. Let us project Eq.~\eqref{f_nu} into its parallel and orthogonal components respectively, with respect to the 4-velocity of the fluid $u^{\mu}$, this is,
\begin{subequations}
\begin{eqnarray}
    u^{\mu}\nabla^{\nu}T_{\mu \nu} &=& u^{\mu}J_{\mu}\, ,\label{par} \\
    h^{\mu \beta}\nabla^{\nu}T_{\mu \nu} &=& h^{\mu \beta}J_{\mu}\, ,\label{per}
\end{eqnarray}
\end{subequations}
where $T_{\mu \nu}$ is the energy-momentum tensor for a perfect fluid,
\begin{equation}
    T_{\mu \nu} = \rho u_{\mu}u_{\nu} + p(g_{\mu \nu} + u_{\mu}u_{\nu})\, ,
    \label{emt_pf}
\end{equation}
and the metric tensor $h_{\beta \mu}$ is given by
\begin{equation}
    h_{\beta \mu} = g_{\beta \mu} + u_{\beta}u_{\mu}\, ,
    \label{hmunu}
\end{equation}
and then, it is orthogonal to the 4-velocity $u^{\mu}$, this is, $u^{\mu}h_{\beta \mu} = 0$, where $u^{\mu}$ obeys $g_{\mu \nu}u^{\mu}u^{\nu}=-1$. In particular, by choosing comoving observers, we have
\begin{equation}
    u^{\mu} = \left( \frac{1}{\sqrt{\mid g_{00}\mid}}\, , 0\, , 0\, , 0  \right)\, .
    \label{4vel}
\end{equation}
Replacing~\eqref{emt_pf} into~\eqref{par} and~\eqref{per} we have
\begin{subequations}
\begin{eqnarray}
    u^{\mu}\nabla_{\mu}\rho + (\rho +p)\nabla_{\mu}u^{\mu} &=& -u^{\mu}J_{\mu}\, , \\
    h^{\mu \beta}\nabla_{\mu}p + (\rho+p)u^{\mu}\nabla_{\mu}u^{\beta} &=& h^{\mu \beta}J_{\mu}\, ,
\end{eqnarray}
\end{subequations}

As we have considered in Section~\ref{back}, baryons and $cdm$ behave as non-relativistic matter, and then they are pressureless. Then, the equations from above get reduced as,
\begin{subequations}
\begin{eqnarray}
    u^{\mu}\nabla_{\mu}\rho + \rho\nabla_{\mu}u^{\mu} &=& -u^{\mu}J_{\mu}\, ,\label{gen_cons_eq} \\
    \rho u^{\mu}\nabla_{\mu}u^{\beta} &=& h^{\mu \beta}J_{\mu}\, .\label{gen_geo_eq}
\end{eqnarray}
\end{subequations}

Notice that in the case of General Relativity,  $J_{\mu}=0$, and Eq.~\eqref{gen_cons_eq} is the equation for local conservation of matter, whereas the geodesic equation is obtained from Eq.~\eqref{gen_geo_eq},
\begin{equation}
    u^{\mu}\nabla_{\mu}u^{\beta} =u^{\mu}\left( \partial_{\mu}u^{\beta} + \Gamma^{\beta}_{\ \ \mu \alpha}u^{\alpha} \right) = u^{\mu} \partial_{\mu}u^{\beta} + \Gamma^{\beta}_{\ \ \mu \alpha}u^{\mu}u^{\alpha} = 0\, ,\quad \Rightarrow \quad \frac{du^{\beta}}{d\tau}+ \Gamma^{\beta}_{\ \ \mu \alpha}u^{\mu}u^{\alpha} = 0\, ,
    \label{gr_geo_eq}
\end{equation}
where $\Gamma^{\beta}_{\ \ \mu \alpha}$ are the Christoffel symbols, and $\tau$ is the proper time, which play the role of the affine parameter. In Unimodular Gravity, however, we have in general that $J_{\mu}\neq 0$, and the corresponding geodesic equations are given by
\begin{equation}
    u^{\mu}\nabla_{\mu}u^{\beta} = \frac{h^{\mu \beta}J_{\mu}}{\rho}\, ,
    \label{ug_geo_eq}
\end{equation}
and an external force-like term is present due to the non-gravitational interaction encoded in the energy-momentum current violation $J_{\mu}$. Therefore, we announce the following general result:\\

``\textit{a massive test particle in Unimodular Gravity will deviate its trajectory from the standard geodesic equations of General Relativity due to the non-gravitational interactions in the matter sector.}''\\

In the particular case of a cosmological context, a homogeneous and isotropic spacetime is described by the FRW metric. Specifically, in a (flat) FRW universe, the Ricci scalar $R$ and $T=g^{\mu \nu}T_{\mu \nu}$ are respectively given by $R=6(\dot{H}+2H^2) \, ,$ and $T=3p-\rho \Rightarrow T=-\rho$ for dust. Using the Friedmann equations~\eqref{friedmann} and~\eqref{fried_acc} considering as matter content only baryons and $cdm$ ($\rho=\rho_b+\rho_{cdm}$, and $p_b=p_{cdm}=0$), we obtain $R=\kappa^2\rho + 4\Lambda(t)$. Thus, the explicit form of $J_{\mu}$ in this case is,
\begin{equation}
    J_{\mu} = \frac{1}{4}\nabla_{\mu}(R+\kappa^2 T) = \frac{1}{4}\nabla_{\mu}\left[ 4\Lambda(t) \right] \quad \Rightarrow \quad J_{\mu} = \left[\dot{\Lambda}(t)\, , 0\, , 0\, , 0\right]\, ,
    \label{J0}
\end{equation}

Thus, the energy-momentum current violation $J_{\mu}$ is parallel to the 4-velocity $u^{\mu}$. On the other hand, and as we said before, $h^{\mu \beta}$ is orthogonal to the 4-velocity $u^{\mu}$. Therefore, from the right-hand side of Eq.~\eqref{ug_geo_eq} we obtain
\begin{equation}
    h^{\mu \beta}J_{\mu} = h^{0 \beta}J_{0} + h^{i \beta}J_{i} = 0\, ,
\end{equation}
and then, the geodesic equations in Unimodular Gravity given by Eq.~\eqref{ug_geo_eq} obey $u^{\mu}\nabla_{\mu}u^{\beta}=0$. Therefore, we conclude the following:\\

``\textit{in a FRW spacetime, a massive test particle in Unimodular Gravity will follow the same geodesic equations as those in General Relativity. This is a consequence of homogeneity and isotropy.}''\\

This result allow us to use the well-known geodesics in a FRW spacetime in General Relativity, to study the future behavior of the Universe according to the diffusion models in Unimodular Gravity.

\subsection{Analytical solutions at late times}

Now, we want to explore whether exist some finite time $t_s$ for which a future singularity might occur. It has been shown that the timelike geodesic equations for a (flat) FRW metric, parameterized by its proper time $\tau$, can be written as~\cite{FernandezJambrina:2004yy},
\begin{equation}\label{geo_eq}
    \dot{t}^2 = 1 + \frac{P^2}{a(t)}\, , \quad
    \dot{r} = \frac{P_1\cos \phi + P_2\sin \phi}{a(t)}\, , \quad
    \dot{\phi} = \frac{L_3}{a(t)r^2}\, ,
\end{equation}
where $P_1\, , P_2\, , P\, ,$ and $L_3$ are constants of geodesic motion, and the dot indicates derivative with respect to the proper time $\tau$. It can be seen that these equations are divergent when $a(t)=0$, i.e., Big Bang (past singularity) or Big Crunch (future singularity). Thus, while the scale factor does not go to zero, the 4-velocities will remain finite and the spacetime is nonsingular.

Consider the Friedmann equations~\eqref{ez_1},~\eqref{ez_2},~\eqref{ez_3}, and~\eqref{ez_4_gen} at late times, where the only dominant components are $\Omega_{\Lambda}$ ($\Omega_{\Lambda_{eff}}$ for Model 4) and $\Omega_M= \Omega_{b} + \Omega_{cdm}$. Thus, the analytical solutions at late times have the form,
\begin{equation}
    a(t) = \left[ A\sinh^2\left(Bt \right) \right]^C\, ,
    \label{late_time_sol}
\end{equation}
where the constants $A\, , B\, , C$ are given in Table~\ref{parameters_analy_sol} for each model. Notice that these solutions are finite at $t=t_s$, i.e., $a=a_s\neq 0 <\infty$ when $t=t_s$. In this sense, the scale factor~\eqref{late_time_sol} possesses the same properties as those in~\cite{Barrow:2004xh,Barrow:2004hk}, where a power expansion of the scale factor is proposed. 

\begin{table}[h!]
\centering
\begin{tabular}{cccccc}
\hline
\hline
model  & A & B & C \\
\hline
\hline
$\Lambda$CDM & $\frac{\Omega_{M}}{1 - \Omega_{M}}$ & $\frac{3}{2}\sqrt{ 1 - \Omega_{M}}H_0$ & $\frac{1}{3}$ \\
\hline
Model 1 & $\frac{\Omega_{M}(1-\alpha)}{1 - \Omega_{M}(1-\alpha)}$ & $\frac{3}{2}\sqrt{ 1 - \Omega_{M}(1-\alpha)}H_0$ & $\frac{1}{3}$  \\
\hline
Model 2 & $\frac{3\Omega_{M}}{(3+\gamma)(1+z^{\star})^{\gamma} - 3\Omega_M}$ & $\frac{3+\gamma}{2}\sqrt{1 - \frac{3 \Omega_{M}}{(3+\gamma)(1+z^{\star})^{\gamma}}}H_0$  & $\frac{1}{3+\gamma}$  \\
\hline
Model 3 & $\frac{(x_{cdm} + 1)\Omega_{M}}{1 - (x_{cdm} + 1)\Omega_{M}}$ & $\frac{3}{2(x_{cdm} + 1)}\sqrt{ 1 - (x_{cdm} + 1)\Omega_{M}}H_0$  & $\frac{x_{cdm}+1}{3}$  \\
\hline
Model 4 & $\frac{\Omega_{M}}{1 - \Omega_{M}(1-2u_0)}$ & $\frac{3}{2}\sqrt{ 1 - \Omega_{M}(1-2u_0)}H_0$  & $\frac{1}{3}$  \\
\hline
\end{tabular}
\caption{Values of the constants $A\, , B\, , C$ from Eq.~\eqref{late_time_sol} for each model.}
\label{parameters_analy_sol}
\end{table}

A classification of different future singularities can be given by (see~\cite{BeltranJimenez:2016dfc} and references therein),
\begin{itemize}
    \item Big Rip singularity: when $t\rightarrow t_s$, then $a\rightarrow \infty\, , \rho \rightarrow \infty\, , \mid p\mid \rightarrow \infty\, .$
    \item Sudden singularity: when $t\rightarrow t_s$, then $a\rightarrow a_s\, , \rho \rightarrow \rho_s\, , \mid p\mid \rightarrow \infty\, .$
    \item Big Freeze singularity: when $t\rightarrow t_s$, then $a\rightarrow a_s\, , \rho \rightarrow \infty\, , \mid p\mid \rightarrow \infty\, .$
    \item Generalized Sudden singularity: when $t\rightarrow t_s$, then $a\rightarrow a_s\, , \rho \rightarrow \rho_s\, , \mid p\mid \rightarrow p_s\, .$
    \item $\omega$ singularity: when $t\rightarrow t_s$, then $a\rightarrow \infty\, , \rho \rightarrow 0\, , \mid p\mid \rightarrow 0\, , \omega \rightarrow \infty\, .$
\end{itemize}

In our case, the matter content consists of a pressureless fluid, and the energy density $\rho_{cdm}$ remains finite at $t=t_s$ for all diffusion models. Then, a possibility for the diffusion models to lead to a singular spacetime, is through a Generalized Sudden singularity. However, second and higher derivatives of the Hubble parameter must diverge at $t_s$ in order to present unavoidable singularities of this kind~\cite{Barrow:2004xh,Barrow:2004hk,Dabrowski:2013sea}. Taking derivatives of Eq.~\eqref{late_time_sol} it can be seen that $H(t)\, , \dot{H}(t)\, ,$ and higher derivatives will remain finite at $t_s$. Therefore, independently of the values of the diffusion parameters, the very far future cosmological evolution will be similar of that of the $\Lambda$CDM.

In~\cite{BeltranJimenez:2016dfc}, the authors analyze a new kind of singularity, dubbed $Q$-\textit{singularity}, where the interaction term $Q$ between dark energy and dark matter is written in terms of the Hubble parameter and its derivatives (see Eq.(20) in such work). Thus, divergences in $H(t)\, , \dot{H}(t)$ and higher derivatives can be translated into divergences on the interaction term $Q(t)$. Following this idea, in our case the effective cosmological constant can be written from~ Eq.\eqref{friedmann} and Eq.~\eqref{fried_acc} as,
\begin{equation}
    \Lambda(t) = 3H^2(t) + 2\dot{H}(t)\, ,
\end{equation}
where again, we have considered only non relativistic matter. Thus, since $H(t)$ and its derivatives remain nonsingular at finite future time, the diffusion processes due to the non-gravitational interactions in the dark sector we have analyzed in the context of Unimodular Gravity, do not lead to future singularities. This is consistent with our previous analysis in terms of the geodesic completeness.

\bibliographystyle{plunsrt}
\bibliography{references}  

\begin{thebibliography}{100}

\bibitem{Will:2014kxa}
Clifford~M. Will.
\newblock {The Confrontation between General Relativity and Experiment}.
\newblock {\em Living Rev. Rel.}, 17:4, 2014.

\bibitem{Will:2018lln}
Clifford~M. Will.
\newblock {Putting General Relativity to the Test: Twentieth-Century Highlights
  and Twenty-First-Century Prospects}.
\newblock {\em Einstein Stud.}, 14:81--96, 2018.

\bibitem{Riess:1998cb}
Adam~G. Riess,  i~in.
\newblock {Observational evidence from supernovae for an accelerating universe
  and a cosmological constant}.
\newblock {\em Astron. J.}, 116:1009--1038, 1998.

\bibitem{Perlmutter:1998np}
S.~Perlmutter,  i~in.
\newblock {Measurements of Omega and Lambda from 42 high redshift supernovae}.
\newblock {\em Astrophys. J.}, 517:565--586, 1999.

\bibitem{Eisenstein:2005su}
Daniel~J. Eisenstein,  i~in.
\newblock {Detection of the Baryon Acoustic Peak in the Large-Scale Correlation
  Function of SDSS Luminous Red Galaxies}.
\newblock {\em Astrophys. J.}, 633:560--574, 2005.

\bibitem{Parkinson:2012vd}
David Parkinson,  i~in.
\newblock {The WiggleZ Dark Energy Survey: Final data release and cosmological
  results}.
\newblock {\em Phys. Rev.}, D86:103518, 2012.

\bibitem{Aghanim:2018eyx}
N.~Aghanim,  i~in.
\newblock {Planck 2018 results. VI. Cosmological parameters}.
\newblock {\em Astron. Astrophys.}, 641:A6, 2020.

\bibitem{Carturan:2002si}
Daniela Carturan,  Fabio Finelli.
\newblock {Cosmological effects of a class of fluid dark energy models}.
\newblock {\em Phys. Rev. D}, 68:103501, 2003.

\bibitem{Cardone:2005ut}
Vincenzo~F. Cardone, C.~Tortora, A.~Troisi, S.~Capozziello.
\newblock {Beyond the perfect fluid hypothesis for dark energy equation of
  state}.
\newblock {\em Phys. Rev. D}, 73:043508, 2006.

\bibitem{Nojiri:2006zh}
Shin'ichi Nojiri,  Sergei~D. Odintsov.
\newblock {The New form of the equation of state for dark energy fluid and
  accelerating universe}.
\newblock {\em Phys. Lett. B}, 639:144--150, 2006.

\bibitem{Brevik:2007jt}
Iver~H. Brevik, O.G. Gorbunova, A.V. Timoshkin.
\newblock {Dark energy fluid with time-dependent, inhomogeneous equation of
  state}.
\newblock {\em Eur. Phys. J. C}, 51:179--183, 2007.

\bibitem{Linder:2008ya}
Eric~V. Linder,  Robert~J. Scherrer.
\newblock {Aetherizing Lambda: Barotropic Fluids as Dark Energy}.
\newblock {\em Phys. Rev. D}, 80:023008, 2009.

\bibitem{Duan:2011jj}
Xiaoxian Duan, Yichao Li, Changjun Gao.
\newblock {Constraining the Lattice Fluid Dark Energy from SNe Ia, BAO and
  OHD}.
\newblock {\em Sci. China Phys. Mech. Astron.}, 56:1220--1226, 2013.

\bibitem{Bini:2014pmk}
Donato Bini, Andrea Geralico, Daniele Gregoris, Sauro Succi.
\newblock {Dark energy from cosmological fluids obeying a Shan-Chen nonideal
  equation of state}.
\newblock {\em Phys. Rev. D}, 88(6):063007, 2013.

\bibitem{Barrera-Hinojosa:2019yyh}
Cristian Barrera-Hinojosa,  Domenico Sapone.
\newblock {Relativistic effects in the large-scale structure with effective
  dark energy fluids}.
\newblock {\em JCAP}, 03:037, 2020.

\bibitem{Caldwell:1999ew}
R.~R. Caldwell.
\newblock {A Phantom menace?}
\newblock {\em Phys. Lett.}, B545:23--29, 2002.

\bibitem{Caldwell:2003vq}
Robert~R. Caldwell, Marc Kamionkowski, Nevin~N. Weinberg.
\newblock {Phantom energy and cosmic doomsday}.
\newblock {\em Phys. Rev. Lett.}, 91:071301, 2003.

\bibitem{Nojiri:2005sx}
Shin'ichi Nojiri, Sergei~D. Odintsov, Shinji Tsujikawa.
\newblock {Properties of singularities in (phantom) dark energy universe}.
\newblock {\em Phys. Rev.}, D71:063004, 2005.

\bibitem{Feng:2006ya}
Bo~Feng.
\newblock {The quintom model of dark energy}.
\newblock {\em {Proceedings, fifteenth Workshop on General Relativity and
  Gravitation in Japan, JGRG 15, Tokyo Institute of Technology, Tokyo, Japan,
  November 28 - December 2, 2005}}, 2006.

\bibitem{Linder:2007wa}
Eric~V. Linder.
\newblock {The Dynamics of Quintessence, The Quintessence of Dynamics}.
\newblock {\em Gen. Rel. Grav.}, 40:329--356, 2008.

\bibitem{Setare:2008sf}
M.~R. Setare,  E.~N. Saridakis.
\newblock {Quintom dark energy models with nearly flat potentials}.
\newblock {\em Phys. Rev.}, D79:043005, 2009.

\bibitem{Tsujikawa:2013fta}
Shinji Tsujikawa.
\newblock {Quintessence: A Review}.
\newblock {\em Class. Quant. Grav.}, 30:214003, 2013.

\bibitem{Chiba:2012cb}
Takeshi Chiba, Antonio De~Felice, Shinji Tsujikawa.
\newblock {Observational constraints on quintessence: thawing, tracker, and
  scaling models}.
\newblock {\em Phys. Rev.}, D87(8):083505, 2013.

\bibitem{Linde:2015uga}
Andrei Linde.
\newblock {Single-field $\alpha$-attractors}.
\newblock {\em JCAP}, 1505:003, 2015.

\bibitem{Linder:2015qxa}
Eric~V. Linder.
\newblock {Dark Energy from $\alpha$-Attractors}.
\newblock {\em Phys. Rev.}, D91(12):123012, 2015.

\bibitem{Durrive:2018quo}
Jean-Baptiste Durrive, Junpei Ooba, Kiyotomo Ichiki, Naoshi Sugiyama.
\newblock {Updated observational constraints on quintessence dark energy
  models}.
\newblock {\em Phys. Rev.}, D97(4):043503, 2018.

\bibitem{Bag:2017vjp}
Satadru Bag, Swagat~S. Mishra, Varun Sahni.
\newblock {New tracker models of dark energy}.
\newblock {\em JCAP}, 08:009, 2018.

\bibitem{Leon:2018lnd}
Genly Leon, Andronikos Paliathanasis, Jorge~Luis Morales-Martínez.
\newblock {The past and future dynamics of quintom dark energy models}.
\newblock {\em Eur. Phys. J.}, C78(9):753, 2018.

\bibitem{Garcia-Garcia:2018hlc}
Carlos Garc\'\i{}a-Garc\'\i{}a, Eric~V. Linder, Pilar Ru\'\i{}z-Lapuente,
  Miguel Zumalac\'arregui.
\newblock {Dark energy from $\alpha$-attractors: phenomenology and
  observational constraints}.
\newblock {\em JCAP}, 08:022, 2018.

\bibitem{Alestas:2020mvb}
G.~Alestas, L.~Kazantzidis, L.~Perivolaropoulos.
\newblock {$H_0$ tension, phantom dark energy, and cosmological parameter
  degeneracies}.
\newblock {\em Phys. Rev. D}, 101(12):123516, 2020.

\bibitem{Lobo:2008sg}
Francisco S.~N. Lobo.
\newblock {The Dark side of gravity: Modified theories of gravity}.
\newblock 173-204:Research Signpost, ISBN 978--81--308--0341--8, 2009.

\bibitem{Clifton:2011jh}
Timothy Clifton, Pedro~G. Ferreira, Antonio Padilla, Constantinos Skordis.
\newblock {Modified Gravity and Cosmology}.
\newblock {\em Phys. Rept.}, 513:1--189, 2012.

\bibitem{Dimitrijevic:2012kb}
Ivan Dimitrijevic, Branko Dragovich, Jelena Grujic, Zoran Rakic.
\newblock {On Modified Gravity}.
\newblock {\em Springer Proc. Math. Stat.}, 36:251--259, 2013.

\bibitem{Brax:2015cla}
Philippe Brax,  Anne-Christine Davis.
\newblock {Distinguishing modified gravity models}.
\newblock {\em JCAP}, 1510(10):042, 2015.

\bibitem{Joyce:2016vqv}
Austin Joyce, Lucas Lombriser, Fabian Schmidt.
\newblock {Dark Energy Versus Modified Gravity}.
\newblock {\em Ann. Rev. Nucl. Part. Sci.}, 66:95--122, 2016.

\bibitem{Jaime:2018ftn}
Luisa~G. Jaime, Mariana Jaber, Celia Escamilla-Rivera.
\newblock {New parametrized equation of state for dark energy surveys}.
\newblock {\em Phys. Rev.}, D98(8):083530, 2018.

\bibitem{Slosar:2019flp}
Anže Slosar,  i~in.
\newblock {Dark Energy and Modified Gravity}.
\newblock 2019.

\bibitem{Einstein:1917ce}
Albert Einstein.
\newblock {Cosmological Considerations in the General Theory of Relativity}.
\newblock {\em Sitzungsber. Preuss. Akad. Wiss. Berlin (Math. Phys.)},
  1917:142--152, 1917.

\bibitem{Hubble:1929ig}
Edwin Hubble.
\newblock {A relation between distance and radial velocity among extra-galactic
  nebulae}.
\newblock {\em Proc. Nat. Acad. Sci.}, 15:168--173, 1929.

\bibitem{Einstein:1914bx}
Albert Einstein.
\newblock {The Formal Foundation of the General Theory of Relativity}.
\newblock {\em Sitzungsber. Preuss. Akad. Wiss. Berlin (Math. Phys.)},
  1914:1030--1085, 1914.

\bibitem{Einstein:1919gv}
Albert Einstein.
\newblock {Spielen Gravitationsfelder im Aufbau der materiellen
  Elementarteilchen eine wesentliche Rolle?}
\newblock {\em Sitzungsber. Preuss. Akad. Wiss. Berlin (Math. Phys. )},
  1919:349--356, 1919.

\bibitem{Anderson:1971pn}
J.~L. Anderson,  D.~Finkelstein.
\newblock {Cosmological constant and fundamental length}.
\newblock {\em Am. J. Phys.}, 39:901--904, 1971.

\bibitem{Bonder:2018mfz}
Yuri Bonder,  Crist\'obal Corral.
\newblock {Unimodular Einstein--Cartan gravity: Dynamics and conservation
  laws}.
\newblock {\em Phys. Rev. D}, 97(8):084001, 2018.

\bibitem{Corral:2018hxi}
Crist\'obal Corral,  Yuri Bonder.
\newblock {Symmetry algebra in gauge theories of gravity}.
\newblock {\em Class. Quant. Grav.}, 36:045002, 2019.

\bibitem{vanderBij:1981ym}
J.~J. van~der Bij, H.~van Dam, Yee~Jack Ng.
\newblock {The Exchange of Massless Spin Two Particles}.
\newblock {\em Physica}, 116A:307--320, 1982.

\bibitem{Weinberg:1988cp}
Steven Weinberg.
\newblock {The Cosmological Constant Problem}.
\newblock {\em Rev. Mod. Phys.}, 61:1--23, 1989.
\newblock [,569(1988)].

\bibitem{Buchmuller:1988wx}
W.~Buchmuller,  N.~Dragon.
\newblock {Einstein Gravity From Restricted Coordinate Invariance}.
\newblock {\em Phys. Lett.}, B207:292--294, 1988.

\bibitem{Buchmuller:1988yn}
W.~Buchmuller,  N.~Dragon.
\newblock {Gauge Fixing and the Cosmological Constant}.
\newblock {\em Phys. Lett.}, B223:313--317, 1989.

\bibitem{Ng:1990xz}
Y.~Jack Ng,  H.~van Dam.
\newblock {Unimodular Theory of Gravity and the Cosmological Constant}.
\newblock {\em J. Math. Phys.}, 32:1337--1340, 1991.

\bibitem{Smolin:2009ti}
Lee Smolin.
\newblock {The Quantization of unimodular gravity and the cosmological constant
  problems}.
\newblock {\em Phys. Rev. D}, 80:084003, 2009.

\bibitem{Eichhorn:2013xr}
Astrid Eichhorn.
\newblock {On unimodular quantum gravity}.
\newblock {\em Class. Quant. Grav.}, 30:115016, 2013.

\bibitem{Alvarez:2015pla}
E.~\'Alvarez, S.~Gonz\'alez-Mart\'\i{}n, M.~Herrero-Valea, C.P. Mart\'\i{}n.
\newblock {Unimodular Gravity Redux}.
\newblock {\em Phys. Rev. D}, 92(6):061502, 2015.

\bibitem{Bufalo:2015wda}
R.~Bufalo, M.~Oksanen, A.~Tureanu.
\newblock {How unimodular gravity theories differ from general relativity at
  quantum level}.
\newblock {\em Eur. Phys. J. C}, 75(10):477, 2015.

\bibitem{Percacci:2017fsy}
R.~Percacci.
\newblock {Unimodular quantum gravity and the cosmological constant}.
\newblock {\em Found. Phys.}, 48(10):1364--1379, 2018.

\bibitem{Jain:2012gc}
Pankaj Jain, Atul Jaiswal, Purnendu Karmakar, Gopal Kashyap, Naveen~K. Singh.
\newblock {Cosmological implications of unimodular gravity}.
\newblock {\em JCAP}, 11:003, 2012.

\bibitem{Ellis:2013uxa}
George F~R Ellis.
\newblock {The Trace-Free Einstein Equations and inflation}.
\newblock {\em Gen. Rel. Grav.}, 46:1619, 2014.

\bibitem{Barvinsky:2017pmm}
A.O. Barvinsky,  A.~Yu. Kamenshchik.
\newblock {Darkness without dark matter and energy -- generalized unimodular
  gravity}.
\newblock {\em Phys. Lett. B}, 774:59--63, 2017.

\bibitem{Garcia-Aspeitia:2019yni}
Miguel~A. García-Aspeitia, C.~Martínez-Robles, A.~Hernández-Almada, Juan
  Magaña, V.~Motta.
\newblock {Cosmic acceleration in unimodular gravity}.
\newblock {\em Phys. Rev. D}, 99(12):123525, 2019.

\bibitem{Garcia-Aspeitia:2019yod}
Miguel~A. García-Aspeitia, A.~Hernández-Almada, Juan Magaña, V.~Motta.
\newblock {On the birth of the cosmological constant and the reionization era}.
\newblock 12 2019.

\bibitem{Barvinsky:2019qzx}
A.O. Barvinsky,  N.~Kolganov.
\newblock {Inflation in generalized unimodular gravity}.
\newblock {\em Phys. Rev. D}, 100(12):123510, 2019.

\bibitem{Barvinsky:2019agh}
A.O. Barvinsky, N.~Kolganov, A.~Kurov, D.~Nesterov.
\newblock {Dynamics of the generalized unimodular gravity theory}.
\newblock {\em Phys. Rev. D}, 100(2):023542, 2019.

\bibitem{Perez:2020cwa}
Alejandro Perez, Daniel Sudarsky, Edward Wilson-Ewing.
\newblock {Resolving the $H_0$ tension with diffusion}.
\newblock {\em Gen. Rel. Grav.}, 53(1):7, 2021.

\bibitem{Corral:2020lxt}
Cristóbal Corral, Norman Cruz, Esteban González.
\newblock {Diffusion in unimodular gravity: Analytical solutions, late-time
  acceleration, and cosmological constraints}.
\newblock {\em Phys. Rev. D}, 102(2):023508, 2020.

\bibitem{Henneaux:1984ji}
M.~Henneaux,  C.~Teitelboim.
\newblock {THE COSMOLOGICAL CONSTANT AS A CANONICAL VARIABLE}.
\newblock {\em Phys. Lett. B}, 143:415--420, 1984.

\bibitem{Henneaux:1989zc}
M.~Henneaux,  C.~Teitelboim.
\newblock {The Cosmological Constant and General Covariance}.
\newblock {\em Phys. Lett. B}, 222:195--199, 1989.

\bibitem{Kuchar:1991xd}
Karel~V. Kuchar.
\newblock {Does an unspecified cosmological constant solve the problem of time
  in quantum gravity?}
\newblock {\em Phys. Rev. D}, 43:3332--3344, 1991.

\bibitem{Jirousek:2018ago}
Pavel Jirou\v{s}ek,  Alexander Vikman.
\newblock {New Weyl-invariant vector-tensor theory for the cosmological
  constant}.
\newblock {\em JCAP}, 04:004, 2019.

\bibitem{Hammer:2020dqp}
Katrin Hammer, Pavel Jirousek, Alexander Vikman.
\newblock {Axionic cosmological constant}.
\newblock 1 2020.

\bibitem{Ellis:2010uc}
George F.~R. Ellis, Henk van Elst, Jeff Murugan, Jean-Philippe Uzan.
\newblock {On the Trace-Free Einstein Equations as a Viable Alternative to
  General Relativity}.
\newblock {\em Class. Quant. Grav.}, 28:225007, 2011.

\bibitem{Josset:2016vrq}
Thibaut Josset, Alejandro Perez, Daniel Sudarsky.
\newblock {Dark Energy from Violation of Energy Conservation}.
\newblock {\em Phys. Rev. Lett.}, 118(2):021102, 2017.

\bibitem{Amendola:1999dr}
Luca Amendola.
\newblock {Perturbations in a coupled scalar field cosmology}.
\newblock {\em Mon. Not. Roy. Astron. Soc.}, 312:521, 2000.

\bibitem{Amendola:1999er}
Luca Amendola.
\newblock {Coupled quintessence}.
\newblock {\em Phys. Rev. D}, 62:043511, 2000.

\bibitem{Billyard:2000bh}
Andrew~P. Billyard,  Alan~A. Coley.
\newblock {Interactions in scalar field cosmology}.
\newblock {\em Phys. Rev. D}, 61:083503, 2000.

\bibitem{Olivares:2005tb}
German Olivares, Fernando Atrio-Barandela, Diego Pavon.
\newblock {Observational constraints on interacting quintessence models}.
\newblock {\em Phys. Rev. D}, 71:063523, 2005.

\bibitem{Olivares:2007rt}
German Olivares, Fernado Atrio-Barandela, Diego Pavon.
\newblock {Dynamics of Interacting Quintessence Models: Observational
  Constraints}.
\newblock {\em Phys. Rev. D}, 77:063513, 2008.

\bibitem{CalderaCabral:2008bx}
Gabriela Caldera-Cabral, Roy Maartens, L.Arturo Urena-Lopez.
\newblock {Dynamics of interacting dark energy}.
\newblock {\em Phys. Rev. D}, 79:063518, 2009.

\bibitem{LopezHonorez:2010ij}
Laura Lopez~Honorez, Olga Mena, Grigoris Panotopoulos.
\newblock {Higher-order coupled quintessence}.
\newblock {\em Phys. Rev. D}, 82:123525, 2010.

\bibitem{Pan:2012ki}
Supriya Pan, Subhra Bhattacharya, Subenoy Chakraborty.
\newblock {An analytic model for interacting dark energy and its observational
  constraints}.
\newblock {\em Mon. Not. Roy. Astron. Soc.}, 452(3):3038--3046, 2015.

\bibitem{Salvatelli:2013wra}
Valentina Salvatelli, Andrea Marchini, Laura Lopez-Honorez, Olga Mena.
\newblock {New constraints on Coupled Dark Energy from the Planck satellite
  experiment}.
\newblock {\em Phys. Rev. D}, 88(2):023531, 2013.

\bibitem{Bolotin:2013jpa}
Yu.~L. Bolotin, A.~Kostenko, O.A. Lemets, D.A. Yerokhin.
\newblock {Cosmological Evolution With Interaction Between Dark Energy And Dark
  Matter}.
\newblock {\em Int. J. Mod. Phys. D}, 24(03):1530007, 2014.

\bibitem{Pan:2016ngu}
Supriya Pan,  G.S. Sharov.
\newblock {A model with interaction of dark components and recent observational
  data}.
\newblock {\em Mon. Not. Roy. Astron. Soc.}, 472(4):4736--4749, 2017.

\bibitem{CarrilloGonzalez:2017cll}
Mariana Carrillo~González,  Mark Trodden.
\newblock {Field Theories and Fluids for an Interacting Dark Sector}.
\newblock {\em Phys. Rev. D}, 97(4):043508, 2018.
\newblock [Erratum: Phys.Rev.D 101, 089901 (2020)].

\bibitem{Barros:2018efl}
Bruno~J. Barros, Luca Amendola, Tiago Barreiro, Nelson~J. Nunes.
\newblock {Coupled quintessence with a $\Lambda$CDM background: removing the
  $\sigma_8$ tension}.
\newblock {\em JCAP}, 01:007, 2019.

\bibitem{Yang:2019uzo}
Weiqiang Yang, Olga Mena, Supriya Pan, Eleonora Di~Valentino.
\newblock {Dark sectors with dynamical coupling}.
\newblock {\em Phys. Rev. D}, 100(8):083509, 2019.

\bibitem{Benetti:2019lxu}
Micol Benetti, Welber Miranda, Humberto~A. Borges, Cassio Pigozzo, Saulo
  Carneiro, Jailson~S. Alcaniz.
\newblock {Looking for interactions in the cosmological dark sector}.
\newblock {\em JCAP}, 12:023, 2019.

\bibitem{Yang:2020tax}
Weiqiang Yang, Eleonora Di~Valentino, Olga Mena, Supriya Pan.
\newblock {Dynamical Dark sectors and Neutrino masses and abundances}.
\newblock {\em Phys. Rev. D}, 102(2):023535, 2020.

\bibitem{Asghari:2020ffe}
Mahnaz Asghari, Shahram Khosravi, Amir Mollazadeh.
\newblock {Perturbation level interacting dark energy model and its consequence
  on late-time cosmological parameters}.
\newblock {\em Phys. Rev. D}, 101(4):043503, 2020.

\bibitem{Johnson:2020gzn}
Joseph~P. Johnson,  S.~Shankaranarayanan.
\newblock {Cosmological perturbations in the interacting dark sector: Mapping
  fields and fluids}.
\newblock {\em Phys. Rev. D}, 103(2):023510, 2021.

\bibitem{Benetti:2021div}
Micol Benetti, Humberto Borges, Cassio Pigozzo, Saulo Carneiro, Jailson
  Alcaniz.
\newblock {Dark sector interactions and the curvature of the Universe in light
  of Planck's 2018 data}.
\newblock 2 2021.

\bibitem{DiValentino:2017iww}
Eleonora Di~Valentino, Alessandro Melchiorri, Olga Mena.
\newblock {Can interacting dark energy solve the $H_0$ tension?}
\newblock {\em Phys. Rev. D}, 96(4):043503, 2017.

\bibitem{Yang:2018uae}
Weiqiang Yang, Ankan Mukherjee, Eleonora Di~Valentino, Supriya Pan.
\newblock {Interacting dark energy with time varying equation of state and the
  $H_0$ tension}.
\newblock {\em Phys. Rev. D}, 98(12):123527, 2018.

\bibitem{Yang:2018euj}
Weiqiang Yang, Supriya Pan, Eleonora Di~Valentino, Rafael~C. Nunes, Sunny
  Vagnozzi, David~F. Mota.
\newblock {Tale of stable interacting dark energy, observational signatures,
  and the $H_0$ tension}.
\newblock {\em JCAP}, 09:019, 2018.

\bibitem{DiValentino:2019ffd}
Eleonora Di~Valentino, Alessandro Melchiorri, Olga Mena, Sunny Vagnozzi.
\newblock {Interacting dark energy in the early 2020s: A promising solution to
  the $H_0$ and cosmic shear tensions}.
\newblock {\em Phys. Dark Univ.}, 30:100666, 2020.

\bibitem{Pan:2019gop}
Supriya Pan, Weiqiang Yang, Eleonora Di~Valentino, Emmanuel~N. Saridakis,
  Subenoy Chakraborty.
\newblock {Interacting scenarios with dynamical dark energy: Observational
  constraints and alleviation of the $H_0$ tension}.
\newblock {\em Phys. Rev. D}, 100(10):103520, 2019.

\bibitem{Gomez-Valent:2020mqn}
Adri\`a G\'omez-Valent, Valeria Pettorino, Luca Amendola.
\newblock {Update on coupled dark energy and the $H_0$ tension}.
\newblock {\em Phys. Rev. D}, 101(12):123513, 2020.

\bibitem{Lucca:2020zjb}
Matteo Lucca,  Deanna~C. Hooper.
\newblock {Shedding light on dark matter-dark energy interactions}.
\newblock {\em Phys. Rev. D}, 102(12):123502, 2020.

\bibitem{Pan:2020bur}
Supriya Pan, Weiqiang Yang, Andronikos Paliathanasis.
\newblock {Non-linear interacting cosmological models after Planck 2018 legacy
  release and the $H_0$ tension}.
\newblock {\em Mon. Not. Roy. Astron. Soc.}, 493(3):3114--3131, 2020.

\bibitem{Freedman:2017yms}
Wendy~L. Freedman.
\newblock {Cosmology at at Crossroads: Tension with the Hubble Constant}.
\newblock {\em Nat. Astron.}, 1:0169, 2017.

\bibitem{Verde:2019ivm}
L.~Verde, T.~Treu, A.~G. Riess.
\newblock {Tensions between the Early and the Late Universe}.
\newblock {\em {Nature Astronomy 2019}}, 2019.

\bibitem{Aiola:2020azj}
Simone Aiola,  i~in.
\newblock {The Atacama Cosmology Telescope: DR4 Maps and Cosmological
  Parameters}.
\newblock {\em JCAP}, 12:047, 2020.

\bibitem{Abbott:2017smn}
T.M.C. Abbott,  i~in.
\newblock {Dark Energy Survey Year 1 Results: A Precise H0 Estimate from DES
  Y1, BAO, and D/H Data}.
\newblock {\em Mon. Not. Roy. Astron. Soc.}, 480(3):3879--3888, 2018.

\bibitem{Philcox:2020vvt}
Oliver~H.E. Philcox, Mikhail~M. Ivanov, Marko Simonovi\'c, Matias Zaldarriaga.
\newblock {Combining Full-Shape and BAO Analyses of Galaxy Power Spectra: A
  1.6\% CMB-independent constraint on H0}.
\newblock {\em JCAP}, 05:032, 2020.

\bibitem{Riess:2019cxk}
Adam~G. Riess, Stefano Casertano, Wenlong Yuan, Lucas~M. Macri, Dan Scolnic.
\newblock {Large Magellanic Cloud Cepheid Standards Provide a 1\% Foundation
  for the Determination of the Hubble Constant and Stronger Evidence for
  Physics beyond $\Lambda$CDM}.
\newblock {\em Astrophys. J.}, 876(1):85, 2019.

\bibitem{Freedman:2019jwv}
Wendy~L. Freedman,  i~in.
\newblock {The Carnegie-Chicago Hubble Program. VIII. An Independent
  Determination of the Hubble Constant Based on the Tip of the Red Giant
  Branch}.
\newblock 7 2019.

\bibitem{Freedman:2020dne}
Wendy~L. Freedman, Barry~F. Madore, Taylor Hoyt, In~Sung Jang, Rachael Beaton,
  Myung~Gyoon Lee, Andrew Monson, Jill Neeley, Jeffrey Rich.
\newblock {Calibration of the Tip of the Red Giant Branch (TRGB)}.
\newblock 2 2020.

\bibitem{Huang:2019yhh}
Caroline~D. Huang, Adam~G. Riess, Wenlong Yuan, Lucas~M. Macri, Nadia~L.
  Zakamska, Stefano Casertano, Patricia~A. Whitelock, Samantha~L. Hoffmann,
  Alexei~V. Filippenko, Daniel Scolnic.
\newblock {Hubble Space Telescope Observations of Mira Variables in the Type Ia
  Supernova Host NGC 1559: An Alternative Candle to Measure the Hubble
  Constant}.
\newblock 8 2019.

\bibitem{Wong:2019kwg}
Kenneth~C. Wong,  i~in.
\newblock {H0LiCOW \textendash{} XIII. A 2.4 per cent measurement of H0 from
  lensed quasars: 5.3\ensuremath{\sigma} tension between early- and
  late-Universe probes}.
\newblock {\em Mon. Not. Roy. Astron. Soc.}, 498(1):1420--1439, 2020.

\bibitem{Pesce:2020xfe}
D.W. Pesce,  i~in.
\newblock {The Megamaser Cosmology Project. XIII. Combined Hubble constant
  constraints}.
\newblock {\em Astrophys. J. Lett.}, 891(1):L1, 2020.

\bibitem{potter2018calibrating}
Cicely Potter, Joseph~B Jensen, John Blakeslee, Peter Milne, Peter~M Garnavich,
  Peter Brown.
\newblock Calibrating the type ia supernova distance scale using surface
  brightness fluctuations.
\newblock {\em American Astronomical Society Meeting Abstracts\# 232}, wolumen
  232, 2018.

\bibitem{vivien_bonvin_2020_3635517}
Vivien Bonvin,  Martin Millon.
\newblock H0licow h0 tension plotting notebook, Luty 2020.

\bibitem{Calogero:2013zba}
Simone Calogero,  Hermano Velten.
\newblock {Cosmology with matter diffusion}.
\newblock {\em JCAP}, 1311:025, 2013.

\bibitem{Haba:2016swv}
Zbigniew Haba, Aleksander Stachowski, Marek Szyd\l~owski.
\newblock {Dynamics of the diffusive DM-DE interaction -- Dynamical system
  approach}.
\newblock {\em JCAP}, 07:024, 2016.

\bibitem{Benisty:2017rbw}
David Benisty,  E.I. Guendelman.
\newblock {Unified DE--DM with diffusive interactions scenario from scalar
  fields}.
\newblock {\em Int. J. Mod. Phys. D}, 26(12):1743021, 2017.

\bibitem{Benisty:2017eqh}
David Benisty,  E.~I. Guendelman.
\newblock {Interacting Diffusive Unified Dark Energy and Dark Matter from
  Scalar Fields}.
\newblock {\em Eur. Phys. J.}, C77(6):396, 2017.

\bibitem{Benisty:2018oyy}
David Benisty, Eduardo Guendelman, Zbigniew Haba.
\newblock {Unification of dark energy and dark matter from diffusive
  cosmology}.
\newblock {\em Phys. Rev.}, D99(12):123521, 2019.
\newblock [Erratum: Phys. Rev.D101,no.4,049901(2020)].

\bibitem{Lesgourgues:2011re}
Julien Lesgourgues.
\newblock {The Cosmic Linear Anisotropy Solving System (CLASS) I: Overview}.
\newblock 2011.

\bibitem{Pearle:1976ka}
Philip~M. Pearle.
\newblock {Reduction of the State Vector by a Nonlinear Schrodinger Equation}.
\newblock {\em Phys. Rev. D}, 13:857--868, 1976.

\bibitem{Pearle:1988uh}
Philip~M. Pearle.
\newblock {Combining Stochastic Dynamical State Vector Reduction With
  Spontaneous Localization}.
\newblock {\em Phys. Rev. A}, 39:2277--2289, 1989.

\bibitem{Ghirardi:1989cn}
Gian~Carlo Ghirardi, Philip~M. Pearle, Alberto Rimini.
\newblock {Markov Processes in Hilbert Space and Continuous Spontaneous
  Localization of Systems of Identical Particles}.
\newblock {\em Phys. Rev. A}, 42:78--79, 1990.

\bibitem{Bassi:2003gd}
Angelo Bassi,  Gian~Carlo Ghirardi.
\newblock {Dynamical reduction models}.
\newblock {\em Phys. Rept.}, 379:257, 2003.

\bibitem{Perez:2005gh}
Alejandro Perez, Hanno Sahlmann, Daniel Sudarsky.
\newblock {On the quantum origin of the seeds of cosmic structure}.
\newblock {\em Class. Quant. Grav.}, 23:2317--2354, 2006.

\bibitem{Lochan:2012di}
Kinjalk Lochan, Suratna Das, Angelo Bassi.
\newblock {Constraining CSL strength parameter $\lambda$ from standard
  cosmology and spectral distortions of CMBR}.
\newblock {\em Phys. Rev. D}, 86:065016, 2012.

\bibitem{Martin:2012pea}
Jerome Martin, Vincent Vennin, Patrick Peter.
\newblock {Cosmological Inflation and the Quantum Measurement Problem}.
\newblock {\em Phys. Rev. D}, 86:103524, 2012.

\bibitem{Canate:2013isa}
Pedro Cañate, Philip Pearle, Daniel Sudarsky.
\newblock {Continuous spontaneous localization wave function collapse model as
  a mechanism for the emergence of cosmological asymmetries in inflation}.
\newblock {\em Phys. Rev. D}, 87(10):104024, 2013.

\bibitem{Piccirilli:2017mto}
María~Pía Piccirilli, Gabriel León, Susana~J. Landau, Micol Benetti, Daniel
  Sudarsky.
\newblock {Constraining quantum collapse inflationary models with current data:
  The semiclassical approach}.
\newblock {\em Int. J. Mod. Phys. D}, 28(02):1950041, 2018.

\bibitem{Leon:2017yna}
Gabriel León, Abhishek Majhi, Elias Okon, Daniel Sudarsky.
\newblock {Expectation of primordial gravity waves generated during inflation}.
\newblock {\em Phys. Rev. D}, 98(2):023512, 2018.

\bibitem{Leon:2020sqt}
Gabriel Leon,  Maria~Pia Piccirilli.
\newblock {Generation of inflationary perturbations in the continuous
  spontaneous localization model: The second order power spectrum}.
\newblock {\em Phys. Rev. D}, 102(4):043515, 2020.

\bibitem{Adler_2007}
Stephen~L Adler,  Angelo Bassi.
\newblock Collapse models with non-white noises.
\newblock {\em Journal of Physics A: Mathematical and Theoretical},
  40(50):15083--15098, nov 2007.

\bibitem{Adler_2008}
Stephen~L Adler,  Angelo Bassi.
\newblock Collapse models with non-white noises: {II}. particle-density coupled
  noises.
\newblock {\em Journal of Physics A: Mathematical and Theoretical},
  41(39):395308, sep 2008.

\bibitem{Magana:2017nfs}
Juan Magana, Mario~H. Amante, Miguel~A. Garcia-Aspeitia, V.~Motta.
\newblock {The Cardassian expansion revisited: constraints from updated Hubble
  parameter measurements and type Ia supernova data}.
\newblock {\em Mon. Not. Roy. Astron. Soc.}, 476(1):1036--1049, 2018.

\bibitem{Scolnic:2017caz}
D.M. Scolnic,  i~in.
\newblock {The Complete Light-curve Sample of Spectroscopically Confirmed SNe
  Ia from Pan-STARRS1 and Cosmological Constraints from the Combined Pantheon
  Sample}.
\newblock {\em Astrophys. J.}, 859(2):101, 2018.

\bibitem{Chen:2018dbv}
Lu~Chen, Qing-Guo Huang, Ke~Wang.
\newblock {Distance Priors from Planck Final Release}.
\newblock {\em JCAP}, 1902:028, 2019.

\bibitem{Malekjani:2018qcz}
Mohammad Malekjani, Mehdi Rezaei, Iman~A. Akhlaghi.
\newblock {Can Holographic dark energy models fit the observational data?}
\newblock {\em Phys. Rev.}, D98(6):063533, 2018.

\bibitem{Cedeno:2019cgr}
Francisco X.~Linares Cedeño, Ariadna Montiel, Juan~Carlos Hidalgo, Gabriel
  Germán.
\newblock {Bayesian evidence for $\alpha$-attractor dark energy models}.
\newblock {\em JCAP}, 1908(08):002, 2019.

\bibitem{Rivera:2016zzr}
Alexander Bonilla~Rivera,  Jorge~Enrique García-Farieta.
\newblock {Exploring the Dark Universe: constraints on dynamical Dark Energy
  models from CMB, BAO and growth rate measurements}.
\newblock {\em Int. J. Mod. Phys.}, D28(09):1950118, 2019.

\bibitem{Li:2019san}
Xiaolei Li, Arman Shafieloo, Varun Sahni, Alexei~A. Starobinsky.
\newblock {Revisiting Metastable Dark Energy and Tensions in the Estimation of
  Cosmological Parameters}.
\newblock {\em Astrophys. J.}, 887:153, 2019.

\bibitem{Li:2019ypi}
Xiaolei Li,  Arman Shafieloo.
\newblock {A Simple Phenomenological Emergent Dark Energy Model can Resolve the
  Hubble Tension}.
\newblock {\em Astrophys. J.}, 883(1):L3, 2019.
\newblock [Astrophys. J. Lett.883,L3(2019)].

\bibitem{Davari:2019tni}
Zahra Davari, Valerio Marra, Mohammad Malekjani.
\newblock {Cosmological constrains on minimally and non-minimally coupled
  scalar field models}.
\newblock {\em Mon. Not. Roy. Astron. Soc.}, 491(2):1920--1933, 2020.

\bibitem{Montiel:2020rnd}
Ariadna Montiel, J.~I. Cabrera, Juan~Carlos Hidalgo.
\newblock {Improving sampling and calibration of gamma-ray bursts as distance
  indicators}.
\newblock {\em Mon. Not. Roy. Astron. Soc.}, 501(3):3515--3526, 2021.

\bibitem{Audren:2012wb}
Benjamin Audren, Julien Lesgourgues, Karim Benabed, Simon Prunet.
\newblock {Conservative Constraints on Early Cosmology: an illustration of the
  Monte Python cosmological parameter inference code}.
\newblock {\em JCAP}, 1302:001, 2013.

\bibitem{Gelman:1992zz}
Andrew Gelman,  Donald~B. Rubin.
\newblock {Inference from Iterative Simulation Using Multiple Sequences}.
\newblock {\em Statist. Sci.}, 7:457--472, 1992.

\bibitem{stefan_taubenberger_2020_3632967}
Stefan Taubenberger,  Sherry~H. Suyu.
\newblock H0licow distance likelihoods in montepython, Luty 2020.

\bibitem{Suyu:2009by}
S.H. Suyu, P.J. Marshall, M.W. Auger, S.~Hilbert, R.D. Blandford, L.V.E.
  Koopmans, C.D. Fassnacht, T.~Treu.
\newblock {Dissecting the Gravitational Lens B1608+656. II. Precision
  Measurements of the Hubble Constant, Spatial Curvature, and the Dark Energy
  Equation of State}.
\newblock {\em Astrophys. J.}, 711:201--221, 2010.

\bibitem{Suyu:2013kha}
S.H. Suyu,  i~in.
\newblock {Cosmology from gravitational lens time delays and Planck data}.
\newblock {\em Astrophys. J.}, 788:L35, 2014.

\bibitem{Wong:2016dpo}
Kenneth~C. Wong,  i~in.
\newblock {H0LiCOW -- IV. Lens mass model of HE 0435$-$1223 and blind
  measurement of its time-delay distance for cosmology}.
\newblock {\em Mon. Not. Roy. Astron. Soc.}, 465(4):4895--4913, 2017.

\bibitem{Birrer:2018vtm}
S.~Birrer,  i~in.
\newblock {H0LiCOW - IX. Cosmographic analysis of the doubly imaged quasar SDSS
  1206+4332 and a new measurement of the Hubble constant}.
\newblock {\em Mon. Not. Roy. Astron. Soc.}, 484:4726, 2019.

\bibitem{Jee:2019hah}
Inh Jee, Sherry Suyu, Eiichiro Komatsu, Christopher~D. Fassnacht, Stefan
  Hilbert, Léon~V.E. Koopmans.
\newblock {A measurement of the Hubble constant from angular diameter distances
  to two gravitational lenses}.
\newblock 9 2019.

\bibitem{Chen:2019ejq}
Geoff C.-F. Chen,  i~in.
\newblock {A SHARP view of H0LiCOW: $H_{0}$ from three time-delay gravitational
  lens systems with adaptive optics imaging}.
\newblock {\em Mon. Not. Roy. Astron. Soc.}, 490(2):1743--1773, 2019.

\bibitem{Rusu:2019xrq}
Cristian~E. Rusu,  i~in.
\newblock {H0LiCOW XII. Lens mass model of WFI2033 \ensuremath{-} 4723 and
  blind measurement of its time-delay distance and H0}.
\newblock {\em Mon. Not. Roy. Astron. Soc.}, 498(1):1440--1468, 2020.

\bibitem{Camarena:2018nbr}
David Camarena,  Valerio Marra.
\newblock {Impact of the cosmic variance on $H_0$ on cosmological analyses}.
\newblock {\em Phys. Rev. D}, 98(2):023537, 2018.

\bibitem{Betoule:2014frx}
M.~Betoule,  i~in.
\newblock {Improved cosmological constraints from a joint analysis of the
  SDSS-II and SNLS supernova samples}.
\newblock {\em Astron. Astrophys.}, 568:A22, 2014.

\bibitem{Kessler:2016uwi}
Richard Kessler,  Dan Scolnic.
\newblock {Correcting Type Ia Supernova Distances for Selection Biases and
  Contamination in Photometrically Identified Samples}.
\newblock {\em Astrophys. J.}, 836(1):56, 2017.

\bibitem{Jones:2018vbn}
D.O. Jones,  i~in.
\newblock {Should Type Ia Supernova Distances be Corrected for their Local
  Environments?}
\newblock {\em Astrophys. J.}, 867(2):108, 2018.

\bibitem{Suyu:2018vqs}
Sherry~H. Suyu, Tzu-Ching Chang, Frédéric Courbin, Teppei Okumura.
\newblock {Cosmological distance indicators}.
\newblock {\em Space Sci. Rev.}, 214(5):91, 2018.

\bibitem{Martin:2019jye}
Jérôme Martin,  Vincent Vennin.
\newblock {Cosmic Microwave Background Constraints Cast a Shadow On Continuous
  Spontaneous Localization Models}.
\newblock {\em Phys. Rev. Lett.}, 124(8):080402, 2020.

\bibitem{Dowker:2003hb}
Fay Dowker, Joe Henson, Rafael~D. Sorkin.
\newblock {Quantum gravity phenomenology, Lorentz invariance and discreteness}.
\newblock {\em Mod. Phys. Lett. A}, 19:1829--1840, 2004.

\bibitem{Philpott:2008vd}
Lydia Philpott, Fay Dowker, Rafael~D. Sorkin.
\newblock {Energy-momentum diffusion from spacetime discreteness}.
\newblock {\em Phys. Rev. D}, 79:124047, 2009.

\bibitem{Gao:2014nia}
Caixia Gao, Robert~H. Brandenberger, Yifu Cai, Pisin Chen.
\newblock {Cosmological Perturbations in Unimodular Gravity}.
\newblock {\em JCAP}, 09:021, 2014.

\bibitem{Sachs:1967er}
R.K. Sachs,  A.M. Wolfe.
\newblock {Perturbations of a cosmological model and angular variations of the
  microwave background}.
\newblock {\em Astrophys. J.}, 147:73--90, 1967.

\bibitem{Basak:2015swx}
Abhishek Basak, Ophélia Fabre, S.~Shankaranarayanan.
\newblock {Cosmological perturbations of unimodular gravity and general
  relativity are identical}.
\newblock {\em Gen. Rel. Grav.}, 48(10):123, 2016.

\bibitem{Perez:2019gyd}
Alejandro Perez,  Daniel Sudarsky.
\newblock {Black holes, Planckian granularity, and the changing cosmological
  'constant'}.
\newblock 11 2019.

\bibitem{abramowitz1965handbook}
Milton Abramowitz,  Irene~A Stegun.
\newblock Handbook of mathematical functions dover publications.
\newblock {\em New York}, strona 361, 1965.

\bibitem{Gautschi98theincomplete}
Walter Gautschi.
\newblock The incomplete gamma functions since tricomi.
\newblock {\em In Tricomi's Ideas and Contemporary Applied Mathematics, Atti
  dei Convegni Lincei, n. 147, Accademia Nazionale dei Lincei}, strony
  203--237, 1998.

\bibitem{olver2010nist}
Frank~WJ Olver, Daniel~W Lozier, Ronald~F Boisvert, Charles~W Clark.
\newblock {\em NIST handbook of mathematical functions hardback and CD-ROM}.
\newblock Cambridge university press, 2010.

\bibitem{jameson_2016}
G.~J.~O. Jameson.
\newblock The incomplete gamma functions.
\newblock {\em The Mathematical Gazette}, 100(548):298–306, 2016.

\bibitem{adams1922smithsonian}
Edwin~Plimpton Adams,  Richard~Lionel Hippisley.
\newblock {\em Smithsonian mathematical formulae and tables of elliptic
  functions}, wolumen 2672.
\newblock Smithsonian institution, 1922.

\bibitem{1953hft1.book...59E}
A.~{Erdelyi}.
\newblock {\em {Higher Transcendental Functions}}, strona~59.
\newblock 1953.

\bibitem{gradshteyn2014table}
Izrail~Solomonovich Gradshteyn,  Iosif~Moiseevich Ryzhik.
\newblock {\em Table of integrals, series, and products}.
\newblock Academic press, 2014.

\bibitem{hadamard1923lectures}
J~Hadamard.
\newblock Lectures on cauchy's problem in linear partial differential
  equations, yale univ.
\newblock {\em Press. New Haven}, 1923.

\bibitem{van1959introduction}
JG~Van~der Corput.
\newblock Introduction to the neutrix calculus.
\newblock {\em Journal d’Analyse Math{\'e}matique}, 7(1):281--398, 1959.

\bibitem{fisher1976neutrices}
B~Fisher.
\newblock Neutrices and the product of distributions.
\newblock {\em Studia Mathematica}, 57:263--274, 1976.

\bibitem{Ng:2004tk}
Y.Jack Ng,  H.~van Dam.
\newblock {Neutrix calculus and finite quantum field theory}.
\newblock {\em J. Phys. A}, 38:L317, 2005.

\bibitem{Ng:2005er}
Y.Jack Ng,  H.~van Dam.
\newblock {An Application of neutrix calculus to quantum field theory}.
\newblock {\em Int. J. Mod. Phys. A}, 21:297--312, 2006.

\bibitem{CHAUDHRY199499}
M.Aslam Chaudhry,  S.M. Zubair.
\newblock Generalized incomplete gamma functions with applications.
\newblock {\em Journal of Computational and Applied Mathematics}, 55(1):99 --
  123, 1994.

\bibitem{fisher2003defining}
Brian Fisher, Biljana Jolevsaka-Tuneska, Adem Kili{\c{C}}man.
\newblock On defining the incomplete gamma function.
\newblock {\em Integral Transforms and Special Functions}, 14(4):293--299,
  2003.

\bibitem{fisher2004defining}
Brain Fisher.
\newblock On defining the incomplete gamma function $\gamma$ (- m, x-).
\newblock {\em Integral Transforms and Special Functions}, 15(6):467--476,
  2004.

\bibitem{ozccaug2007some}
Emin {\"O}z{\c{c}}a{\u{g}}, Inci Ege, Ha{\c{s}}met G{\"u}r{\c{c}}ay, Biljana
  Jolevska-Tuneska.
\newblock Some remarks on the incomplete gamma function.
\newblock {\em Mathematical Methods in Engineering}, strony 97--108. Springer,
  2007.

\bibitem{fisher2012some}
Brian Fisher,  Adem K{\i}l{\i}cman.
\newblock Some results on the gamma function for negative integers.
\newblock {\em Appl. Math. Inform. Sci}, 6(2):173--176, 2012.

\bibitem{thompson2013algorithm}
Ian Thompson.
\newblock Algorithm 926: Incomplete gamma functions with negative arguments.
\newblock {\em ACM Transactions on Mathematical Software (TOMS)}, 39(2):1--9,
  2013.

\bibitem{ozcaug2016remarks}
Emin {\"O}zca{\u{g}},  {\.I}nci Ege.
\newblock Remarks on polygamma and incomplete gamma type functions.
\newblock {\em Journal of Number Theory}, 169:369--387, 2016.

\bibitem{lin2019incomplete}
Mongkolsery Lin, Brian Fisher, Somsak Orankitjaroen.
\newblock On the incomplete gamma function and its neutrix convolution for
  negative integers.
\newblock {\em Eurasian Mathematical Journal}, 10(1):30--51, 2019.

\bibitem{bohmer1939differenzengleichung}
Paul~Eugen B{\"o}hmer.
\newblock {\em Differenzengleichung und bestimmte Integrale}.
\newblock KF Koehler, 1939.

\bibitem{tricomi1950asymptotische}
FG~Tricomi.
\newblock Asymptotische eigenschaften der unvollst{\"a}ndigen gammafunktion.
\newblock {\em Mathematische Zeitschrift}, 53(2):136--148, 1950.

\bibitem{gautschi1977evaluation}
Walter Gautschi.
\newblock An evaluation procedure for incomplete gamma functions.
\newblock Raport instytutowy, WISCONSIN UNIV MADISON MATHEMATICS RESEARCH
  CENTER, 1977.

\bibitem{gautschi1979computational}
Walter Gautschi.
\newblock A computational procedure for incomplete gamma functions.
\newblock {\em ACM Transactions on Mathematical Software (TOMS)},
  5(4):466--481, 1979.

\bibitem{thukral2014factorials}
Ashwani~K Thukral.
\newblock Factorials of real negative and imaginary numbers-a new perspective.
\newblock {\em SpringerPlus}, 3(1):658, 2014.

\bibitem{Hawking:1973uf}
S.~W. Hawking,  G.~F.~R. Ellis.
\newblock {\em {The Large Scale Structure of Space-Time}}.
\newblock Cambridge Monographs on Mathematical Physics. Cambridge University
  Press, 2 2011.

\bibitem{Wald:1984rg}
Robert~M. Wald.
\newblock {\em {General Relativity}}.
\newblock Chicago Univ. Pr., Chicago, USA, 1984.

\bibitem{FernandezJambrina:2004yy}
L.~Fernandez-Jambrina,  Ruth Lazkoz.
\newblock {Geodesic behaviour of sudden future singularities}.
\newblock {\em Phys. Rev. D}, 70:121503, 2004.

\bibitem{FernandezJambrina:2006hj}
L.~Fernandez-Jambrina,  R.~Lazkoz.
\newblock {Classification of cosmological milestones}.
\newblock {\em Phys. Rev. D}, 74:064030, 2006.

\bibitem{FernandezJambrina:2008dt}
L.~Fernandez-Jambrina,  Ruth Lazkoz.
\newblock {Singular fate of the universe in modified theories of gravity}.
\newblock {\em Phys. Lett. B}, 670:254--258, 2009.

\bibitem{Barrow:2004xh}
John~D. Barrow.
\newblock {Sudden future singularities}.
\newblock {\em Class. Quant. Grav.}, 21:L79--L82, 2004.

\bibitem{Barrow:2004hk}
John~D. Barrow.
\newblock {More general sudden singularities}.
\newblock {\em Class. Quant. Grav.}, 21:5619--5622, 2004.

\bibitem{BeltranJimenez:2016dfc}
Jose Beltr\'an~Jim\'enez, Diego Rubiera-Garcia, Diego S\'aez-G\'omez, Vincenzo
  Salzano.
\newblock {Cosmological future singularities in interacting dark energy
  models}.
\newblock {\em Phys. Rev. D}, 94(12):123520, 2016.

\bibitem{Dabrowski:2013sea}
Mariusz~P. Dabrowski, Konrad Marosek, Adam Balcerzak.
\newblock {Standard and exotic singularities regularized by varying constants}.
\newblock {\em Mem. Soc. Ast. It.}, 85(1):44--49, 2014.

\end{thebibliography}


\end{document}